\documentclass[prd,nofootinbib,shownopacs,preprint]{revtex4}
\usepackage{graphicx}
\usepackage{epsfig}
\usepackage{amsmath}
\usepackage{amsfonts}
\usepackage{amssymb}
\usepackage{url}
\usepackage{hyperref}
\usepackage{subfigure}
\usepackage{hhline}
\usepackage{color}
\usepackage{bm}
\usepackage{cancel}
\newcommand{\beqa}{\begin{eqnarray}}
\newcommand{\eeqa}{\end{eqnarray}}
\newcommand{\beq}{\begin{equation}}
\newcommand{\eeq}{\end{equation}}

\newcommand{\nn}{\nonumber}
\newcommand{\bmt}{\begin{pmatrix}}
\newcommand{\emt}{\end{pmatrix}}
\usepackage[toc,page]{appendix}
\usepackage{comment}
\newcommand{\be}{\begin{equation}}
\newcommand{\ee}{\end{equation}}
\newcommand{\bea}{\begin{eqnarray}}
\newcommand{\eea}{\end{eqnarray}}

\begin{document}
\title{Probing new physics in semileptonic $\Lambda_b$ decays}

\author{Atasi Ray$^a$}
\email{atasiray92@gmail.com}
\author{Suchismita Sahoo$^b$}
\email{suchismita8792@gmail.com}

\author{Rukmani Mohanta$^a$}
\email{rmsp@uohyd.ernet.in}
\affiliation{$^a$School of Physics,  University of Hyderabad, Hyderabad-500046,  India\\
$^b$Theoretical Physics Division, Physical Research Laboratory, Ahmedabad-380009, India}
\begin{abstract}
In recent times, several hints of lepton non-universality  have  been observed in semileptonic $B$ meson decays, both in the charged-current ($b \to c l \bar \nu_l$) and 
neutral-current ($b \to s ll $) transitions.   
Motivated by these intriguing results, we perform a model independent analysis of  the  semileptonic $\Lambda_b$ decays   involving the  quark level transitions $b \to (u,c) l \nu_l$, in order to scrutinize the nature of new physics. We constrain the new parameter space by using the  measured branching ratios  of $B_{c, u}^+ \to \tau^+ \nu_\tau$, $B \to \pi \tau \nu_\tau$ processes and the existing experimental results on $R_{D^{(*)}}$, $R_{J/\psi}$ and $R_{\pi}^l$  parameters.  Using the constrained parameters,  we estimate the branching ratios, forward backward asymmetries, hadron and lepton polarization asymmetries of the $\Lambda_b \to (\Lambda_c, p) l \nu_l$  processes.  Moreover, we also examine whether there could be any lepton universality violation in these  decay modes. 
\end{abstract}
\maketitle
\section{Introduction}
Though the Standard Model (SM) is considered as the most fundamental theory describing almost all the phenomena of particle physics, still it is unable to shed light on some of the open issues,  like matter-antimatter asymmetry, dark matter, dark energy,  etc., which eventually necessitates to probe the physics beyond  it.  In this respect, the rare decays of $B$ mesons involving the flavor changing neutral current (FCNC) transitions play an important role in the quest for new physics (NP). Even though the SM gauge interactions are lepton flavor universal,  the  violation of lepton universality  has been observed in various  semileptonic $B$ decays.   Recently, the  LHCb Collaboration has reported a spectacular discrepancy of $1.9\sigma~(3.3\sigma)$ \cite{Lees:2012xj, Lees:2013uzd, Huschle:2015rga, Abdesselam:2016cgx, Aaij:2015yra, HFLAV16} and $2\sigma$ \cite{Aaij:2017tyk} on  the lepton non-universality (LNU)  parameters $R_{D^{(*)}}={\rm Br}(\bar B \to \bar D^{(*)} \tau \bar \nu_\tau)/{\rm Br}(\bar B \to \bar D^{(*)} l \bar \nu_l)$ and $R_{J/\psi}={\rm Br}(B_c \to J/\psi  \tau \bar \nu_\tau)/{\rm Br}(B_c \to J/\psi l \bar  \nu_l) $ respectively from their  corresponding SM values.   Analogous LNU parameters are also observed in $b \to s ll$ processes i.e.,  $R_{K^{(*)}}={\rm Br}(\bar B \to \bar K^{(*)} \mu^+ \mu^-)/{\rm Br}(\bar B \to \bar K^{(*)} e^+ e^-)$  with    discrepancies of  $2.6\sigma~(2.2-2.4)\sigma$  \cite{Aaij:2014ora, Aaij:2017vbb}.   The SM predictions as well as the  corresponding experimental values of   various LNU parameters along with their  deviations are presented in Table \ref{LNU-data}\,. 
\begin{table}[htb]
\begin{center}
\caption{List of measured  lepton non-universality parameters.} \label{LNU-data}
\begin{tabular}{|c|c|c|c|}
\hline

LNU parameters~& ~Experimental value~&~ SM prediction ~&~ Deviation~\\
\hline
 $R_K|_{q^2\in [1, 6]~{\rm GeV}^2}  $ ~&~ $0.745^{+0.090}_{-0.074} \pm 0.036 $ \cite{Aaij:2014ora}~&~ $1.003\pm 0.0001$ \cite{Bobeth:2007dw}~ &~ $2.6\sigma$ \\
\hline
 $R_{K^*}|_{q^2\in [0.045,1.1]~{\rm GeV}^2} $~ & ~$0.66^{+0.11}_{-0.07} \pm 0.03 $ \cite{Aaij:2017vbb}~& ~$0.92\pm 0.02$ \cite{Capdevila:2017bsm}~&~$ 2.2\sigma$ \\
 \hline
  $R_{K^*}|_{q^2\in [1.1, 6]~{\rm GeV}^2} $~ & ~$0.69^{+0.11}_{-0.07} \pm 0.05$ \cite{Aaij:2017vbb}~& ~$1.00 \pm 0.01$ \cite{Capdevila:2017bsm}~&~$ 2.4 \sigma$ \\
 \hline
$R_D $ ~& ~$0.391\pm 0.041 \pm 0.028$ \cite{HFLAV16}~& ~$0.300 \pm 0.008$ \cite{Na:2015kha}~ & ~$1.9\sigma$~\\
\hline
 $R_{D^*}$ ~& ~ $0.316\pm0.016\pm 0.010$ \cite{HFLAV16}~ &~$ 0.252 \pm 0.003$ \cite{Fajfer:2012vx, Fajfer:2012jt}~ &~$ 3.3 \sigma$\\
 \hline
 $ R_{J/\psi} $ ~&~$ 0.71\pm 0.17 \pm 0.184$ \cite{Aaij:2017tyk}~ &~ $0.289\pm 0.01$ \cite{Wen-Fei:2013uea, Ivanov:2005fd}~&~ $2\sigma$\\
 \hline
\end{tabular}
\end{center}
\end{table}

In addition, another discrepancy in $b \to u l \bar \nu_l$ transition   is also noticed in the measured ratio
\bea
R_\pi^l = \frac{\tau_{B^0}}{\tau_{B^-}}\frac{{\rm Br}(B^- \to \tau^-  \bar \nu_\tau)}{{\rm Br}(B^0 \to \pi^+ l^- \bar \nu_l)}, ~~~l=e, \mu\,,
\eea
where $\tau_{B^0}~(\tau_{B^-})$ is the life time of $B^0~(B^-)$ meson.  
Using the experimental measured values of the branching ratios of $B_u^- \to \tau^- \bar \nu_\tau$ and  $B^0 \to \pi^+ l^- \bar \nu_l$ decay processes  
\bea
&&{\rm Br}(B_u^- \to \tau^- \bar \nu_\tau)|^{\rm Expt}=(1.09\pm 0.24)\times 10^{-4}, \\
&&{\rm Br}( B^0 \to \pi^+ l^- \bar \nu_l) |^{\rm Expt} =(1.45\pm 0.05)\times 10^{-4}, 
\eea
with $\tau_{B^-}/\tau_{B^0}=1.076\pm 0.004$ from \cite{Patrignani:2016xqp}, one can obtain  
\bea
R_\pi^l |^{\rm Expt} =0.699\pm 0.156,
\eea
which has also nearly $1 \sigma$ deviation from its SM value $R_\pi^l |^{\rm SM}=0.583 \pm 0.055$. It is generally argued that,  compared to the first two generations, the processes involving the third generation leptons are more sensitive to NP due to their reasonably large mass.  As the LNU parameters  are the ratio of branching fractions, the uncertainties arising due to the CKM matrix elements and  hadronic form factors are expected to be  reduced, as they  cancelled out in the ratio. Hence, these deviations of various LNU parameters hint towards the possible interplay of new physics  in an ambiguous manner.

 On the other hand, around  $20\%$  of  the total number of hadrons produced at LHCb are $\Lambda_b$  baryon \cite{Aaij:2011jp, Aaij:2014jyk}, and hence the study of $\Lambda_b$ becomes quite interesting in these days.  The  $b \to q l \bar \nu_l ~(q=u,c)$ quark level transitions can be probed in both $B$ and $\Lambda_b$ decays. Thus, as in $B$ decays one can also scrutinize the presence of  lepton universality violation in the corresponding semileptonic baryon decays $\Lambda_b \to (\Lambda_c, p) l \bar \nu_l$  to corroborate the results from $B$ sector and thus, to  probe the structure of NP.  The heavy-heavy  and heavy-light semileptonic decays of  baryons can serve as an additional source for the determination of the Cabbibo-Kobayashi-Maskawa (CKM) matrix elements $V_{qb}$ \cite{Aaij:2015bfa, Fiore:2015cmx, Patrignani:2016xqp, Hsiao:2017umx}. In the literature \cite{Woloshyn:2014hka, Wu:2015yqa, Shivashankara:2015cta, Gutsche:2015rrt, Gutsche:2015mxa, Detmold:2015aaa, Dutta:2015ueb, Pervin:2005ve, Faustov:2016yza, Datta:2017aue, Li:2016pdv, DiSalvo:2018ngq, Bernlochner:2018kxh}, the baryonic decay modes mediated by $b \to (u,c) l\bar \nu_l$ quark level transitions are studied both in model dependent and independent approaches. The analysis of  $\Lambda_b \to \Lambda_c \tau \bar \nu_\tau$ decay in the context of SM and various NP couplings are performed in \cite{Shivashankara:2015cta}. In  Ref. \cite{Gutsche:2015mxa}, the  SM hadron and lepton polarization asymmetries are computed  in the covariant confined quark model.  The  precise lattice QCD calculation of $\Lambda_b \to (\Lambda_c, p)$ form factors and the investigation of semileptonic  baryonic $b \to (u,c) l\bar \nu_l$  processes are performed in \cite{Detmold:2015aaa}.  Ref. \cite{DiSalvo:2018ngq} investigates the impact of five possible new physics  interactions, adopting five different form factors of $\Lambda_b \to \Lambda_c \tau \bar \nu_\tau$ decay mode.   Considering various real NP couplings,  the differential decay distributions, forward-backward asymmetries  and the ratios of branching fractions of these  baryonic decay modes are investigated  in  \cite{Dutta:2015ueb}.  In this work, we intend to  analyse the effect of complex new couplings on  $\Lambda_b \to (\Lambda_c, p) l \bar \nu_l$  decay processes in a model independent way. The main goal of this work is to check the possible existence of  lepton universality violation in baryonic decays.  The new coefficients  are constrained by using the branching ratios of $B_{u,c} \to \tau \bar \nu_\tau $, $B \to \pi \tau \bar \nu_\tau$ processes and the experimental data on $R_{D^{(*)}}, R_{J/\psi}, R_\pi^l$  ratios. We then compute the branching ratios, forward-backward asymmetries, lepton and hadron polarization asymmetries of these baryonic decay modes. We also check the LNU parameters by using the constrained new couplings.  The main difference between our approach and the previous analyses in \cite{Shivashankara:2015cta,   Datta:2017aue} is that,  we  investigate the impact of individual complex new couplings on all the angular  observables including the lepton and hadron polarization asymmetries. We use the updated experimental limits on $R_{D^{(*)}}, R_\pi^l$  ratios including new $R_{J/\psi}$ parameter to  constrain the allowed parameter space.

  The outline of our paper is  follows. In section II, we  present the general effective Lagrangian of  $b \to (u,c) l\nu_l$ processes in presence of NP, and  the necessary theoretical framework for analysing  these processes. The constraints on new parameter space  associated with $b \to (u,c) l \bar \nu_l$ transitions are  computed  from the experimental data on $R_{D^{(*)}}, R_{J/\psi}, R_\pi^l$, ${\rm Br}(B_{c,u} \to \tau \bar \nu_\tau)$ and ${\rm Br}(B \to \pi \tau \bar \nu_\tau)$ observables in section III.  In section IV, we discuss the branching ratios and all the physical angular observables of  $\Lambda_b \to (\Lambda_c, p) l \bar \nu_l$  processes. Our findings are summarized in section V. 
  
\section{Theoretical framework}

The most general effective Lagrangian  associated with  $B_1 \to B_2 l \bar{\nu_l}$ decay processes, where $B_1=\Lambda_b,~B_2=\Lambda_c, p$ mediated by the quark level transition $b\to q  l \bar{\nu_l},~(q=u,c)$ is given by \cite{Bhattacharya:2011qm, Cirigliano:2009wk}
\begin{eqnarray}
\mathcal L_{\rm eff} &=&
-\frac{4\,G_F}{\sqrt{2}}\,V_{q b}\,\Bigg\{(1 + V_L)\,\bar{l}_L\,\gamma_{\mu}\,\nu_L\,\bar{q}_L\,\gamma^{\mu}\,b_L +
V_R\,\bar{l}_L\,\gamma_{\mu}\,\nu_L\,\bar{q}_R\,\gamma^{\mu}\,b_R 
\nn \\
&&+
S_L\,\bar{l}_R\,\nu_L\,\bar{q}_R\,b_L +
S_R\,\bar{l}_R\,\nu_L\,\bar{q}_L\,b_R + 
T_L\,\bar{l}_R\,\sigma_{\mu\nu}\,\nu_L\,\bar{q}_R\,\sigma^{\mu\nu}\,b_L  \Bigg\} + {\rm h.c.}\,,\label{ham}
\end{eqnarray}
where  $G_F$ denotes the Fermi constant, $V_{q b} $ are the CKM matrix elements and  $q(l)_{L ,R}=P_{L,R}{\hspace*{0.1 true cm}}q(l)$ are the chiral quark(lepton) fields with $P_{L,R}=(1\mp \gamma_5)/2$ as the projection operators. Here $V_{L,R}, S_{L,R},T_L $ represent the vector, scalar and tensor type NP couplings, which are zero in the SM. 

 In the presence of NP,  the double differential decay distribution for $B_1 \to B_2 l \bar{\nu}_l$ processes with respect to $q^2$ and $\cos \theta_l$ ($\theta_l$ is the angle between the directions of parent $ B_1$ baryon and the $l^-$ in the dilepton rest frame) is given as \cite{Shivashankara:2015cta, Li:2016pdv}
 \bea \label{decayrate}
\frac{d\Gamma}{dq^2}&=& =N \left(1-\frac{m_l^2}{q^2}\right)^2\left[A_1+\frac{m_l^2}{q^2}A_{2}+2A_3+\frac{1}{4} A_4 +\frac{4m_l}{\sqrt{q^2}}\left( A_5+A_6 \right)+A_7\right]\,,
\eea
where
\bea
A_1&=&2\sin^2\theta_l\big(H_{\frac{1}{2},0}^2+H_{-\frac{1}{2},0}^2\big)+(1-\cos\theta_l)^2
H_{\frac{1}{2},+}^2+(1+\cos\theta_l)^2H_{-\frac{1}{2},-}^2\,,\nn\\
A_2&=&2\cos^2\theta_l\big(H_{\frac{1}{2},0}^2+H_{-\frac{1}{2},0}^2\big)+\sin^2\theta_l
\big(H_{\frac{1}{2},+}^2+H_{-\frac{1}{2},-}^2\big)+2\big(H_{\frac{1}{2},t}^2+H_{-\frac{1}{2},t}^2\big)\nonumber\\&&-4\cos\theta_l \big(H_{\frac{1}{2},0}H_{\frac{1}{2},t}+H_{-\frac{1}{2},0} H_{-\frac{1}{2},t}\big)\ \,,\nn\\
A_3&=&H^{SP^2}_{\frac{1}{2},0}+H^{SP^2}_{-\frac{1}{2},0}\,,\nn \\
A_4&=&\frac{m_l^2}{q^2}\Big[2\sin^2\theta_l\big(H_{\frac{1}{2},+,-}^{T^2}+H_{\frac{1}{2},0,t}^{T^2}+
H_{-\frac{1}{2},+,-}^{T^2}+H_{-\frac{1}{2},0,t}^{T^2}+2H_{\frac{1}{2},+,-}^TH_{\frac{1}{2},0,t}^T+
2H_{-\frac{1}{2},+,-}^TH_{-\frac{1}{2},0,t}^T\big)\nn\\&&+(1+\cos\theta_l)^2\big(H_{-\frac{1}{2},0,-}^{T^2}+H_{-\frac{1}{2},-,t}^{T^2}+
2H_{-\frac{1}{2},0,-}^T H_{-\frac{1}{2},-,t}^T\big)\nn\\&&+(1-\cos\theta_l)^2\big(H_{\frac{1}{2},+,0}^{T^2}+H_{\frac{1}{2},+,t}^{T^2}+
2H_{\frac{1}{2},+,0}^TH_{\frac{1}{2},+,t}^T\big)\Big]\nn \\ &&+2\cos^2\theta_l\big(H_{\frac{1}{2},+,-}^{T^2}\!+\!H_{\frac{1}{2},0,t}^{T^2}
+H_{-\frac{1}{2},+,-}^{T^2}\!+H_{-\frac{1}{2},0,t}^{T^2}+
2H_{\frac{1}{2},+,-}^TH_{\frac{1}{2},0,t}^T+2H_{-\frac{1}{2},+,-}^TH_{-\frac{1}{2},0,t}^T\big)\nn\\&&+\sin^2\theta_l\big(
H_{\frac{1}{2},+,0}^{T2}\!+\!H_{\frac{1}{2},+,t}^{T^2}+H_{-\frac{1}{2},0,-}^{T2}+H_{-\frac{1}{2},-,t}^{T^2}
+2H_{\frac{1}{2},+,0}^TH_{\frac{1}{2},+,t}^T+2H_{-\frac{1}{2},0,-}^TH_{-\frac{1}{2},-,t}^T\big)\,,\nn\\
A_5&=&-\cos\theta_\ell\big(H_{\frac{1}{2},0}H^{SP}_{\frac{1}{2},0}+H_{-\frac{1}{2},0}H^{SP}_{-\frac{1}{2},0}\big)
+\big(H_{\frac{1}{2},t}H^{SP}_{\frac{1}{2},0}+H_{-\frac{1}{2},t}H^{SP}_{-\frac{1}{2},0}\big)\,,\nn \\
A_6&=&\frac{\cos^2\theta_l}{2}\big(H_{\frac{1}{2},0}H^T_{\frac{1}{2},+,-}+
H_{\frac{1}{2},0}H^T_{\frac{1}{2},0,t}+H_{-\frac{1}{2},0}H^T_{-\frac{1}{2},+,-}+
H_{-\frac{1}{2},0}H^T_{-\frac{1}{2},0,t}\big)\nn\\&&-\frac{\cos\theta_l}{2}\big(H_{\frac{1}{2},t}H^T_{\frac{1}{2},+,-}+
H_{\frac{1}{2},t}H^T_{\frac{1}{2},0,t}+H_{-\frac{1}{2},t}H^T_{-\frac{1}{2},+,-}+
H_{-\frac{1}{2},t}H^T_{-\frac{1}{2},0,t}\big)\nn\\&&+\frac{(1-\cos\theta_l)^2}{4}\big(H_{\frac{1}{2},+}H^T_{\frac{1}{2},+,0}
+H_{\frac{1}{2},+}H^T_{\frac{1}{2},+,t}\big)\nn\\&&+\frac{(1+\cos\theta_l)^2}{4}\big(H_{-\frac{1}{2},-}H^T_{-\frac{1}{2},0,-}
+H_{-\frac{1}{2},-}H^T_{-\frac{1}{2},-,t}\big)\nn\\&&+\frac{\sin^2\theta_l}{4}\big(H_{\frac{1}{2},+}H^T_{\frac{1}{2},+,0}+
H_{\frac{1}{2},+}H^T_{\frac{1}{2},+,t}+H_{-\frac{1}{2},-}H^T_{-\frac{1}{2},0,-}+
H_{-\frac{1}{2},-}H^T_{-\frac{1}{2},-,t}\nn\\
&&+2H_{\frac{1}{2},0}H^T_{\frac{1}{2},+,-}+2H_{\frac{1}{2},0}H^T_{\frac{1}{2},0,t}
+2H_{-\frac{1}{2},0}H^T_{-\frac{1}{2},+,-}+2H_{-\frac{1}{2},0}H^T_{-\frac{1}{2},0,t}\big)\,,\nn \\
A_7&=&-2\cos\theta_l\Big(H^{SP}_{\frac{1}{2},0}H^T_{\frac{1}{2},+,-}+H^{SP}_{\frac{1}{2},0}
H^T_{\frac{1}{2},0,t}+H^{SP}_{-\frac{1}{2},0}H^T_{-\frac{1}{2},+,-}
+H^{SP}_{-\frac{1}{2},0}H^T_{-\frac{1}{2},0,t}\Big)\,,
\eea
with 
\bea
H^{VA}_{\lambda_{\Lambda_c},\lambda}&=&H^V_{\lambda_{\Lambda_c},\lambda}-H^A_{\lambda_{\Lambda_c},\lambda},~~ H_{\lambda_{\Lambda_{c}},\lambda_{w}}^V=H_{-\lambda_{\Lambda_{c}},-\lambda_{w}}^V,~~
H_{\lambda_{\Lambda_{c}},\lambda_{w}}^A=-H_{-\lambda_{\Lambda_{c}},-\lambda_{w}}^A, \nn\\
H^{SP}_{\lambda_{\Lambda_c},\lambda=0}&=&H^S_{\lambda_{\Lambda_c},\lambda=0}+H^P_{\lambda_{\Lambda_c},\lambda=0}, ~~H^{S}_{\lambda_{\Lambda_{c}},\lambda_{NP}} =  H^{S}_{-\lambda_{\Lambda_{c}},-\lambda_{NP}},~~~
H^{P}_{\lambda_{\Lambda_{c}},\lambda_{NP}} =  -H^{P}_{-\lambda_{\Lambda_{c}},-\lambda_{NP}},\nn \\
H^{T}_{\lambda_{\Lambda_{c}},\lambda,\lambda^\prime}&=&-H^{T}_{\lambda_{\Lambda_{c}},\lambda^\prime,\lambda},~~~~~H_{\lambda_{\Lambda_{c}},\lambda,\lambda}^T=0.
\eea
and 
\bea
N=\frac{G_F^2 |V_{q b}|^2 q^2  \sqrt{\lambda(M_{B_1}^2, M_{B_2}^2, q^2)}}{2^{10} \pi^3 M_{B_1}^3}\,,~~~~~\lambda(a,b,c) = a^2+b^2+c^2-2(ab+bc+ca).
\eea
 Here $M_{B_{1(2)}}$ and $m_l$ are the masses of $B_{1(2)}$ baryon and charged leptons respectively. 
 The helicity amplitudes in terms of the various form factors and the NP couplings are given as \cite{Shivashankara:2015cta, Li:2016pdv}
\bea
H_{\frac{1}{2}\,0}^V &=& \left(1+V_L+V_R \right)\,\frac{\sqrt{Q_-}}{\sqrt{q^2}}\,\Big[(M_{B_1} + M_{B_2})\,f_1(q^2) - q^2\,f_2(q^2)\Big]\,, \nonumber \\
H_{\frac{1}{2}\,0}^A &=& \left(1+V_L-V_R\right)\,\frac{\sqrt{Q_+}}{\sqrt{q^2}}\,\Big[(M_{B_1} - M_{B_2})\,g_1(q^2) + q^2\,g_2(q^2)\Big]\,, \nonumber \\
H_{\frac{1}{2}\,+}^V &=& \left(1+V_L+V_R\right)\,\sqrt{2\,Q_-}\,\Big[-f_1(q^2) + (M_{B_1} + M_{B_2})\,f_2(q^2)\Big]\,, \nonumber \\
H_{\frac{1}{2}\,+}^A &=& \left(1+V_L-V_R\right)\,\sqrt{2\,Q_+}\,\Big[-g_1(q^2) - (M_{B_1} - M_{B_2})\,g_2(q^2)\Big]\,, \nonumber \\
H_{\frac{1}{2}\,t}^V &=& \left(1+V_L+V_R\right)\,\frac{\sqrt{Q_+}}{\sqrt{q^2}}\,\Big[(M_{B_1} - M_{B_2})\,f_1(q^2) + q^2\,f_3(q^2)\Big]\,, \nonumber \\
H_{\frac{1}{2}\,t}^A &=& \left(1+V_L-V_R\right)\,\frac{\sqrt{Q_-}}{\sqrt{q^2}}\,\Big[(M_{B_1} + M_{B_2})\,g_1(q^2) - q^2\,g_3(q^2)\Big]\,, \nn \\
H_{\frac{1}{2}\,0}^{S} &=& \left(S_L+S_R\right)\,\frac{\sqrt{Q_+}}{m_b - m_{q}}\,\Big[(M_{B_1} - M_{B_2})\,f_1(q^2) + q^2\,f_3(q^2)\Big]\,, \nonumber \\
H_{\frac{1}{2}\,0}^{P} &=&\left(S_L-S_R\right)\,\frac{\sqrt{Q_-}}{m_b + m_{q}}\,\Big[(M_{B_1} + M_{B_2})\,g_1(q^2) - q^2\,g_3(q^2)\Big]\,,\nn \\
H_{\frac{1}{2},+,0}^T&=&-T_L\sqrt{\frac{2}{q^2}}\left(f_T\sqrt{Q_+}(M_{B_1} - M_{B_2})+g_T\sqrt{Q_-}(M_{B_1} + M_{B_2})\right)\,,\nn \\
H_{\frac{1}{2},+,-}^T&=&-T_L\left(f_T\sqrt{Q_+}+g_T\sqrt{Q_-}\right)\,,\nn \\
H_{\frac{1}{2},+,t}^T&=&T_L\Big[-\sqrt{\frac{2}{q^2}}\left(f_T\sqrt{Q_-}(M_{B_1} + M_{B_2}) +g_T\sqrt{Q_+}(M_{B_1} - M_{B_2})\right)\nn \\ &&
+\sqrt{2q^2}\left(f_T^V\sqrt{Q_-}-g_T^V\sqrt{Q_+}\right)\Big]\,,\nn \\
H_{\frac{1}{2},0,t}^T&=&T_L\Big [-f_T\sqrt{Q_-}-g_T\sqrt{Q_+}+f_T^V\sqrt{Q_-}(M_{B_1} + M_{B_2})-g_T^V\sqrt{Q_+}(M_{B_1} - M_{B_2})\nn \\&&
+f_T^S\sqrt{Q_-}Q_++g_T^S\sqrt{Q_+}Q_-\Bigg ]\,,\nn \\
H_{-\frac{1}{2},+,-}^T&=&T_L\Big [f_T\sqrt{Q_+}-g_T\sqrt{Q_-}\Big]\,,\nn\\
H_{-\frac{1}{2},0,-}^T&=&T_L\Big[\sqrt{\frac{2}{q^2}}\left(f_T\sqrt{Q_+}(M_{B_1} - M_{B_2})--g_T\sqrt{Q_-}(M_{B_1} + M_{B_2})\right)\Big]\,,\nn
\eea
\bea
H_{-\frac{1}{2},-,t}^T&=&T_L \Big [-\sqrt{\frac{2}{q^2}}\left(f_T\sqrt{Q_-}(M_{B_1} + M_{B_2})-g_T\sqrt{Q_+}(M_{B_1} - M_{B_2})\right)\nn \\&&
+\sqrt{2q^2}\left(f_T^V\sqrt{Q_-}+g_T^V\sqrt{Q_+}\right)\Big ]\,,\nn\\
H_{-\frac{1}{2},0,t}^T&=&T_L\Big [-f_T\sqrt{Q_-}+g_T\sqrt{Q_+}+f_T^V\sqrt{Q_-}(M_{B_1} + M_{B_2})+
+g_T^V\sqrt{Q_+}(M_{B_1} - M_{B_2})\nn \\&&+f_T^S\sqrt{Q_-}Q_+-g_T^S\sqrt{Q_+}Q_- \Big ]\,.
\eea
where  $ Q_\pm =\left(M_{B_1}\pm M_{B_2}\right)^2-q^2$ and $f_i^{(a)}, g_i^{(b)}, ~(i=1,2,3, T~ \&~ a,b=V,S)$ are the various form factors. 
After integrating out $\cos \theta_l$ in Eqn. (\ref{decayrate}), one can obtain the $q^2$ dependent differential decay rate. 
Besides the branching ratios, other interesting observables in these decay modes are
\begin{itemize}
\item Forward-backward asymmetry parameter:
\bea
A_{FB}(q^2)=\left ( \int_{-1}^0 d \cos \theta_l \frac{d^2 \Gamma}{d q^2 d \cos \theta_l}- \int_0^1 d \cos \theta_l \frac{d^2 \Gamma}{d q^2 d \cos \theta_l}\right )\Big {/}\frac{d \Gamma}{d q^2}\,.
\eea

\item Convexity parameter:
\bea \label{CFl}
C_F^l(q^2) = \frac{1}{ d \Gamma/d q^2} \frac{\,d^2}{d (\cos\theta_l) ^2\,}\Bigg(\frac{d^2 \Gamma}{dq^2 d\cos\theta_l}\Bigg). 
\eea
\item Longitudinal hadron polarization asymmetry parameter:
\bea
P_L^{h}(q^2)=\frac{{\rm d}\Gamma^{\lambda_2=1/2}/{\rm d}q^2-
    {\rm d}\Gamma^{\lambda_2=-1/2}/{\rm d}q^2}{{\rm d}\Gamma/{\rm d}q^2}\,,
\label{Pol_had}
\eea
where ${\rm d}\Gamma^{\lambda_2=\pm 1/2}$ are the individual helicity-dependent differential decay rates, whose detailed expressions are given in Appendix A \cite{Li:2016pdv}.

\item Longitudinal lepton polarization asymmetry parameter:
\bea
P_L^{\tau}(q^2)=\frac{{\rm d}\Gamma^{\lambda_{\tau}=1/2}/{\rm d}q^2-
    {\rm d}\Gamma^{\lambda_{\tau}=-1/2}/{\rm d}q^2}{{\rm d}\Gamma/{\rm d}q^2}\,,
\label{Pol_lept}
\eea 
where ${\rm d}\Gamma^{\lambda_2=\pm 1/2}$ are the individual helicity-dependent differential decay rates, whose detailed expressions are given in Appendix A \cite{Li:2016pdv}.

\item Lepton non-universality parameter:
\bea
&& R_{B_2}=\frac{{\rm Br}(B_1 \to B_2\tau^- \bar{\nu}_\tau)}{{\rm Br}( B_1 \to B_2\, l^- \bar{\nu}_l)} \,, ~~~l=e,\mu.
\eea
\item 
The LHCb Collaboration has  measured the ratio of the partially integrated decay rates of $\Lambda_b^0 \to p\,\mu\,\bar \nu_l$   over  the $\Lambda_b^0 \to \Lambda_c^+\,\mu\,\bar \nu_l$ process as
\begin{eqnarray}
R_{\Lambda_c\,p}^{\mu} &=& \int_{15\,{\rm GeV^2}}^{q^2_{\rm max}}\frac{d\Gamma(\Lambda_b \to p\,\mu\,\bar \nu_l)}{dq^2}\,dq^2 \Bigg / \int_{7\,{\rm GeV^2}}^{q^2_{\rm max}}\frac{d\Gamma(\Lambda_b \to \Lambda_c\,\mu\,\bar \nu_l)}{dq^2}\,dq^2\nn \\ 
&=& (1.00 \pm 0.04 \pm 0.08) \times 10^{-2}
\end{eqnarray}
 and put  constraint on the ratio $|V_{ub}|/|V_{cb}| = 0.083 \pm 0.004 \pm 0.004$~\cite{Aaij:2015bfa}. Similarly, we define the following parameter, to investigate if there is any possible role of NP 
 \bea
 R_{\Lambda_c \,p}^{\tau} &=& \int_{15\,{\rm GeV^2}}^{q^2_{\rm max}}\frac{d\Gamma(\Lambda_b \to p\,\tau\,\bar \nu_\tau)}{dq^2}\,dq^2 \Bigg / \int_{7\,{\rm GeV^2}}^{q^2_{\rm max}}\frac{d\Gamma(\Lambda_b \to \Lambda_c \,\tau\,\bar \nu_\tau)}{dq^2}\,dq^2.
 \eea
\end{itemize}  
 
 \section{Constraints on new couplings}
 
 After assembling the expressions for all the  interesting observables in  presence of NP, we now proceed to constrain the new coefficients by using the experimental bounds on ${\rm Br}(B_{u,c} \to \tau \bar \nu_\tau)$, ${\rm Br}(B \to\pi \tau \bar \nu_\tau)$,  $R_\pi^l$, $R_{D^{(*)}}$ and $R_{J/\psi}$  parameters.   In this analysis,  the new Wilson coefficients are considered as complex.  We further assume that only one new coefficient to present at a  time and accordingly compute the  allowed parameter space of these couplings.

The branching ratios of $B_q \to l\bar \nu_l$ processes in the presence of NP couplings are given by \cite{Biancofiore:2013ki}
\bea
{\rm Br}(B_q \to l \bar \nu_l) &=&
\frac{G_F^2\,|V_{qb}|^2}{8\,\pi}\,\tau_{B_q} f_{B_q}^2\,m_l^2\,M_{B_q}\,\Big(1 - \frac{m_l^2}{M_{B_q}^2}\Big)^2\,  \nn \\ &\times&
\Big |\left( 1+V_L-V_R\right) - \frac{M_{B_q}^2}{m_l\,(m_b + m_q)}\,\left(S_L-S_R\right) \Big |^2,
\eea
where $M_{B_q}$ is the mass of  $B_q$ meson. By using the masses of all the particles, lifetime of $B_q$ meson, CKM matrix elements from  \cite{Patrignani:2016xqp} and decay constants $f_{B_u}=190.5\pm 4.2$ MeV, $f_{B_c}=489\pm 4\pm 3$ MeV from \cite{Aoki:2013ldr, Chiu:2007km}, the branching ratios of $B_{u,c}^+ \to \tau^+ \nu_\tau$ processes in the SM are found to be
\bea \label{butau-SM}
&&{\rm Br}(B_u^+ \to \tau^+ \nu_\tau)|^{\rm SM}=(8.48\pm 0.5)\times 10^{-5},\\
&&{\rm Br}(B_c^+ \to \tau^+ \nu_\tau)|^{\rm SM}=(3.6\pm 0.14)\times 10^{-2}\,.
\eea
Using the current world average of the $B_c$ lifetime, the  upper limit on the branching ratio of $B_c^+ \to \tau^+ \nu_\tau$ process  is  \cite{Akeroyd:2017mhr}
\bea
{\rm Br}(B_c^+ \to \tau^+ \nu_\tau)\lesssim 30\%.
\eea   
The branching ratios of $B_q \to P l\bar \nu_l$   ($P=\pi, D$ ) are given as  \cite{Sakaki:2013bfa, Tanaka:2012nw}
\bea
\label{pilnu}
\frac{d{\rm Br}(B_q \to P l\bar \nu_l)}{dq^2} &=&\tau_{B_q} {G_F^2 |V_{qb}|^2 \over 192\pi^3 M_{B_q}^3} q^2 \sqrt{\lambda_P(q^2)} \left( 1 - {m_l^2 \over q^2} \right)^2 \times \nn \\ && \biggl\{ \biggr. |1 + V_L + V_R|^2 \left[ \left( 1 + {m_l^2 \over2q^2} \right) H_{0}^{2} + {3 \over 2}{m_l^2 \over q^2} \, H_{t}^{2} \right] \nn \\                                                      &&+ {3 \over 2} |S_L + S_R|^2 \, H_S^{2} + 8|T_L|^2 \left( 1+ {2m_l^2 \over q^2} \right) \, H_T^{2} \nn \\ &&+ 3{\rm Re}[ ( 1 + V_L + V_R ) (S_L^* + S_R^* ) ] {m_l \over \sqrt{q^2}} \, H_S H_{t} \nn \\
 &&- 12{\rm Re}[ ( 1 + V_L + V_R ) T_L^* ] {m_l \over \sqrt{q^2}} \, H_T H_{0} \biggl.\biggr\} \,,
\eea
where the helicity amplitudes in terms of form factors $(F_{0,+})$ are expressed as
\bea
&&H_0 = \sqrt{ \frac{\lambda(M_{B_q}^2,\,m_{P}^2,\,q^2)}{q^2}}\,F_{+}(q^2), ~~~~H_t = \frac{M_{B_q}^2 - M_P^2}{\sqrt{q^2}}\,F_0(q^2)\,, \nn \\
&&H_S=\frac{M_{B_q}^2 - M_P^2}{m_b - m_{q}}\,F_0(q^2)\,~~~~~~~~~~~~~~H_T=-\frac{\sqrt{\lambda_P(q^2)}}{M_{B_q}+M_P}F_T(q^2).
\eea
Using the values of the  $B \to \pi$ form factors  from \cite{Khodjamirian:2011ub, Bourrely:2008za, Boyd:1994tt, Boyd:1995cf},    the obtained branching ratios of $B_q \to \pi l \nu_l$ processes, in the SM  are given as 
\bea \label{Bpil-SM}
&&{\rm Br}( B^0 \to \pi^+ \mu^- \bar \nu_\mu) |^{\rm SM}= (1.35\pm 0.10)\times 10^{-4},\\
&&{\rm Br}( B^0 \to \pi^+ \tau^- \bar \nu_\tau) |^{\rm SM}=(9.40 \pm 0.75)\times 10^{-5}.
\eea
It should be noted that, the branching ratio of the muonic channel agrees reasonably well  with the experimental value as given in Eqn. (3), whereas the tau-channel is within its  current experimental limit  \cite{Patrignani:2016xqp} 
\bea
{\rm Br}( B^0 \to \pi^+ \tau^- \bar \nu_\tau) |^{\rm Expt} < 2.5\times 10^{-4}.
\eea
The branching ratios of $B_q \to V l\bar \nu_l$, where $V=D^*, J/\psi$,  are given as  \cite{Sakaki:2013bfa, Tanaka:2012nw}
\bea
\label{vlnu1}
 {d{\rm Br}(\bar B \to V l\bar \nu_l) \over dq^2} &=& \tau_{B_q} {G_F^2 |V_{qb}|^2 \over 192\pi^3 M_{B_q}^3} q^2 \sqrt{\lambda_V(q^2)} \left( 1 - {m_l^2 \over q^2} \right)^2 \times \nn \\ && \biggl\{ \biggr. 
      ( |1 + V_L|^2 + |V_R|^2 ) \left[ \left( 1 + {m_l^2 \over2q^2} \right) \left( H_{V,+}^2 + H_{V,-}^2 + H_{V,0}^2 \right) + {3 \over 2}{m_l^2 \over q^2} \, H_{V,t}^2 \right] \nn \\ &&  - 2{\rm Re}[(1 + V_L) V_R^*] \left[ \left( 1 + {m_l^2 \over 2q^2} \right) \left( H_{V,0}^2 + 2 H_{V,+} H_{V,-} \right) + {3 \over 2}{m_l^2 \over q^2} \, H_{V,t}^2 \right] \nn \\ &&  + {3 \over 2} |S_L - S_R|^2 \, H_S^2 + 8|T_L|^2 \left( 1+ {2m_l^2 \over q^2} \right) \left( H_{T,+}^2 + H_{T,-}^2 + H_{T,0}^2  \right) \nn \\ &&  + 3{\rm Re}[ ( 1 + V_L - V_R ) (S_L^* - S_R^* ) ] {m_l \over \sqrt{q^2}} \, H_S H_{V,t} \nn \\ &&  - 12{\rm Re}[ (1 + V_L) T_L^* ] {m_l \over \sqrt{q^2}} \left( H_{T,0} H_{V,0} + H_{T,+} H_{V,+} - H_{T,-} H_{V,-} \right) \nn \\ &&  + 12{\rm Re}[V_R T_L^* ] {m_l \over \sqrt{q^2}} \left( H_{T,0} H_{V,0} + H_{T,+} H_{V,-} - H_{T,-} H_{V,+} \right) \biggl.\biggr\} \,,
\eea
where $H_{V, \pm}$, $H_{V, 0}$, $H_{V, t}$  and $H_{S}$ are the  hadronic amplitudes \cite{Sakaki:2013bfa, Tanaka:2012nw}.
\begin{figure}[htb]
\centering
\includegraphics[scale=0.5]{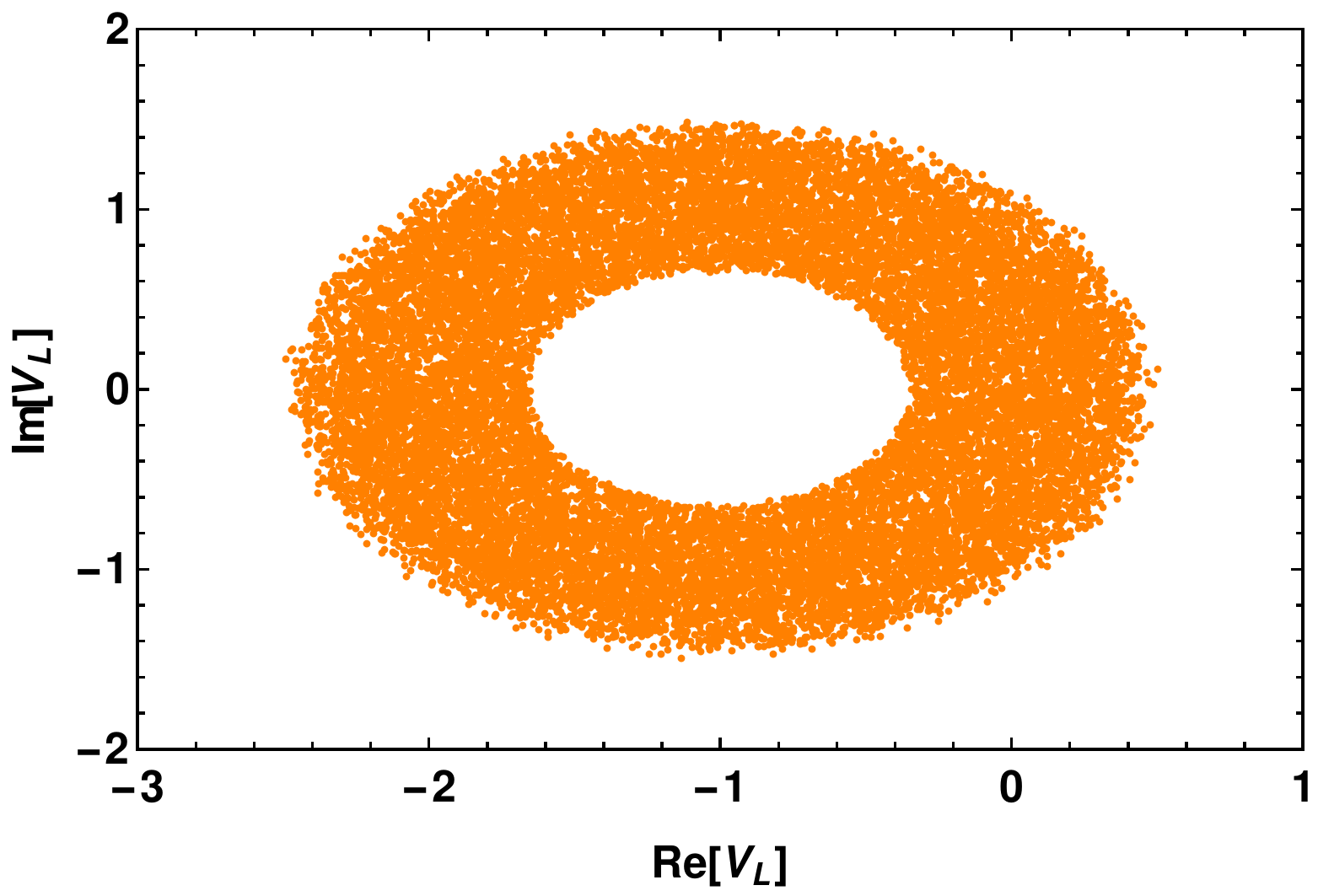}
\quad
\includegraphics[scale=0.5]{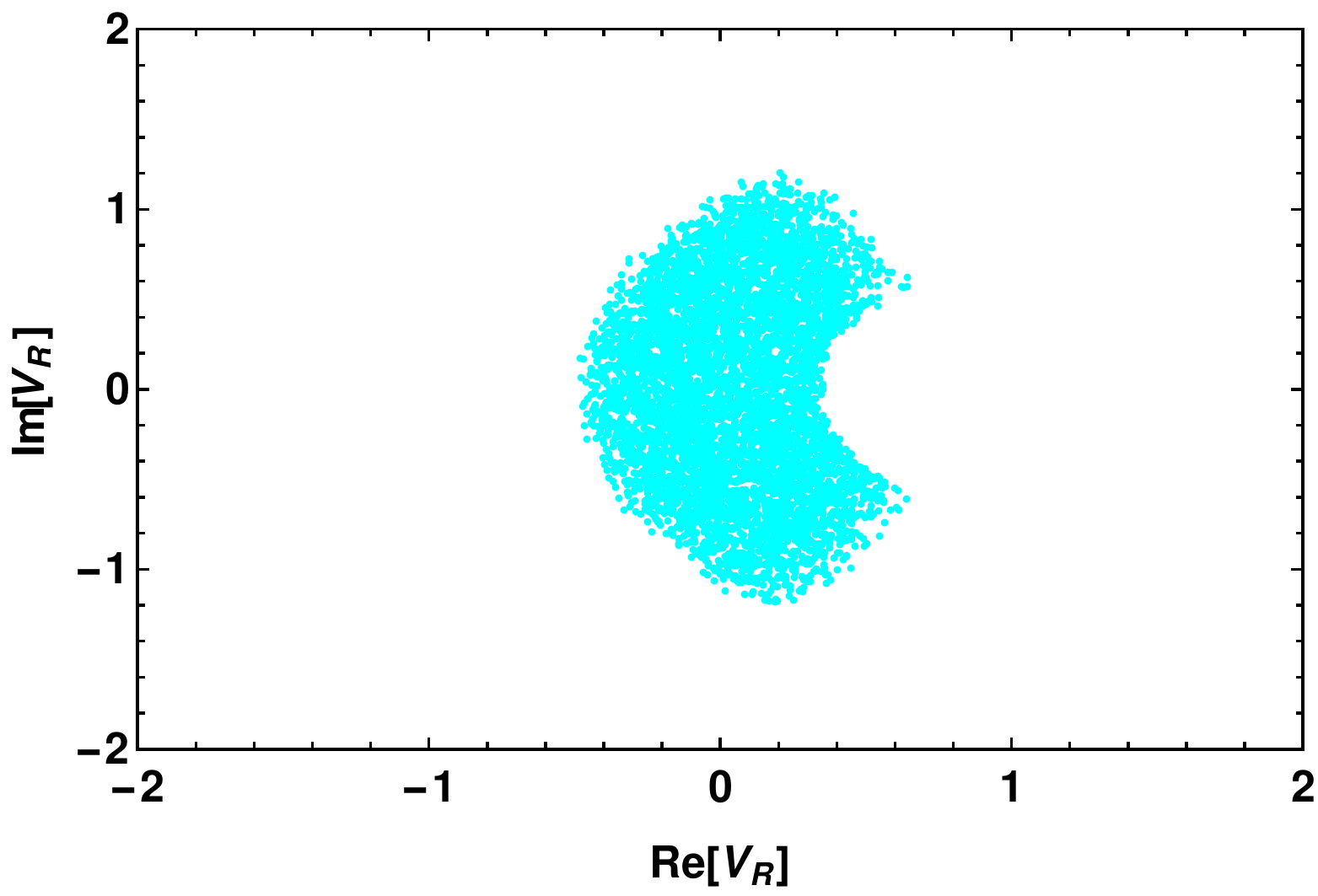}
\quad
\includegraphics[scale=0.5]{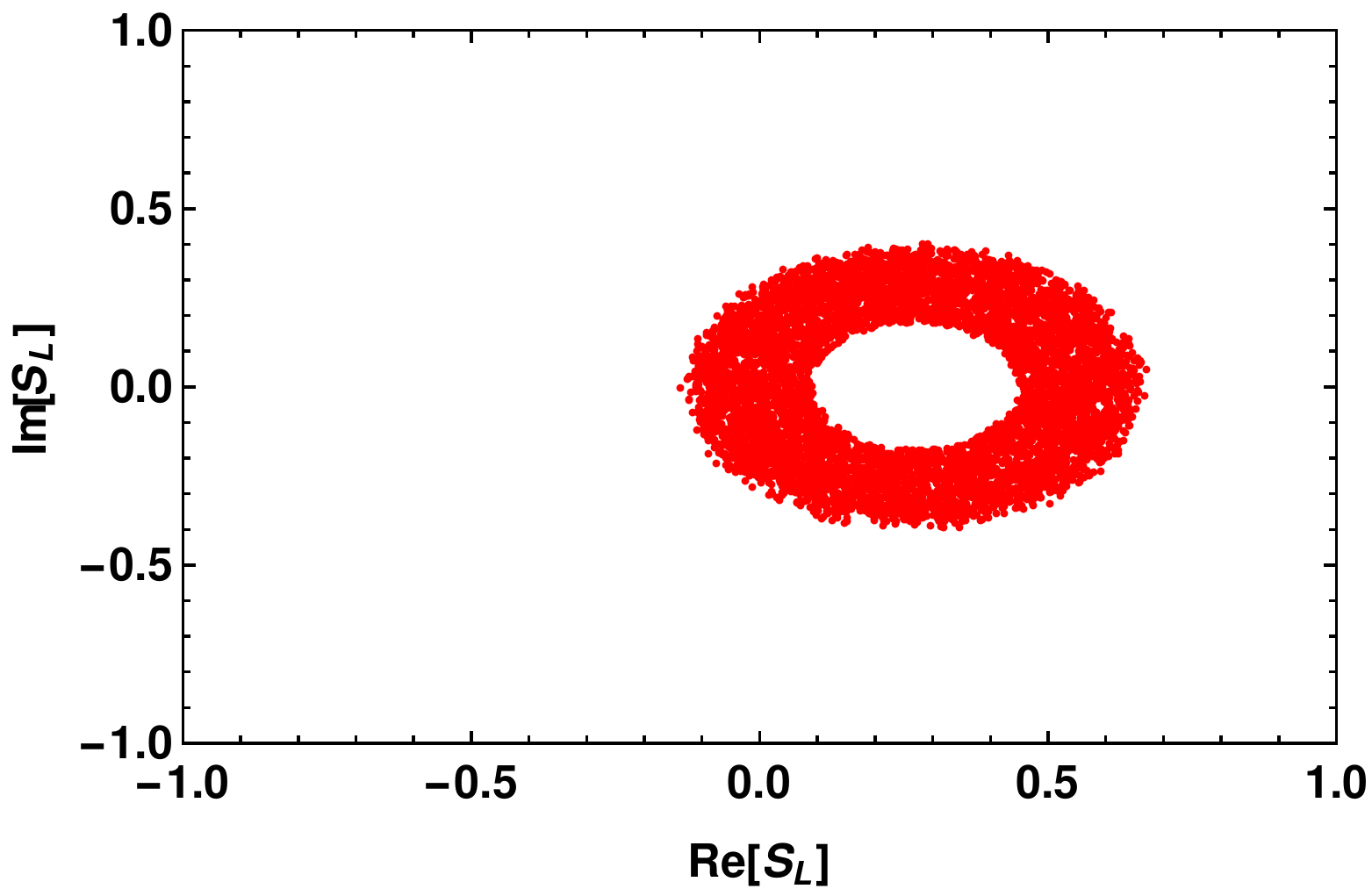}
\quad
\includegraphics[scale=0.5]{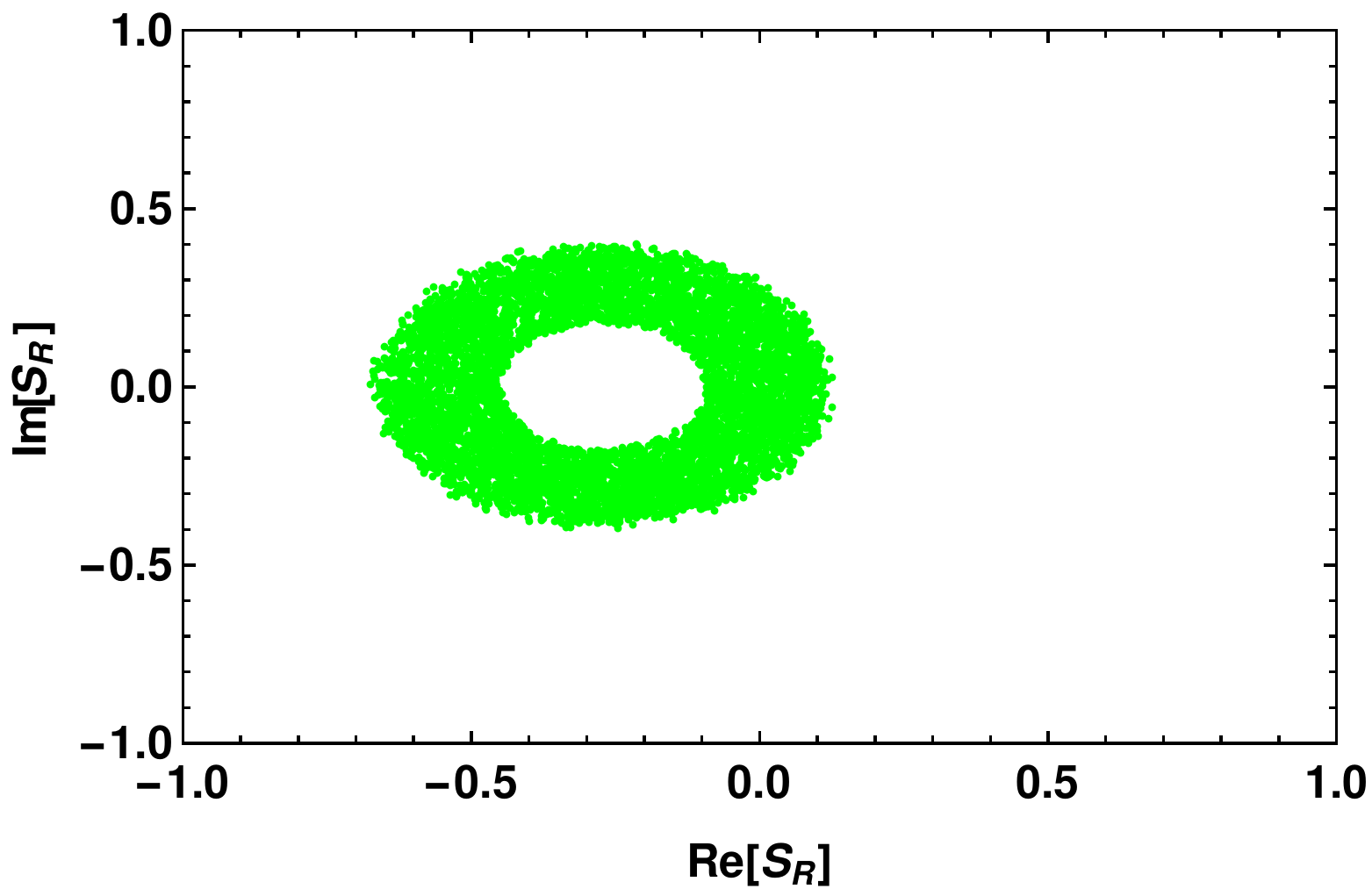}
\quad
\includegraphics[scale=0.5]{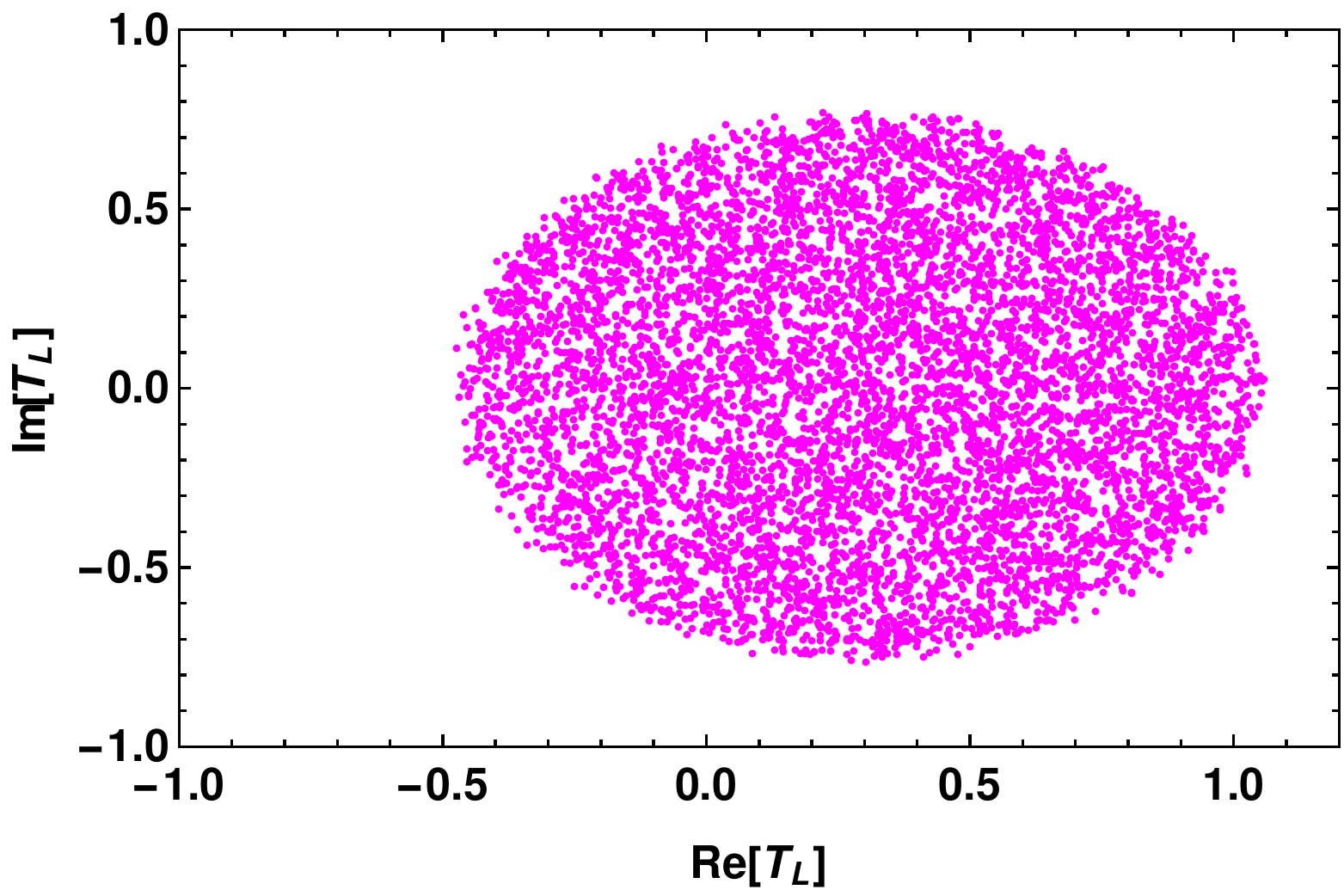}
\caption{Constraints on  $V_L$ (top-left panel), $V_R$ (top-right panel), $S_L$ (middle-left panel), $S_R$ (middle-right panel) and $T_L$ (bottom panel)  coefficients associated with $b \to u \tau \bar \nu_\tau$ transitions,  obtained from  ${\rm Br(B_{u}^+ \to \tau^+ \nu_\tau)}$, ${\rm Br(B \to \pi \tau \bar \nu_\tau)}$, $R_\pi^l$ observables.  Here the constraint on $T_L$ coupling is obtained  from ${\rm Br(B \to \pi \tau \bar \nu_\tau)}$ experimental data.} \label{con-bulnu}
\end{figure}
\begin{figure}[htb]
\centering
\includegraphics[scale=0.5]{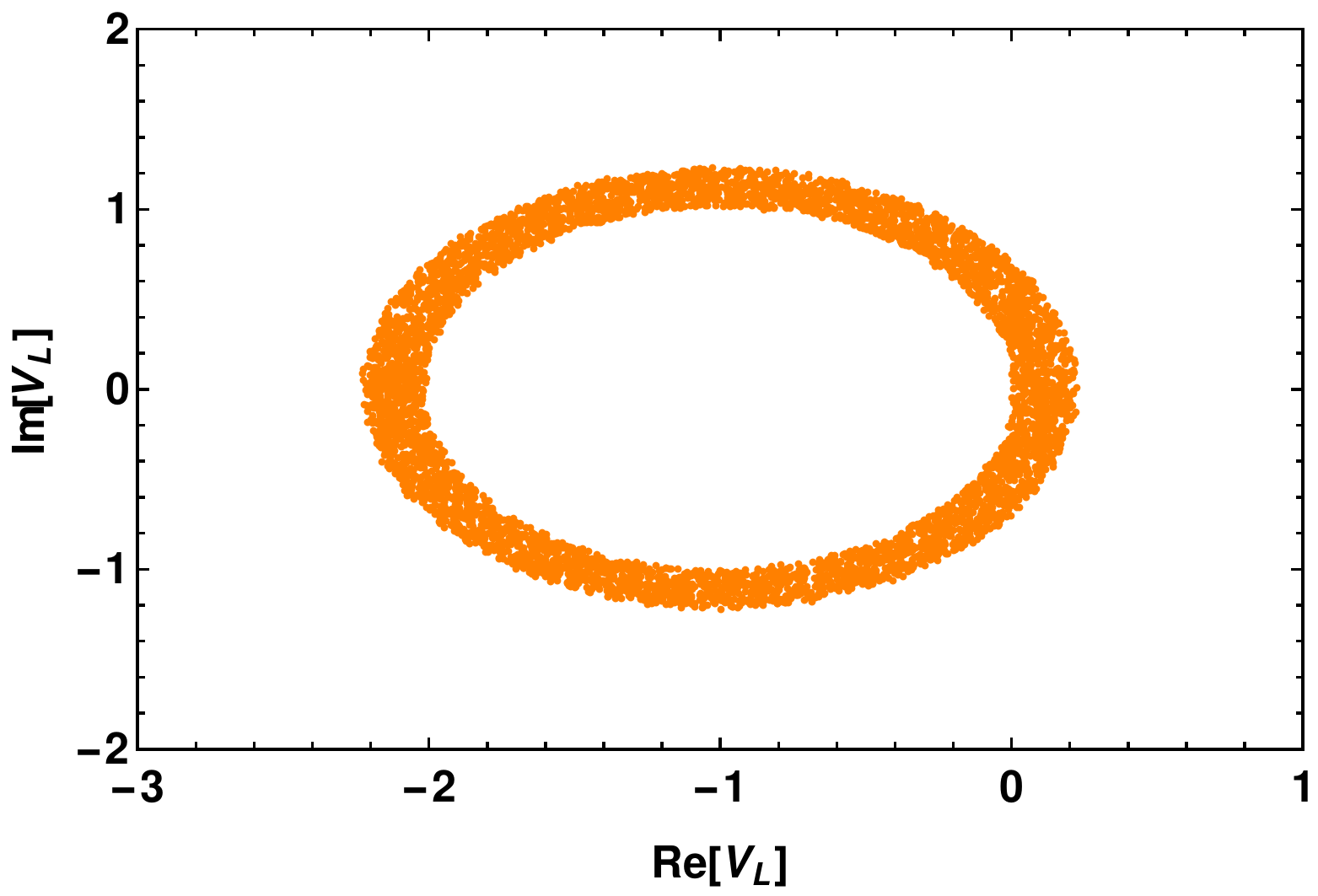}
\quad
\includegraphics[scale=0.5]{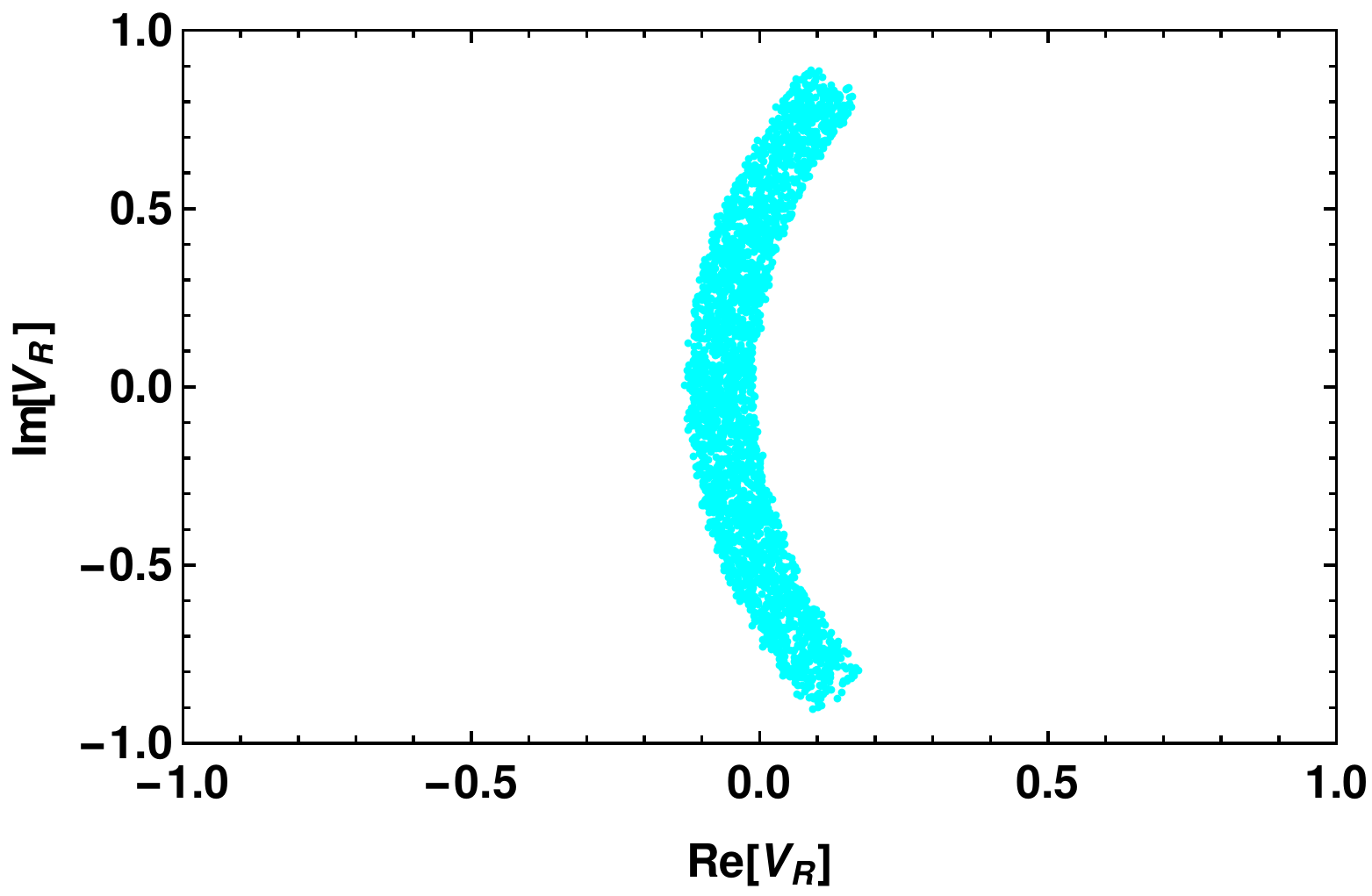}
\quad
\includegraphics[scale=0.5]{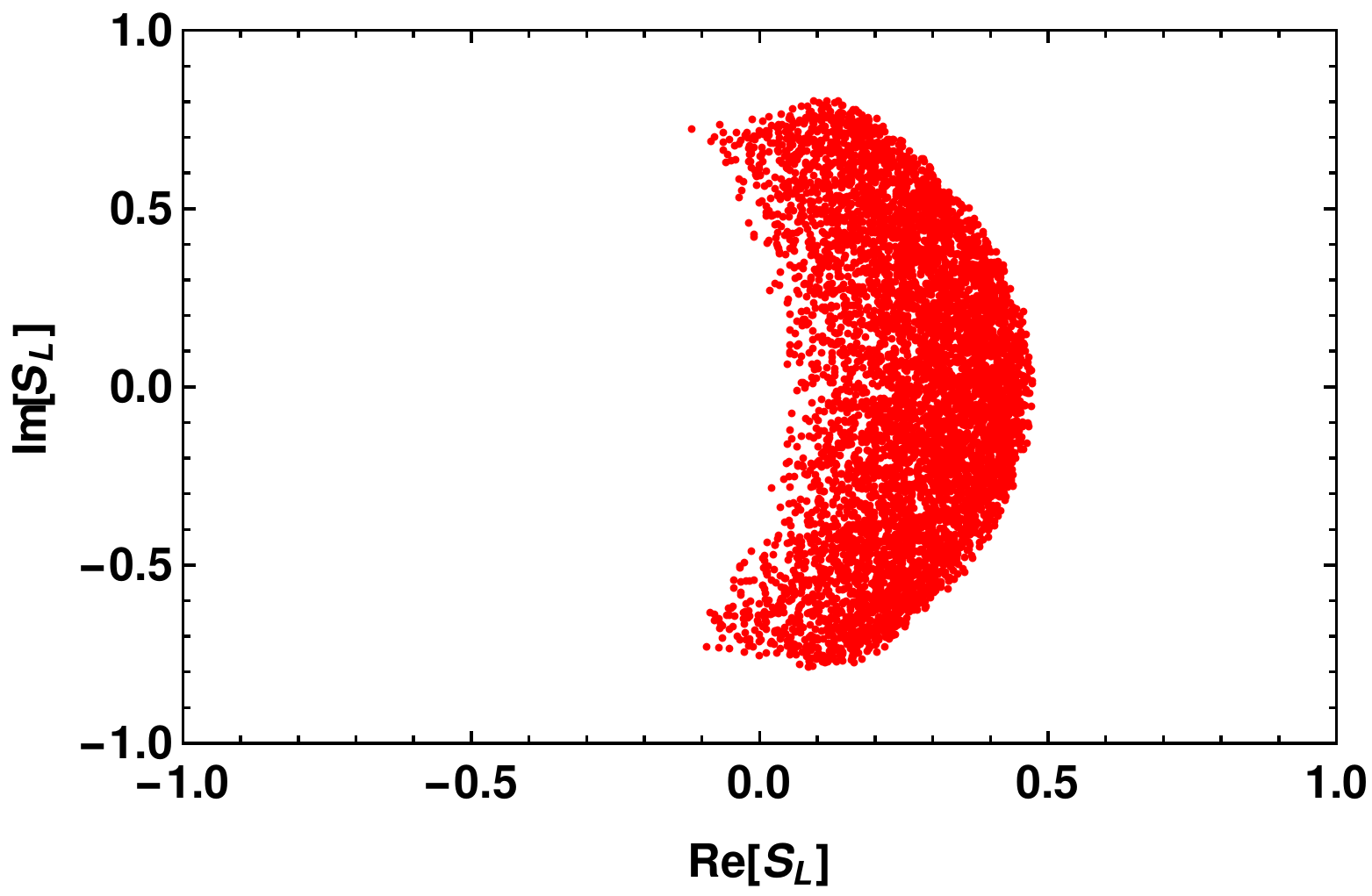}
\quad
\includegraphics[scale=0.5]{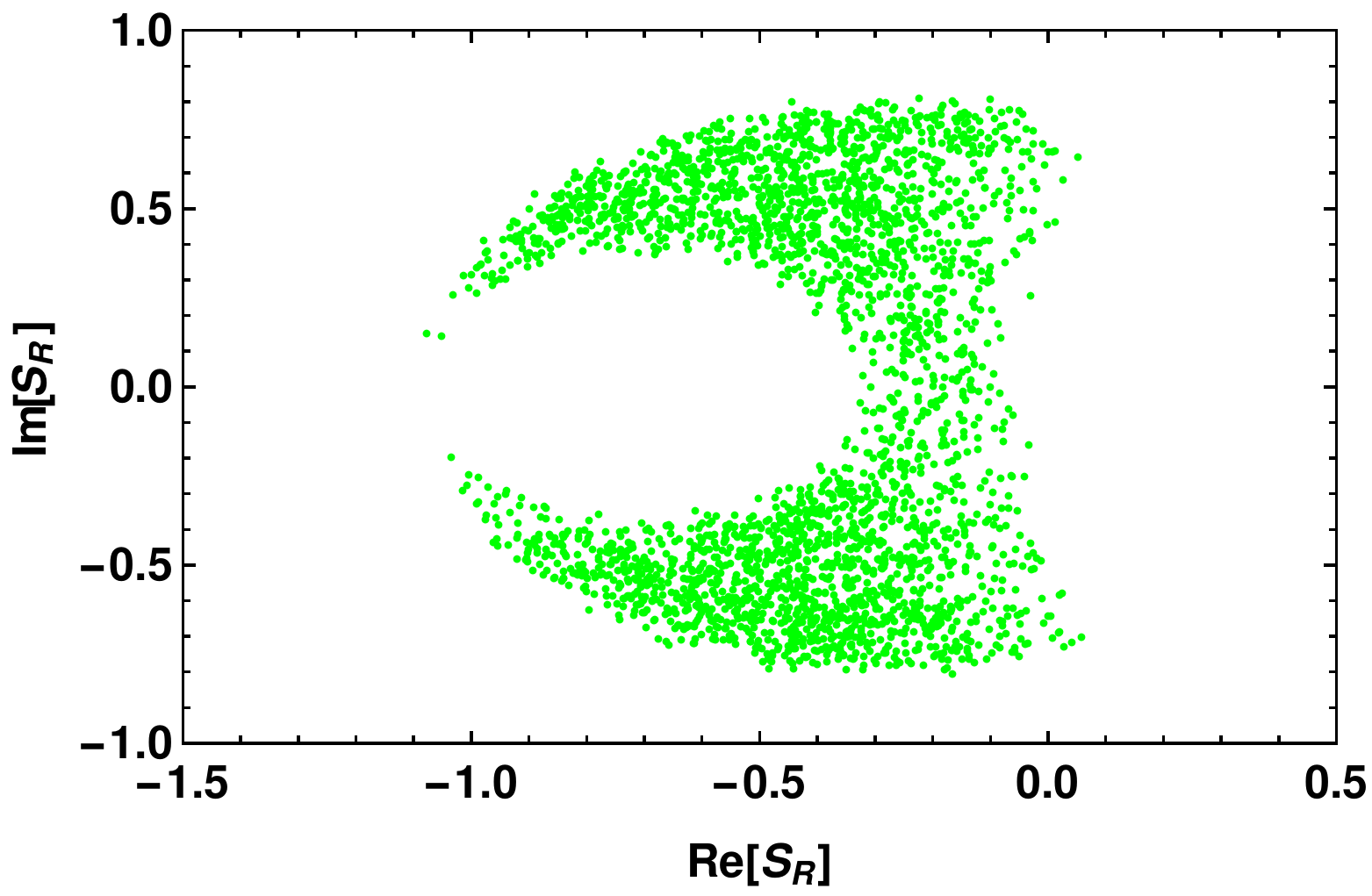}
\quad
\includegraphics[scale=0.5]{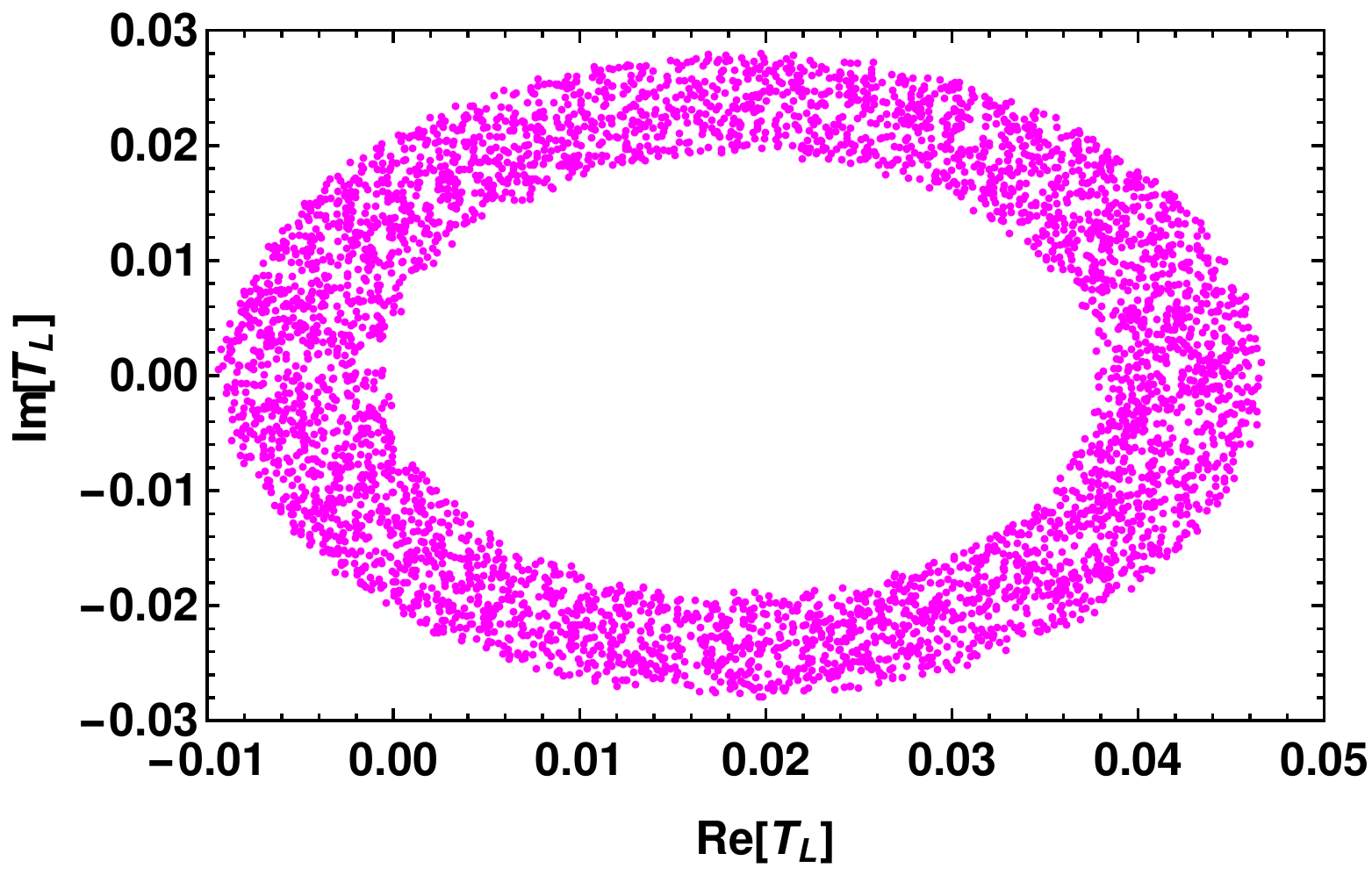}
\caption{Constraints on  $V_L$ (top-left panel), $V_R$ (top-right panel), $S_L$ (middle-left panel), $S_R$ (middle-right panel) and $T_L$ (bottom panel) new coefficients associated with $b \to c \tau \bar \nu_\tau$ transitions,  obtained from  ${\rm Br(B_{c}^+ \to \tau^+ \nu_\tau)}$, $R_{D^{(*)}}$ and $R_{J/\psi}$ observables. Here the constraint on $T_L$ coupling is obtained  from $R_{D^{(*)}}$ experimental data.} \label{con-bclnu}
\end{figure}

In this analysis, we consider the new physics contribution  to third generation lepton only and the couplings with light leptons are assumed to be SM like. By allowing  only one coefficient at a time, we  constrain its real and imaginary parts by comparing the theoretically predicted  values of ${\rm Br}(B_u^+ \to \tau^+ \nu_\tau)$ and $R_\pi^l$  with their corresponding $3\sigma$  range of observed experimental results for $b \to u \tau \bar \nu_\tau$ transitions. 
We have also used  the upper limit of the branching ratio of $B^0 \to \pi^+ \tau^- \bar \nu_\tau$ process. In Fig. \ref{con-bulnu}\,, we show the constraints on real and imaginary parts of new coefficients $V_L$ (top-left panel), $V_R$ (top-right panel), $S_L$ (middle-left panel) and $S_R$ (middle-right panel)  obtained from the ${\rm Br}(B_u^+ \to \tau^+ \nu_\tau)$, ${\rm Br}(B^0 \to \pi^+ \tau^- \bar \nu_\tau)$ and $R_\pi^l$ observables.  Since the branching ratio of $B_u^+ \to \tau^+ \nu_\tau$ process does not receive any contribution from tensor operator, the allowed region of real and imaginary parts of tensor coupling $(T_L)$ obtained only from the upper limit on ${\rm Br}(B^0 \to \pi^+ \tau^- \bar \nu_\tau)$, and is presented in the bottom panel of this figure.  Now imposing the extrema conditions,
the allowed range of the new couplings associated with $b \to u \tau \bar \nu_\tau$  transition  are presented in Table \ref{Tab:con}\,. For the case of $b \to c \tau \bar \nu_\tau$ decay processes, the constraints on the real and imaginary parts of individual   $V_L$ (top-left panel), $V_R$ (top-right panel), $S_L$ (middle-left panel) and $S_R$ (middle-right panel) coefficients obtained from $R_{D^{(*)}}$ and $R_{J/\psi}$ parameters are shown in Fig. \ref{con-bclnu}\,.  Till now, there is no precise determination of the form factors associated with tensorial operators for $B_c \to J/\psi l \bar \nu_l$ process  both from the theoretical and experimental sides. In addition, the leptonic $B_c$ meson decays do not receive any  contribution from tensor coupling.   Therefore, the constraints on $T_L$ coupling is obtained 
from the experimental data on $R_{D^{(*)}}$ ,  which is shown in 
  the bottom panel of Fig. \ref{con-bclnu}.    In Table \ref{Tab:con}\,, we have presented the allowed values of $({\rm Re}[V_{L(R)}]-{\rm Im}[V_{L(R)}])$ and $({\rm Re}[S_{L(R)}]-{\rm Im}[S_{L(R)}])$ coefficients,   which are compatible with the $3\sigma$ range of the experimental data. 

  The constraints on these parameters are obtained earlier from various $B$ decays in Refs.  \cite{Fajfer:2012vx, Fajfer:2012jt, Shivashankara:2015cta, Dutta:2015ueb, Biancofiore:2013ki, Tanaka:2012nw, Ivanov:2017mrj,Tran:2018kuv, Ivanov:2016qtw}.  Our analysis is similar to   Refs. \cite{Shivashankara:2015cta, Datta:2017aue}. In  Ref. \cite{Shivashankara:2015cta}, the authors have considered the  couplings to be complex and constrained the new coefficients associated with $b \to c \tau \bar \nu_\tau$ from $R_{D^{(*)}}$  data. However, they have  not includeed the tensor couplings in their analysis, and found that the effects produced by the pseudoscalar coefficient  are larger than those obtained from the scalar coefficient. In Ref. \cite{Dutta:2015ueb}, the author  assumed the couplings as real and computed the allowed parameter space by comparing the $R_{D^{(*)}}$, $R_\pi^l$ parameters with their corresponding $3\sigma$ experimental data.   
In \cite{Ivanov:2016qtw}, the authors have considered the covariant confined quark model and studied the effect of new physics in the $\bar{B^0 }\to D^* \tau^- \bar{\nu_\tau}$. They took the new coefficients as complex and constrained them using the experimental values of $R_D $ and $R_{D^*}$ within their $2\sigma$ range. 
Recently, the decay process $B_c \to (J/ \psi) \tau \nu_\tau$ has been studied, in the covariant confined quark model \cite{Tran:2018kuv}, where the  parameter space is constrained by using the experimental values of $R_D, R_{D^*}, R_{J/\psi}$ within $2\sigma$ range.  The new coefficients are considered to be complex and their best fit values are  $ V_L=-1.05+i1.15, V_R=0.04+i0.60, T_L=0.38-i0.06$.
Though our analysis is similar to these approaches, but we get more severe bounds on the phases and strengths of the couplings due to additional constraints from ${\rm Br}(B_c \to \tau \nu_\tau)$ and $R_{J/\psi}$ parameters for $b \to c \tau \bar \nu_\tau$ case and from ${\rm Br}(B_u \to \tau \nu_\tau)$ and ${\rm Br}(B \to  \pi \tau \nu_\tau)$ observables for $b \to u \tau \bar \nu_\tau$ process.

\begin{table}[htb]
\centering
\caption{Allowed ranges of the new coefficients.} \label{Tab:con}
\begin{tabular}{|c|c|c|c|}
\hline
Decay processes~&~New coefficients~&~Minimum value~&~Maximum Value~\\
\hline
\hline
$b \to u \tau\bar \nu_\tau$~&~$({\rm Re}[V_L], {\rm Im}[V_L]) $~&~$(-2.489,-1.5)$~&~$(0.504,1.48)$\\
~&~$({\rm Re}[V_R], {\rm Im}[V_R]) $~&~$(-0.478,-1.185)$~&~$(0.645,1.198)$\\
~&~$({\rm Re}[S_L], {\rm Im}[S_L]) $~&~$(-0.136,-0.396)$~&~$(0.672,0.398)$\\
~&~$({\rm Re}[S_R], {\rm Im}[S_R]) $~&~$(-0.6743,-0.398)$~&~$(0.1265,0.398)$\\
~&~$({\rm Re}[T_L], {\rm Im}[T_L]) $~&~$(-0.473,-0.773)$~&~$(1.07,0.773)$\\
\hline
$b \to c \tau\bar \nu_\tau$~&~$({\rm Re}[V_L], {\rm Im}[V_L]) $~&~$(-2.224,-1.228)$~&~$(0.225,1.225)$\\
~&~$({\rm Re}[V_R], {\rm Im}[V_R]) $~&~$(-0.129,-0.906)$~&~$(0.173,0.89)$\\
~&~$({\rm Re}[S_L], {\rm Im}[S_L]) $~&~$(-0.116,-0.788)$~&~$(0.474,0.8)$\\
~&~$({\rm Re}[S_R], {\rm Im}[S_R]) $~&~$(-1.076,-0.809)$~&~$(0.06,0.807)$\\
~&~$({\rm Re}[T_L], {\rm Im}[T_L]) $~&~$(-0.0094,-0.028)$~&~$(0.0467,0.028)$\\
\hline
\end{tabular}
\end{table}

\section{Numerical analysis and discussion}

In this section, we present the numerical results for semileptonic $\Lambda_b$  decay modes with third generation leptons in the final state. The masses of all the particles and the lifetime of $\Lambda_b$   are taken from \cite{Patrignani:2016xqp}.  The $q^2$ dependence of the helicity form factors $(f_{+, \perp,0},~ g_{+, \perp,0}, h_{+,\perp}, \widetilde{h}_{+,\perp})$ in the lattice QCD calculation
can be parametrized as \cite{Detmold:2015aaa, Datta:2017aue}
\bea
f_i(q^2)=\frac{1}{1-q^2/(m_{\rm pole}^f)^2}\Big [ a_0^f+a_1^f z(q^2) \Big ],~~~(i=+,\perp,0)
\eea 
where $m_{\rm pole}^f$ is the pole mass and 
\bea
z(q^2)=\frac{\sqrt{t_+-q^2}-\sqrt{t_+-t_0}}{\sqrt{t_+-q^2}+\sqrt{t_+-t_0}}\;,
\eea
with $t_\pm =(M_{B_1}\pm M_{B_2})^2$.  The values of the parameters $m_{\rm pole}^f$, $a_{0,1}^f$ associated with  (axial)vector and (pseudo)scalar form factors $(f_{+, \perp,0},~ g_{+, \perp,0})$ are taken from \cite{Detmold:2015aaa}.  In the lattice QCD approach, the  $m_{\rm pole}^f$, $a_{0,1}^f$ parameters linked to tensor form factors $(h_{+,\perp}, \widetilde{h}_{+,\perp})$ of $\Lambda_b \to \Lambda_c l \bar \nu_l$ process  are computed in \cite{Datta:2017aue}. However,    currently  no lattice results are available  on the tensor form factors associated with $\Lambda_b \to p l \bar \nu_l$ process. Hence, we relate the tensor form factors of $\Lambda_b \to p l \bar \nu_l$ decay mode with its  (axial)vector form factors by using the HQET relations as \cite{Feldmann:2011xf, Li:2016pdv, Chen:2001zc},
\bea
f_T=g_T=f_1=\frac{(M_{B_1}+M_{B_2})^2f_+- q^2f_\perp}{(M_{B_1}+M_{B_2})^2-q^2}\,, \quad
f_T^V=g_T^V=f_T^S=g_T^S=0\,.
\eea
The detailed relation between the helicity form factors $(f_{+, \perp,0},~ g_{+, \perp,0}, h_{+,\perp}, \widetilde{h}_{+,\perp})$ with other various hadronic form factors $(f_{1,2,3},~ g_{1,2,3},~f_T,~g_T,~f_T^{V(S)},~g_T^{V(S)})$ are listed in Appendix B \cite{Feldmann:2011xf}.
Using all these input parameters, the predicted  branching ratios  of $\Lambda_b \to (\Lambda_{c},p)  \mu \bar \nu_\mu$ processes in the SM are  given by 
\bea \label{br-mu-SM}
&&{\rm Br}(\Lambda_b \to p \mu^- \bar \nu_\mu)|^{\rm SM}=(4.31\pm 0.345)\times 10^{-4},\nn \\
&& {\rm Br}(\Lambda_b \to \Lambda_c \mu^- \bar \nu_\mu)|^{\rm SM}=(4.994\pm 0.4)\times 10^{-2},
\eea
which are in reasonable agreement with the corresponding experimental data \cite{Patrignani:2016xqp} 
\bea 
&&{\rm Br}(\Lambda_b \to p \mu^- \bar \nu_\mu)=(4.1\pm 1.0)\times 10^{-4},\nn \\
&& {\rm Br}(\Lambda_b \to \Lambda_c l^- \bar \nu_l)=\left (6.2_{-1.3}^{+1.4}\right )\times 10^{-2}.
\eea
The  values of the forward-backward asymmetries in these channels are found to be
\bea \label{fb-mu-SM}
\langle A_{FB}^\mu \rangle |_{\Lambda_b \to p}^{\rm SM} =0.316\pm 0.025,  ~~~~
\langle A_{FB}^\mu \rangle |_{\Lambda_b \to \Lambda_c}^{\rm SM} =0.19\pm 0.0152.
\eea
In Eqn. (\ref{br-mu-SM}\,, ~\ref{fb-mu-SM}),  the theoretical uncertainties are mainly due to the uncertainties associated with  the CKM matrix elements and the form factor parameters. After having idea on all the required input parameters and the allowed parameter space of new couplings,  we now proceed
to discuss various new physics scenarios and their
impact on $\Lambda_b \to (\Lambda_c, p) \tau \bar \nu_\tau$ decay modes in a model independent way.
\subsection{Scenario A: Only $V_L$ coefficient}
In this  scenario, we assume that the additional   new physics contribution to the SM result is coming only from the coupling associated with the left-handed vector like quark currents i.e., $V_L\neq 0$ and $V_R, S_{L, R}, T_L=0$.  Since in this case,  the NP operator has the same Lorentz structure as the SM operator,  the
SM decay rate  gets modified by the factor $|1+V_L|^2$. Imposing $3\sigma$ constraint on Br($B_{u,c}^+ \to \tau^+ \nu_\tau$), Br($B^0 \to \pi^+ \tau^- \bar \nu_\tau$), $R_\pi^l$, $R_{D^{(*)}}$ and $R_{J/\psi}$ observables,  the allowed parameter space of $V_L$ couplings associated with $b \to (u,c) \tau \nu_\tau$ are  shown in Figs. \ref{con-bulnu} and \ref{con-bclnu} respectively.  
\begin{figure}[htb]
\centering
\includegraphics[scale=0.55]{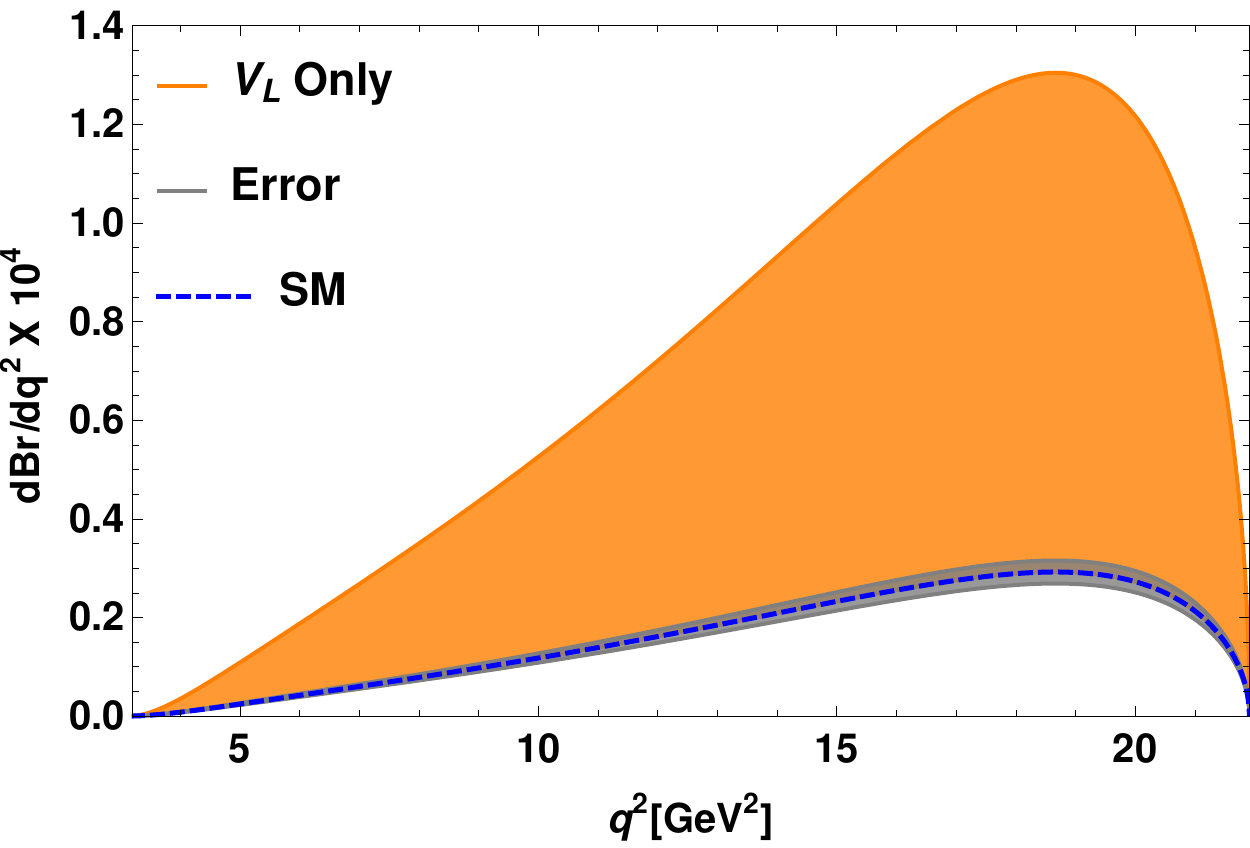}
\quad
\includegraphics[scale=0.55]{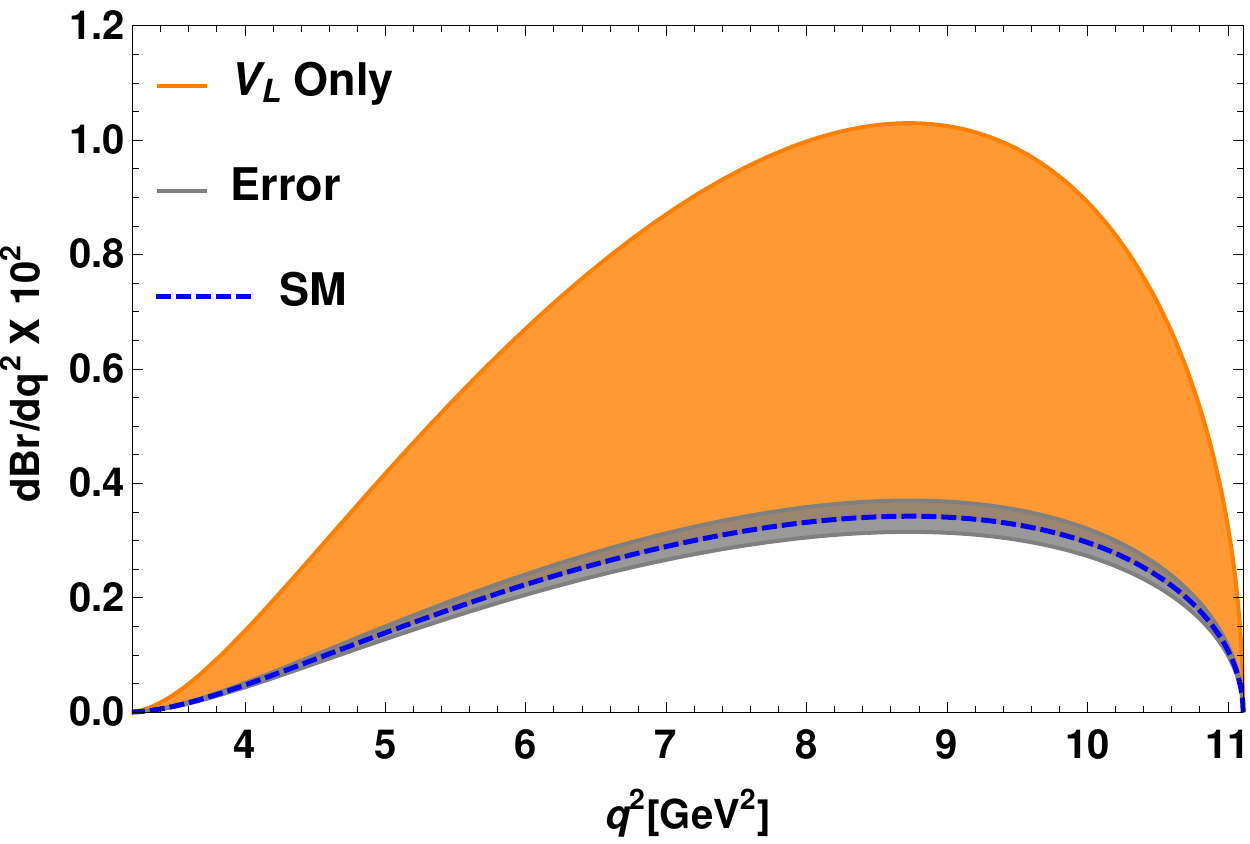}
\caption{The $q^2$ variation of branching ratio of $\Lambda_b \to p \tau^-  \bar \nu_\tau $ (left panel) and $\Lambda_b \to \Lambda_c^+ \tau^-  \bar \nu_\tau$ (right panel) processes in the presence of only $V_L$ new coefficient. Here the orange bands represent the new physics contribution. Blue dashed lines stand for the SM and the theoretical uncertainties arising due to the input parameters are presented in grey color.} \label{br-VL}
\end{figure}
Using the minimum and maximum  values on real and imaginary parts of  $V_L$ coefficient from  Table \ref{Tab:con}\,,  we present the differential branching ratios of $\Lambda_b \to p \tau^- \bar \nu_\tau$ (left panel) and   $\Lambda_b \to \Lambda_c \tau^- \bar \nu_\tau$ (right panel) processes with respect to $q^2$ in Fig. \ref{br-VL}\,. In these figures,  the blue dashed lines represent the SM contribution, the orange bands are due to the presence of new $V_L$  coefficient and the grey bands stand for the theoretical uncertainties  associated with the input parameters  like form factors, CKM matrix elements etc. The branching ratios of $\Lambda_b \to (\Lambda_c, p) \tau^- \bar \nu_\tau$ deviate significantly  from their corresponding SM values due to the NP contribution. 
In addition to the decay rate, other interesting observables, which can  be used to probe new physics, are the zero crossing of the forward-backward asymmetry and the convexity parameters.  
From Eqn. (\ref{CFl}), one can notice that the convexity parameter  depends only on the $V_{L, R}$ and $T_L$ couplings. 
The  values for forward-backward asymmetries of $\Lambda_b \to (\Lambda_{c},p)  \tau \bar \nu_\tau$ processes in the SM are 
\bea
&&\langle A_{FB}^\tau \rangle |_{\Lambda_b \to p}^{\rm SM} =0.115\pm 0.0092\;, ~~~
\langle A_{FB}^\tau \rangle |_{\Lambda_b \to \Lambda_c}^{\rm SM} =-0.09 \pm 0.007\;,
\eea
and the corresponding values for the convexity parameters are 
\bea
\langle C_{F}^\tau \rangle |_{\Lambda_b \to p}^{\rm SM} =-0.157\pm 0.013\;,  ~~~~
\langle C_{F}^\tau \rangle |_{\Lambda_b \to \Lambda_c}^{\rm SM} =-0.098\pm 0.008\;.
\eea
We found no  deviation  from  SM results for the forward-backward asymmetry and convexity parameters   due to the presence of $V_L$ coefficient.
\begin{figure}[htb]
\centering
\includegraphics[scale=0.55]{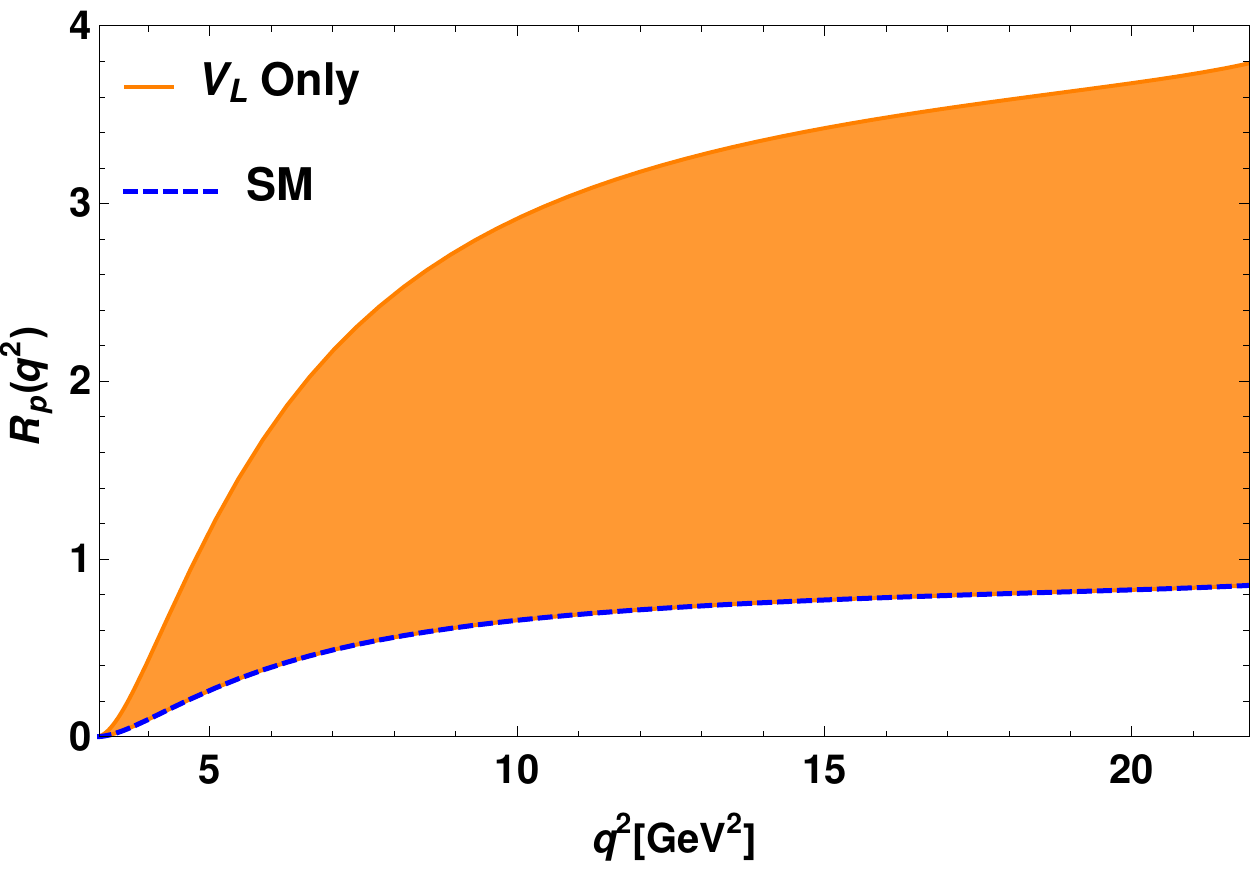}
\quad
\includegraphics[scale=0.55]{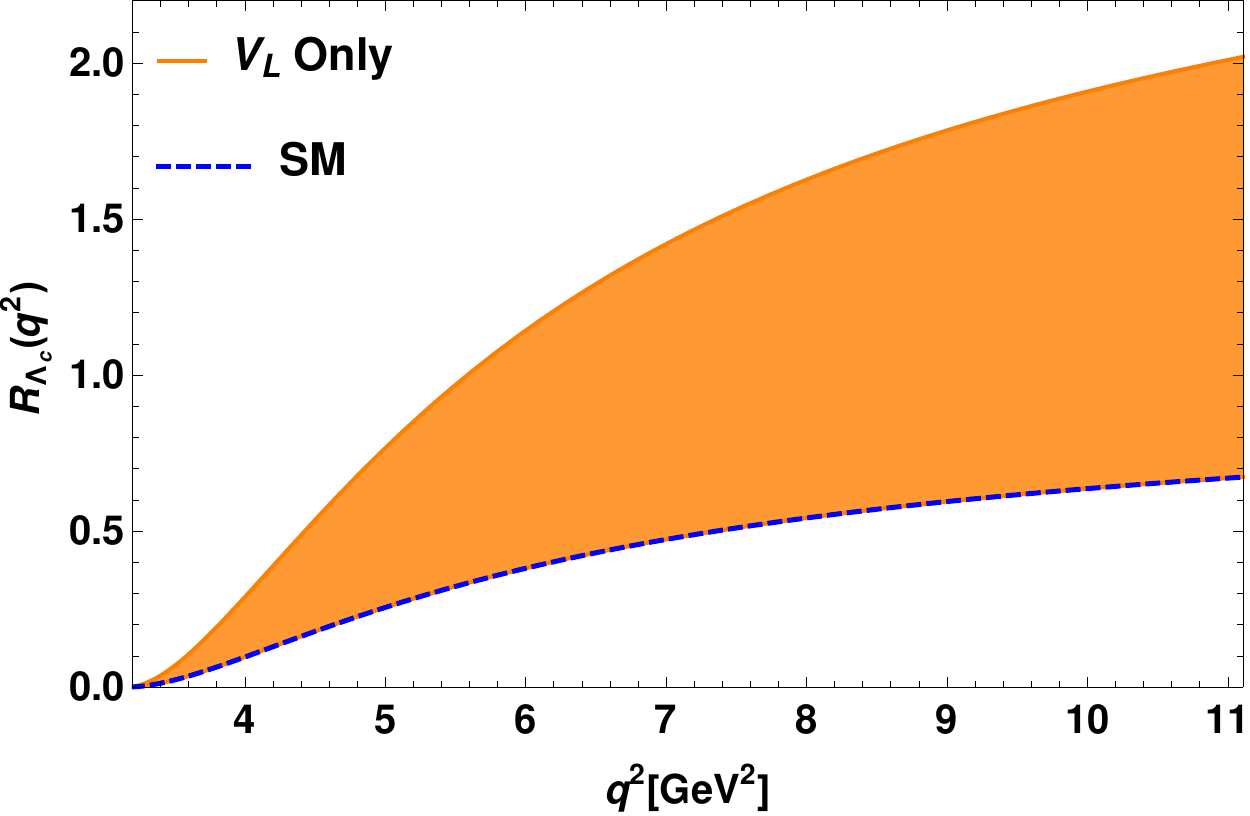}
\caption{The  variation of  $R_{p}$ (left panel) and $R_{\Lambda_c}$ (right  panel) LNU parameters with respect to $q^2$ in the presence of only $V_L$ new coefficient.} \label{LNU-VL}
\end{figure}
In  Fig. \ref{LNU-VL}\,,  left (right) panel  depicts the $q^2$  variation of lepton universality violating parameters $R_p~(R_{\Lambda_c})$. We observe that the NP contribution coming from the  $V_L$ coupling  has significant  impact on  $R_p$ and $R_{\Lambda_c}$ parameters. The  variation of $R_{\Lambda_c p}^\tau$ parameter with $q^2$ for this case,  is presented in the left panel of Fig. \ref{LNU-VLVR}\,. The numerical values of the branching ratios and the LNU parameters for both the SM and the $V_L$-type NP scenario are given in Table \ref{VLVR:Tab}\,. Besides the
branching ratios,  forward-backward asymmetry and LNU  parameters of $\Lambda_b \to (\Lambda_c, p) \tau \bar \nu_\tau$ processes, the NP effects can
also be observed in the hadron and lepton polarization asymmetries. 
However,  no deviation has been found in the presence of $V_L$ coupling  from their corresponding SM results.

\subsection{Scenario B: Only $V_R$ coefficient}
\begin{figure}[htb] 
\centering
\includegraphics[scale=0.55]{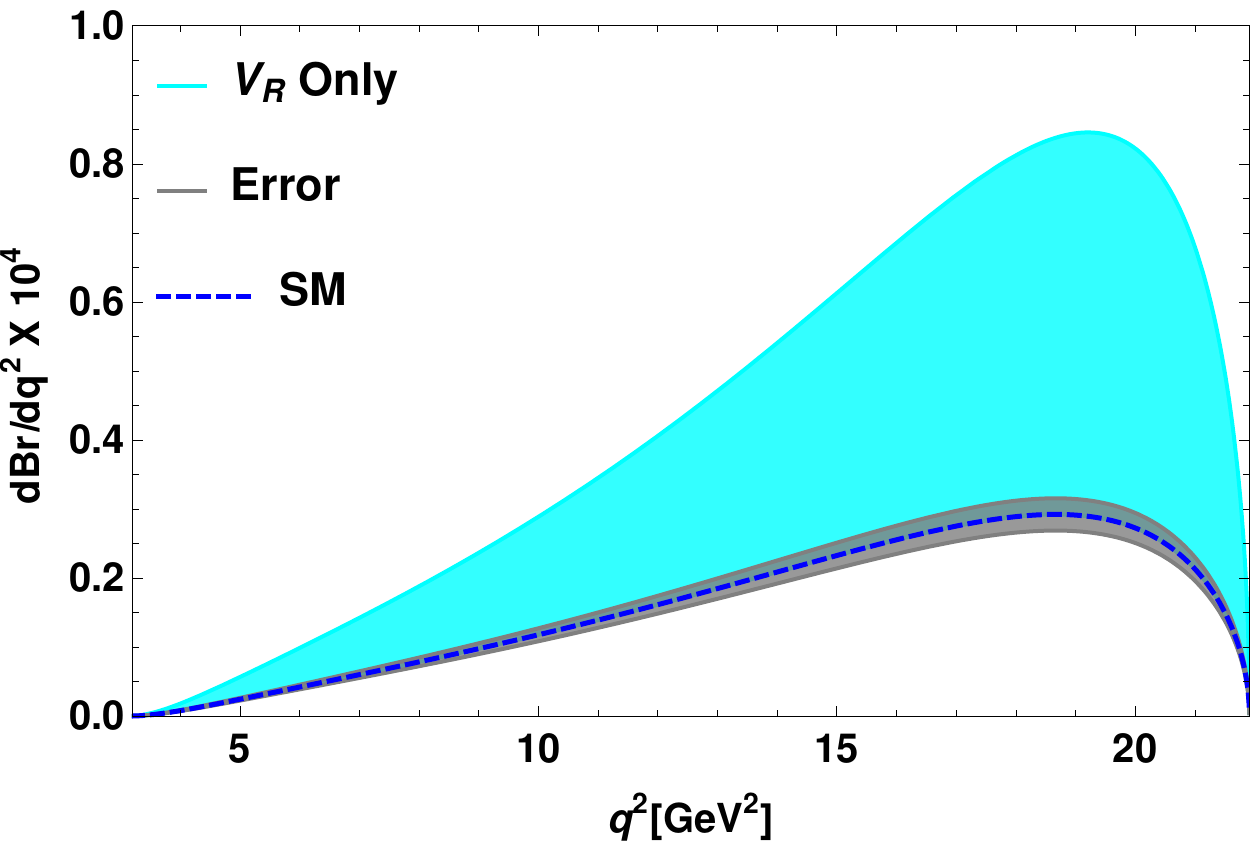}
\quad
\includegraphics[scale=0.55]{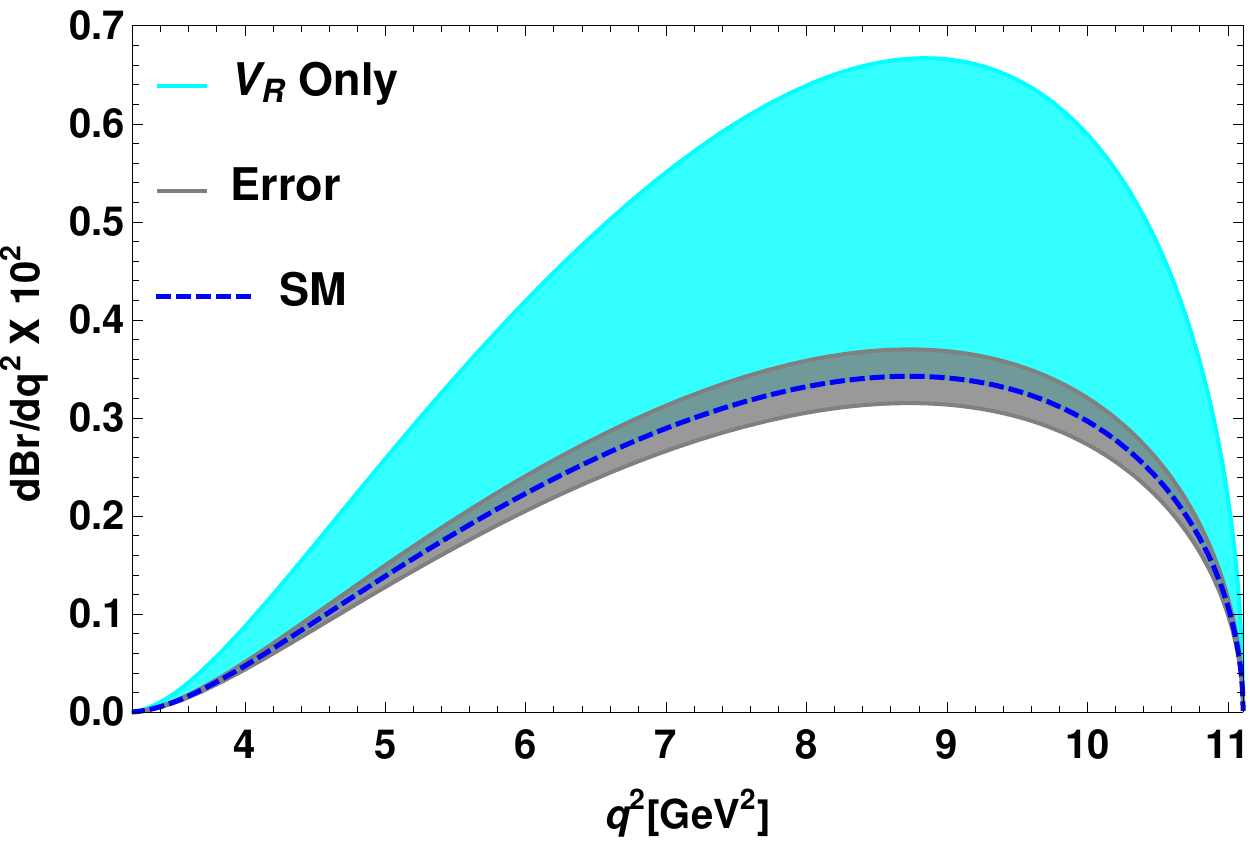}
\quad
\includegraphics[scale=0.55]{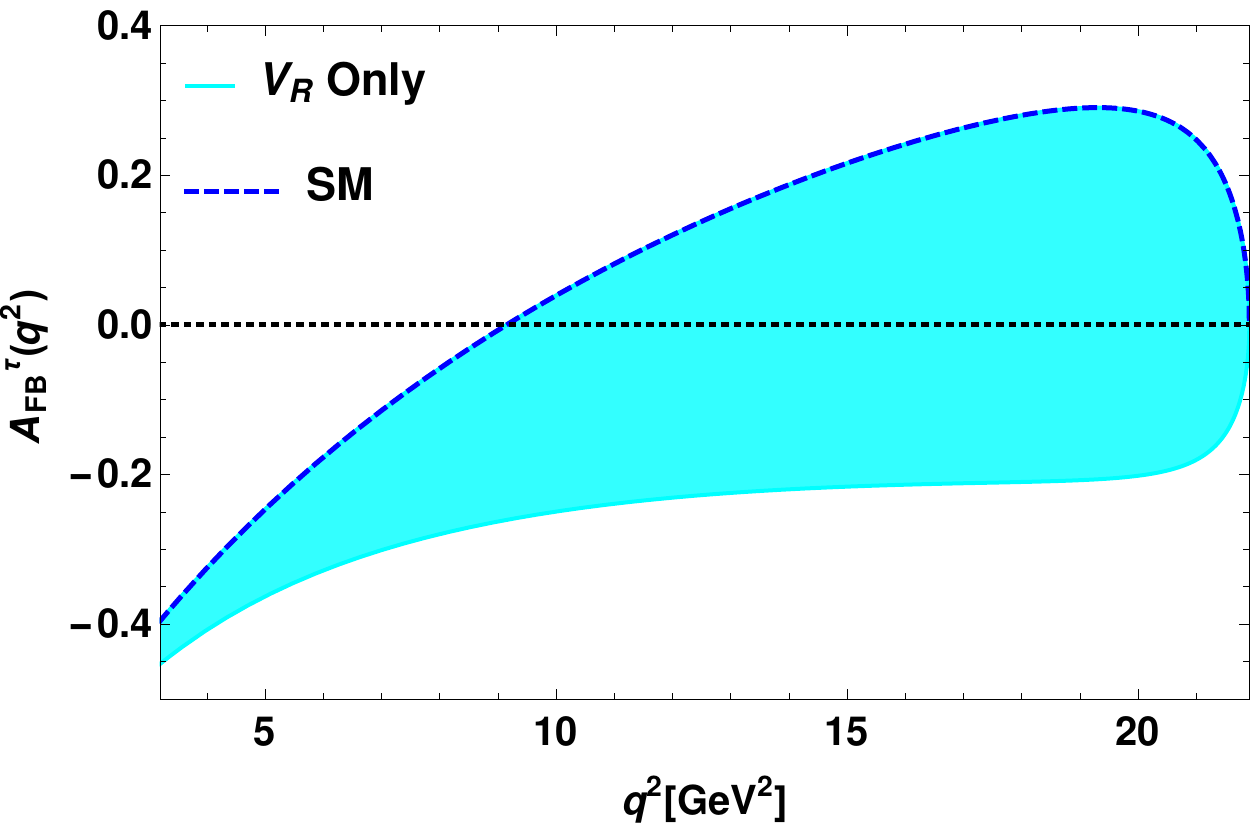}
\quad
\includegraphics[scale=0.55]{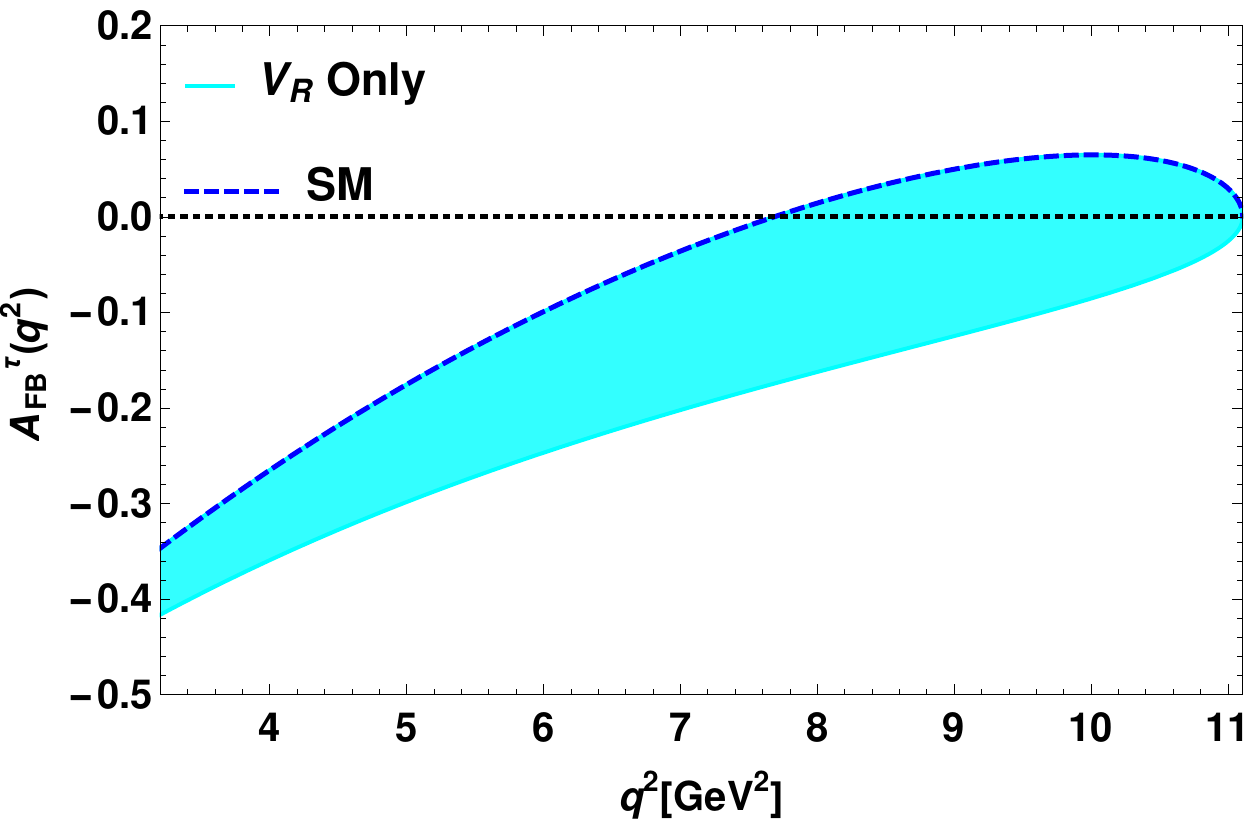}
\quad
\includegraphics[scale=0.55]{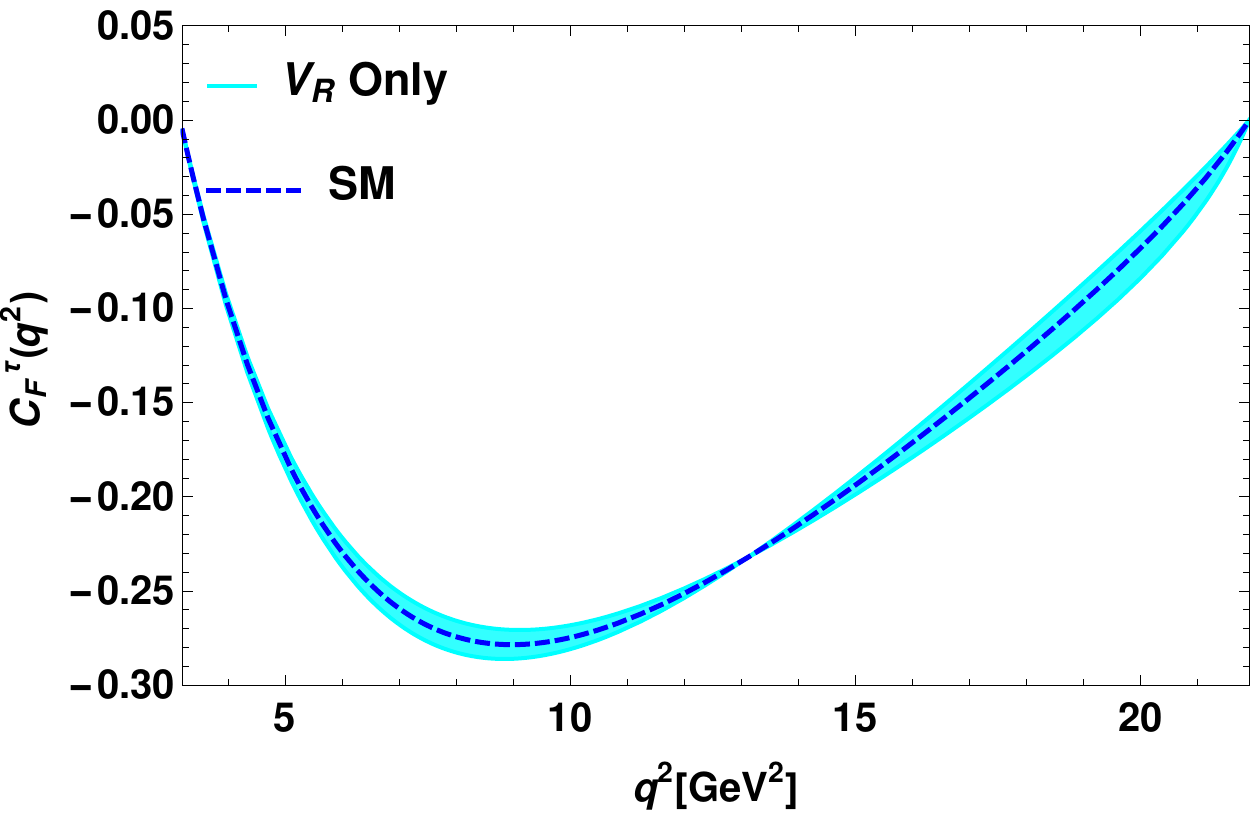}
\quad
\includegraphics[scale=0.55]{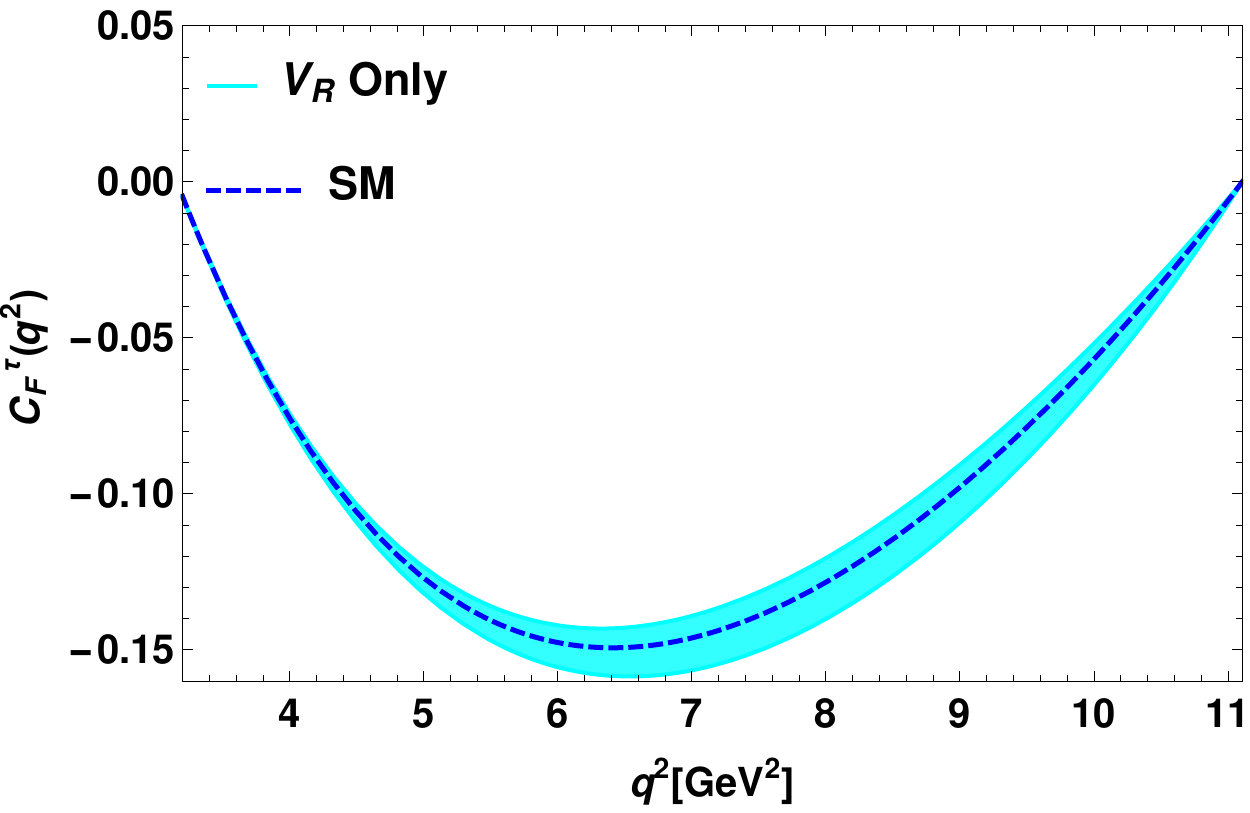}
\caption{Top panel represents the $q^2$ variation of branching ratio of $\Lambda_b \to p \tau^-  \bar \nu_\tau$ (left panel) and $\Lambda_b \to \Lambda_c^+ \tau^-  \bar \nu_\tau$ (right panel) for only $V_R$ new coefficient. The corresponding plots of forward backward asymmetry and the convexity parameters are shown in the middle and bottom panels respectively. Here cyan bands are due to the additional new physics contribution coming from only $V_R$ coefficient. }\label{Br-VR}
\end{figure}

\begin{figure}[htb]
\centering
\includegraphics[scale=0.55]{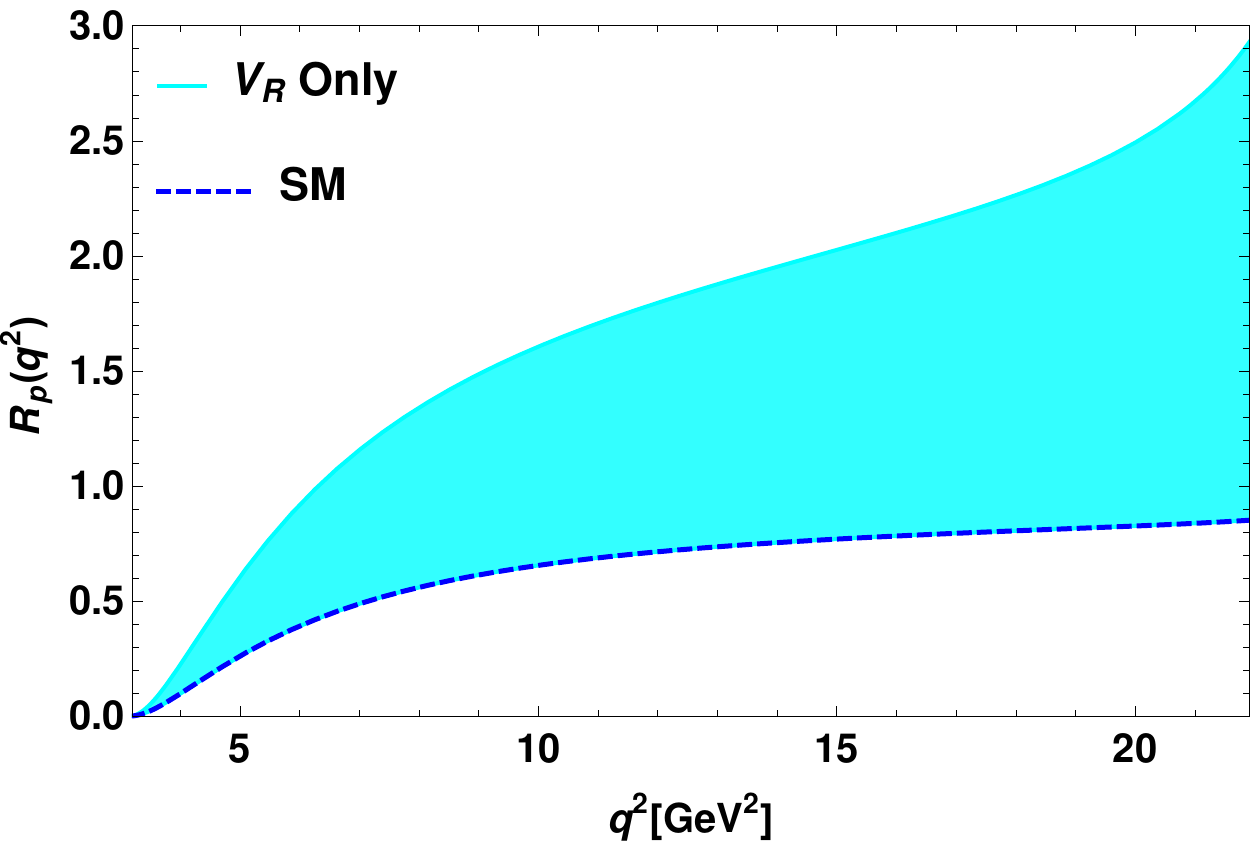}
\quad
\includegraphics[scale=0.55]{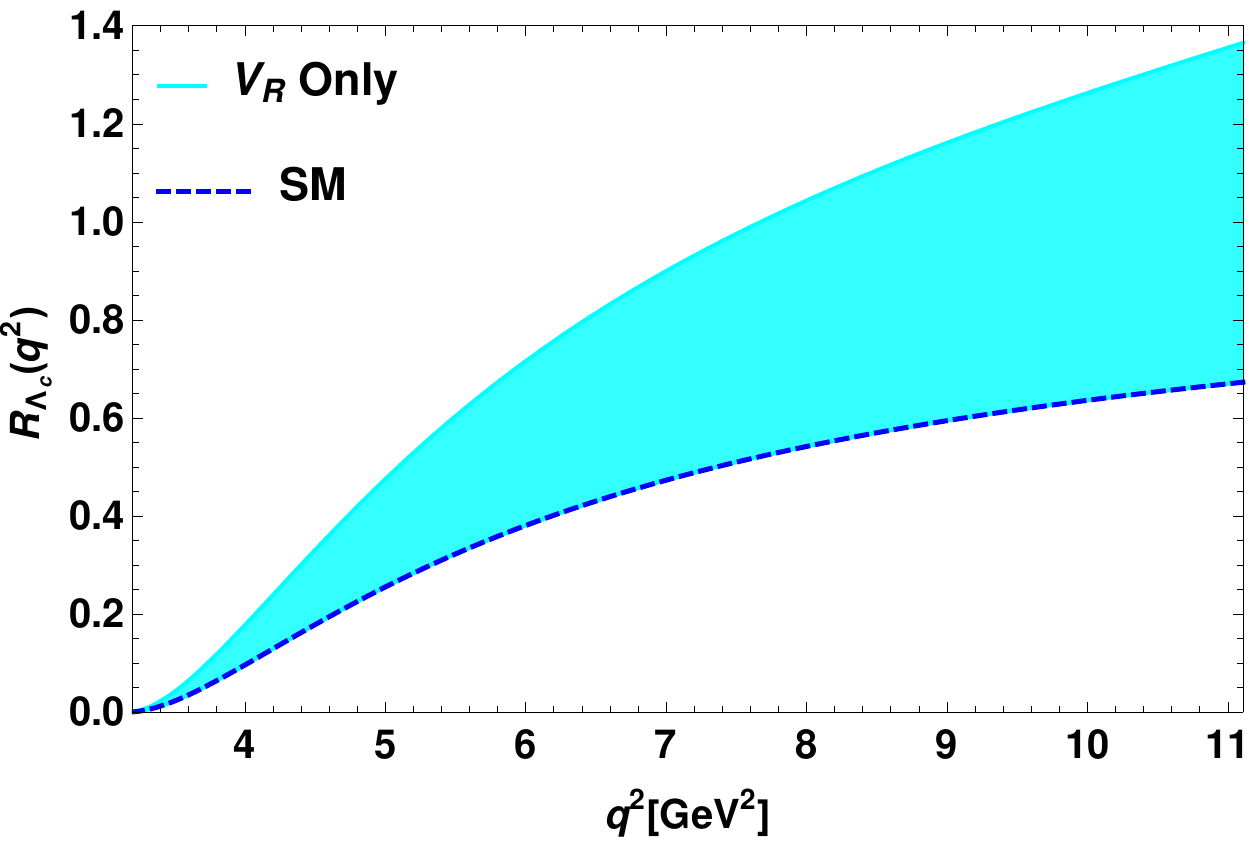}
\caption{The  variation of  $R_{p}$ (left panel) and $R_{\Lambda_c}$ (right  panel) LNU parameters with respect to $q^2$ in the presence of only $V_R$ new coefficient.} \label{LNU-VR}
\end{figure}

\begin{figure}[htb]
\centering
\includegraphics[scale=0.43]{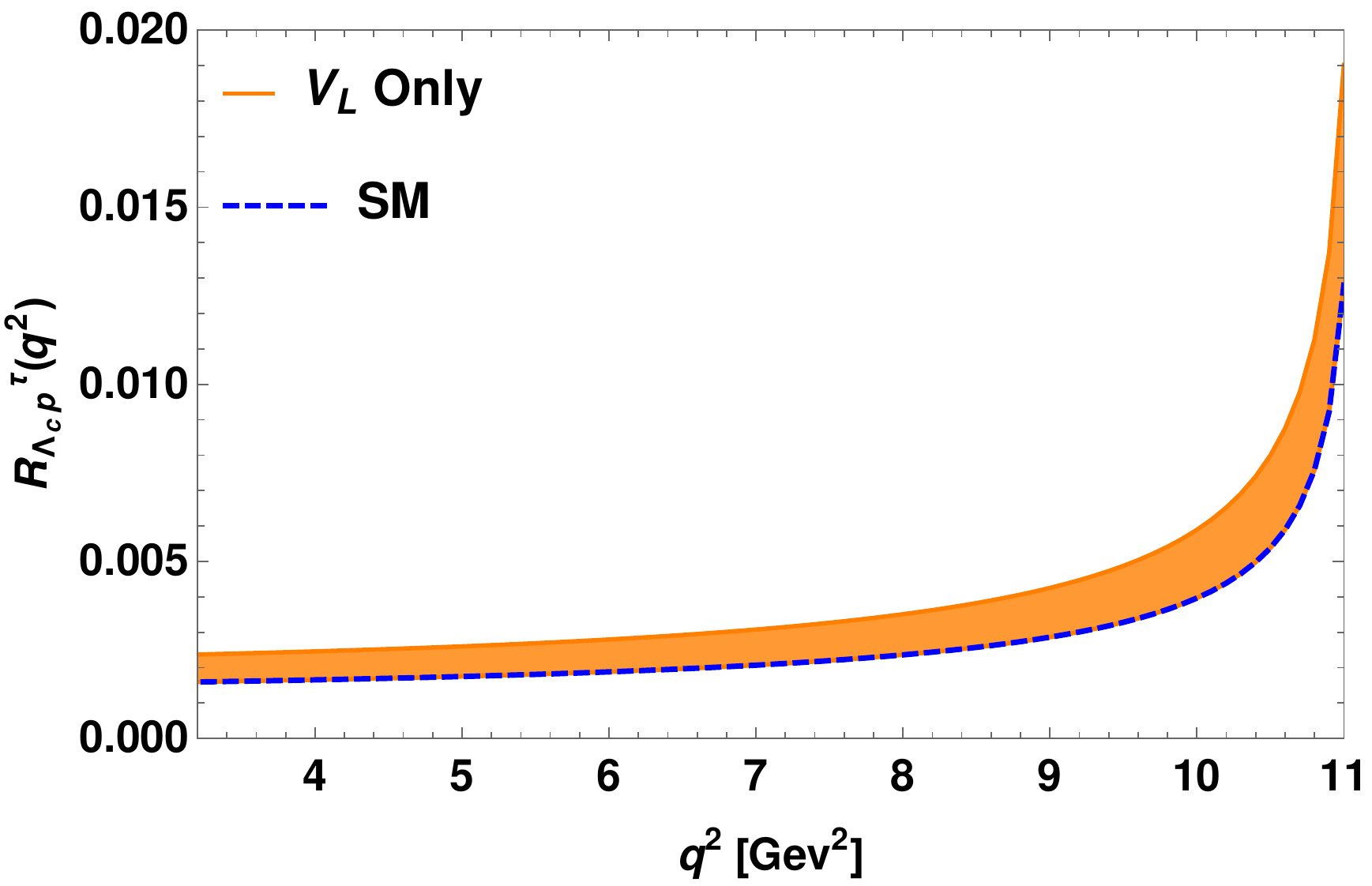}
\quad
\includegraphics[scale=0.43]{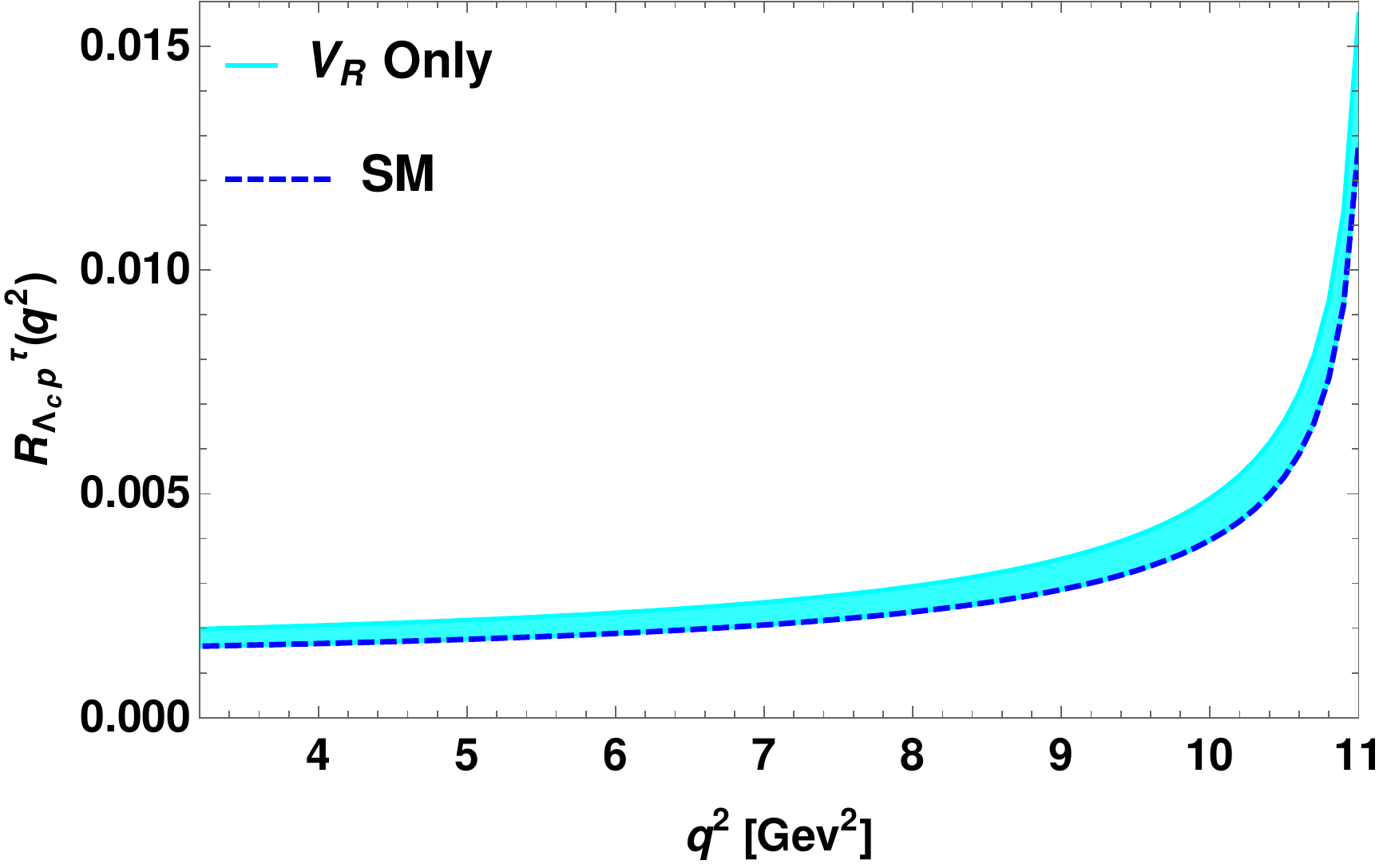}
\caption{The  variation of  $R_{\Lambda_c p}^\tau$ parameter with respect to $q^2$ in the presence of only $V_L$ (left panel) and $V_R$ (right panel) new coefficients.} \label{LNU-VLVR}
\end{figure}

Here, we assume that only the new $V_R$ coefficient
is present in addition to the SM contribution, in the effective Lagrangian (\ref{ham}). To investigate the effect of NP coming from $V_R$ coefficient, we first  constrain the  new coefficient by imposing $3\sigma$ experimental bound on the $b \to (u,c) \tau \bar \nu_\tau$ anomalies.  Using the values from Table \ref{Tab:con}\,, we show the plots for the branching ratios of $\Lambda_b \to p~(\Lambda_c) \tau \bar \nu_\tau$ process in the top-left panel (top-right panel) of Fig. \ref{Br-VR}\,. In these figures,  the cyan bands are due to the additional contribution from $V_R$ coefficient. We notice significant deviation in the branching ratios from their corresponding SM results. The predicted values of the branching ratios  for $V_R$ coefficient are presented in Table \ref{VLVR:Tab}\,.   Apart from branching ratios, we are also interested to  see the effect of this new coefficient on various $q^2$ dependent observables.  The $q^2$ variation of the forward backward asymmetry and the convexity parameters for $\Lambda_b \to p \tau^-  \bar \nu_\tau$ (left) and $\Lambda_b \to \Lambda_c \tau^-  \bar \nu_\tau$ (right) decay processes are depicted in the middle and bottom panels of Fig. \ref{Br-VR}\,, respectively. The deviation of convexity parameters from their SM prediction are quite noticeable  in these plots. In the presence of $V_R$ coefficient, the numerical values of the $C_F^\tau$ parameters are 
\bea
\langle C_{F}^\tau \rangle |_{\Lambda_b \to p}^{V_R} =-0.169\to -0.147\;,  ~~~~
\langle C_{F}^\tau \rangle |_{\Lambda_b \to \Lambda_c}^{V_R} =-0.105\to -0.094\;.
\eea
 The effect of  $V_R$ coefficient is found to be rather significant  on  the forward-backward asymmetry observables of both  $\Lambda_b \to p(\Lambda_c) \tau^-  \bar \nu_\tau$ decay modes and the corresponding numerical values are 
\bea
&&\langle A_{FB}^\tau \rangle |_{\Lambda_b \to p}^{ V_R} =-0.248\to 0.115\;, ~~~
\langle A_{FB}^\tau \rangle |_{\Lambda_b \to \Lambda_c}^{V_R } =-0.23 \to -0.09\;.
\eea 
Left and right panels of Fig. \ref{LNU-VR}, depict the variation of  $R_p$ and $R_{\Lambda_c}$  parameters   with respect to $q^2$.  Though there are no experimental limits on these parameters,  significant deviation  from their SM values are noticed in the scenario with only $V_R$ coupling. The right panel of Fig. \ref{LNU-VLVR}  represents the $q^2$ variation of $R_{\Lambda_c p}^\tau$ parameter.   The corresponding numerical values are listed in Table \ref{VLVR:Tab}\,. 
\begin{table}[htb]
\centering
\caption{The predicted values of branching ratios and lepton non-universality parameters of $\Lambda_b  \to (\Lambda_c, p) \tau \bar \nu_\tau$  processes in the SM and in the presence of only $V_{L,R}$  coefficients.} \label{VLVR:Tab}
\begin{tabular}{|c|c|c|c|}
\hline
Observables~&~SM prediction~&~Values for $V_L$ coupling~&~Values for $V_R$ coupling~\\
\hline
\hline

${\rm Br}(\Lambda_b \to p \tau^- \bar \nu_\tau)$~&~$(2.98\pm 0.238)\times 10^{-4}$ ~&~$(0.298-1.34)\times 10^{-3}$ ~&~$(2.98-8.17)\times 10^{-4}$\\
$R_p$~&~$0.692$~&~$0.692-3.09$~&~$0.692-1.895$\\
\hline
${\rm Br}(\Lambda_b \to \Lambda_c^+ \tau^- \bar \nu_\tau)$~&~$(1.76\pm 0.14)\times 10^{-2}$ ~&~$(1.76-5.29)\times 10^{-2}$ ~&~$(1.76-3.4)\times 10^{-2}$\\
 
$R_{\Lambda_c}$~&~$0.353$~&~$0.353-1.06$~&~$0.353-0.68$\\
\hline
$R_{\Lambda_c p}$~&~$(1.693\pm 0.19)\times 10^{-2}$~&~$(1.693-2.533)\times 10^{-2}$~&~ $(1.693-2.4)\times 10^{-2}$\\
\hline
\end{tabular}
\end{table}
\begin{figure}[h]
\centering
\includegraphics[scale=0.55]{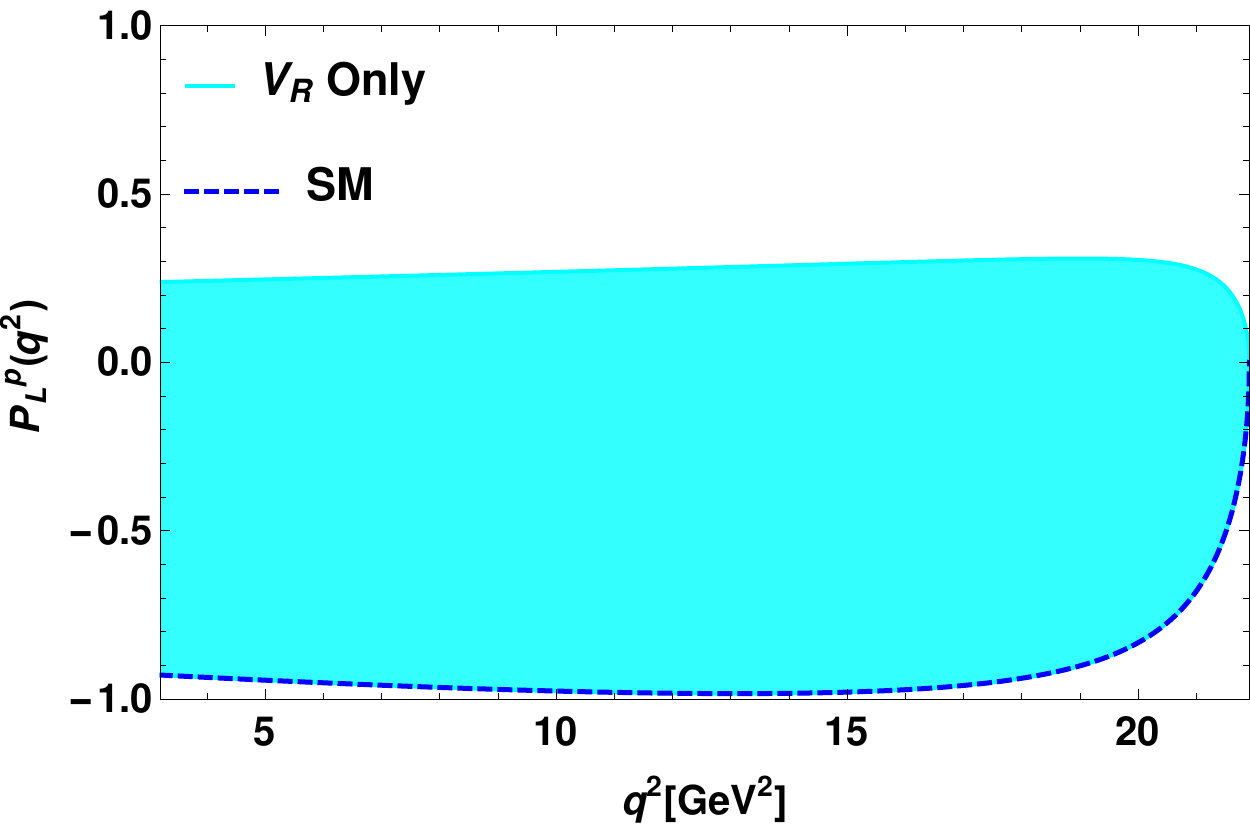}
\quad
\includegraphics[scale=0.55]{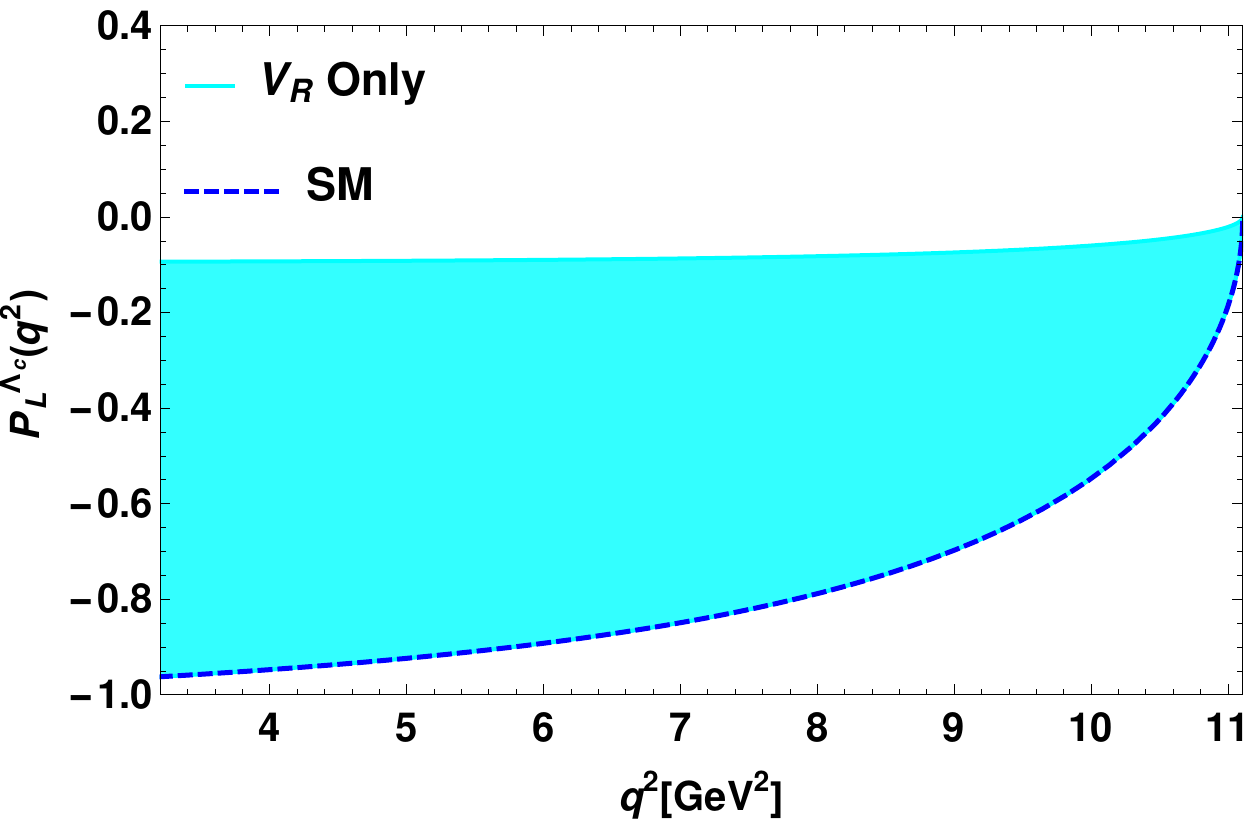}\\
\includegraphics[scale=0.55]{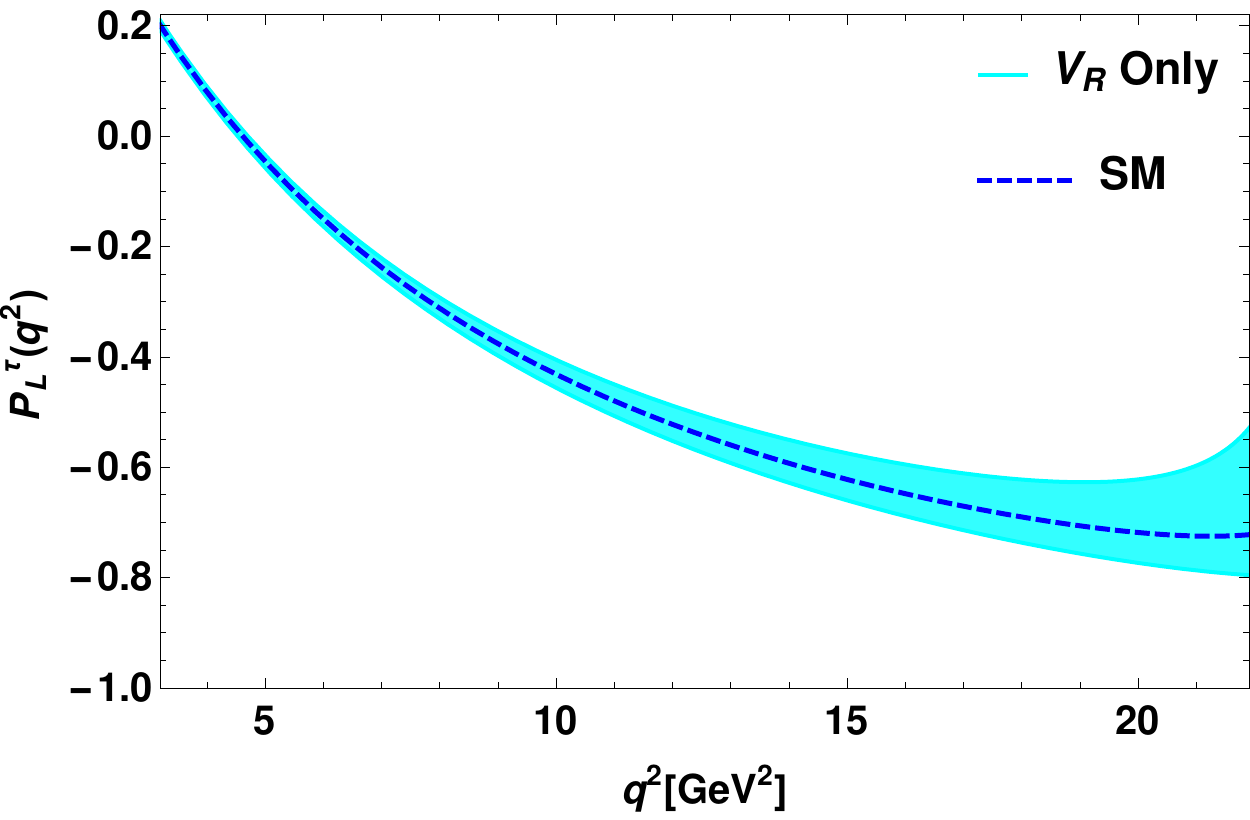}
\quad
\includegraphics[scale=0.55]{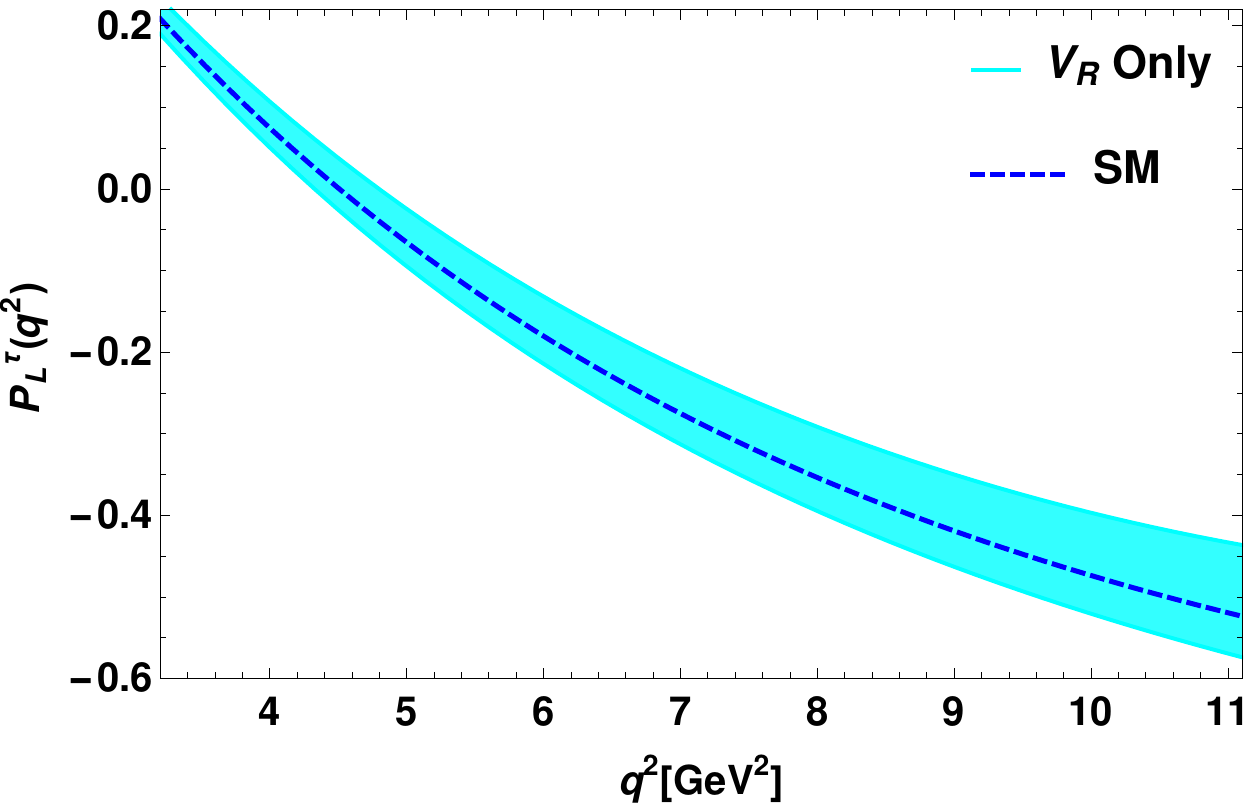}\\
\caption{The plots in the left panel represent the longitudinal  polarizations of daughter light baryon $p$ (left-top panel) and the charged $\tau$ lepton (left-bottom)     with respect to $q^2$ for only $V_R$ coefficient. The corresponding   plots for $\Lambda_b \rightarrow \Lambda_c \tau^- \bar \nu_\tau$ mode are shown in the right panel.} \label{hp-VR}
\end{figure}

Though the presence of  $V_L$ coefficient has no effect on the lepton and hadron polarization asymmetries of  $b \to (u,c) \tau \bar \nu_\tau$ decay modes,  the  $V_R$ coefficient has significant impact on these parameters.  In the top panel of Fig. \ref{hp-VR}\,, the distribution of the longitudinal  polarization  components of the daughter baryon $p$ (left panel) and $\Lambda_c$ (right panel) are shown both in the SM and in the  presence of only $V_R$ coefficient, and the corresponding plots for the charged $\tau$ lepton   are presented in the bottom  panel. The integrated values of  the hadron longitudinal polarization asymmetry parameters in the
full physical phase space are 
\bea
&&\langle P_L^p \rangle |_{\Lambda_b \to p}^{\rm SM} =-0.897\;, \qquad \qquad \qquad
\langle P_L^{\Lambda_c} \rangle |_{\Lambda_b \to \Lambda_c}^{\rm SM} =-0.797\;,\\
&&\langle P_L^p \rangle |_{\Lambda_b \to p}^{ V_R ~\rm Only} =-0.897\to 0.276\;, \qquad
\langle P_L^{\Lambda_c} \rangle |_{\Lambda_b \to \Lambda_c}^{\rm V_R ~\rm Only} =-0.797 \to -0.068\;,
\eea
and the corresponding numerical values for the charged lepton $\tau$,  are
\bea
&&\langle P_L^\tau \rangle |_{\Lambda_b \to p}^{\rm SM} =-0.514\;, \qquad \qquad \qquad ~~
\langle P_L^{\tau} \rangle |_{\Lambda_b \to \Lambda_c}^{\rm SM} =-0.207\;,\\
&&\langle P_L^\tau \rangle |_{\Lambda_b \to p}^{ V_R } =-0.577\to -0.433\;, \qquad
\langle P_L^{\tau} \rangle |_{\Lambda_b \to \Lambda_c}^{V_R} =-0.25 \to -0.146\;.
\eea

\subsection{Scenario C: Only $S_L$ coefficient}
\begin{figure}[htb]
\centering
\includegraphics[scale=0.55]{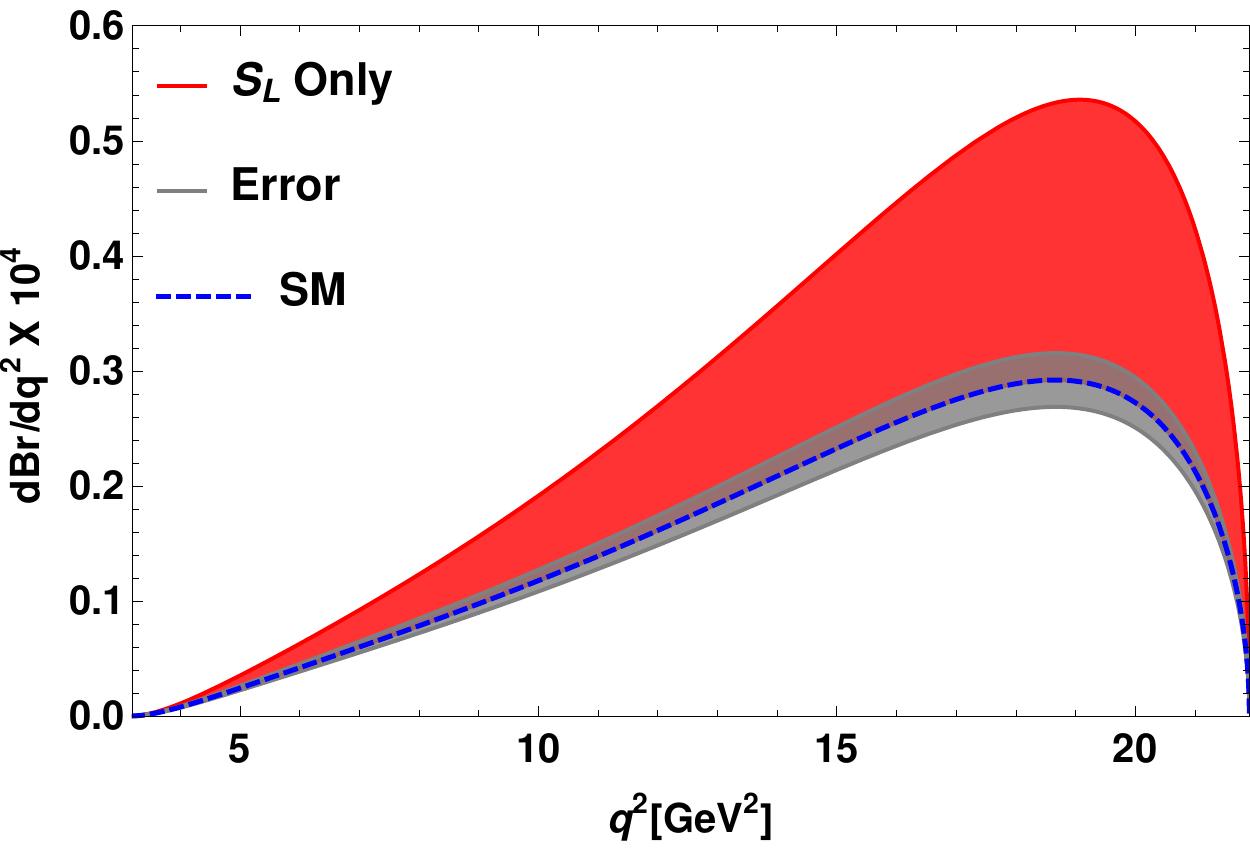}
\quad
\includegraphics[scale=0.55]{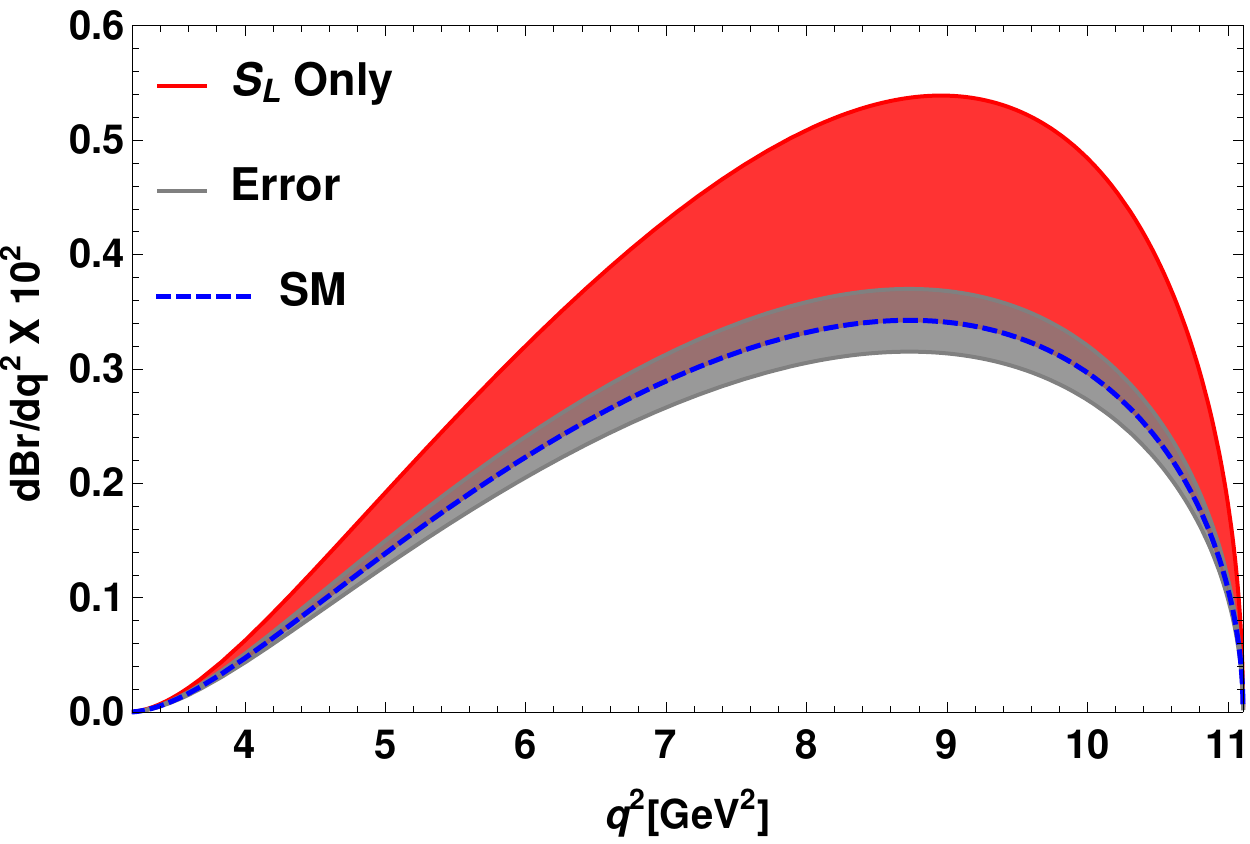}
\quad
\includegraphics[scale=0.55]{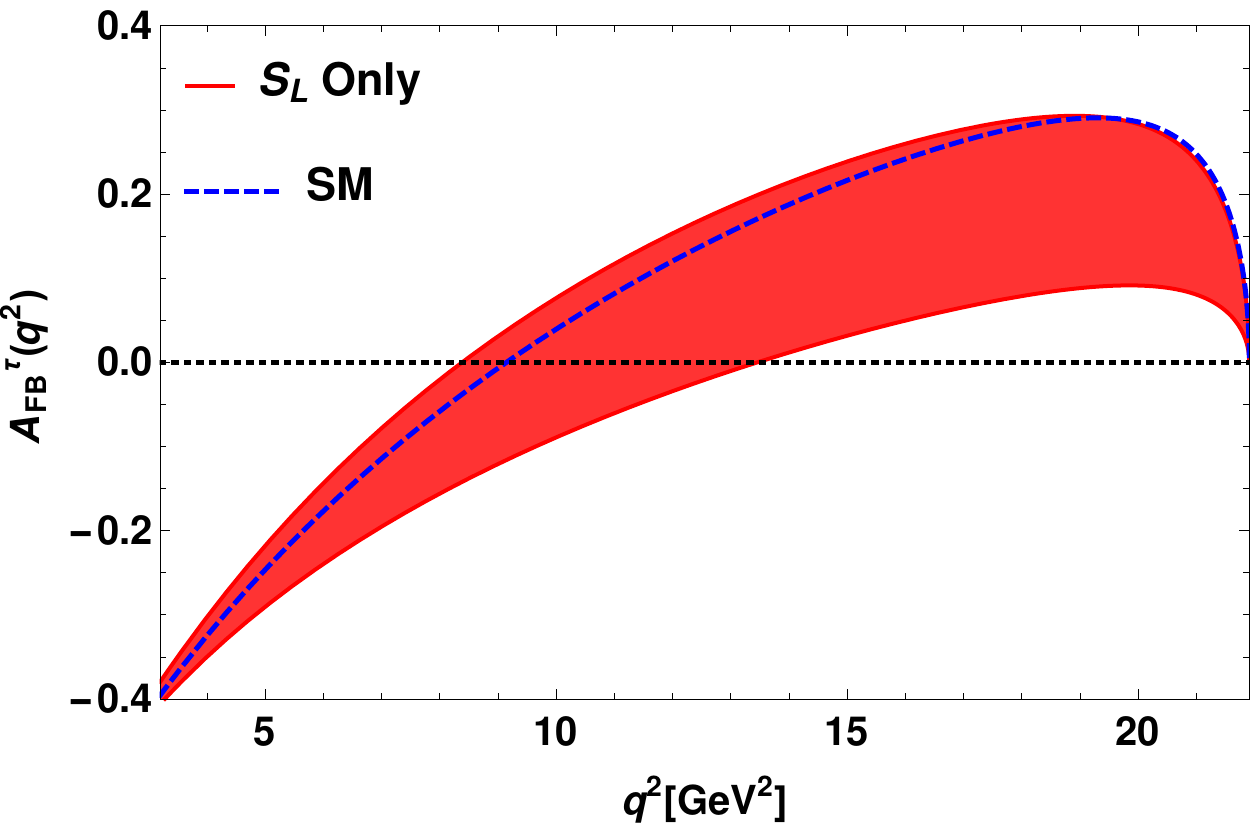}
\quad
\includegraphics[scale=0.55]{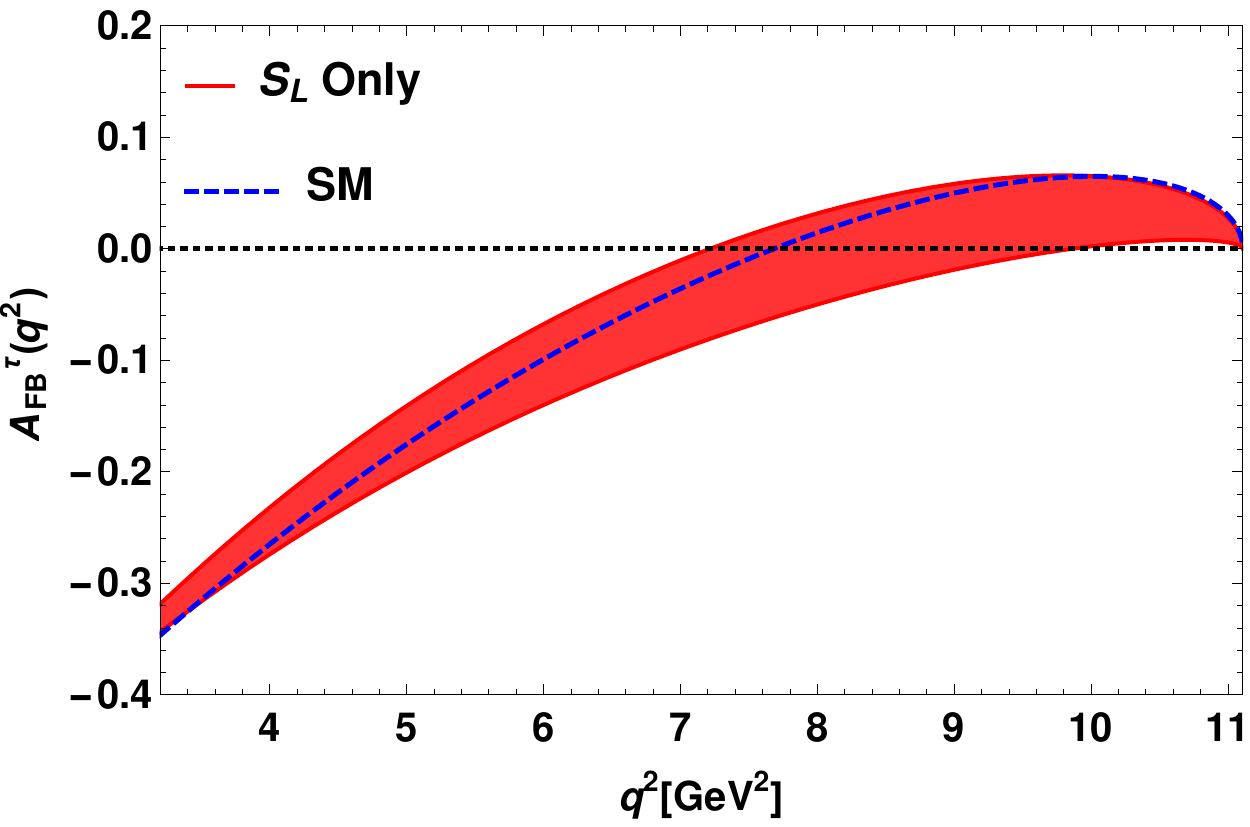}
\caption{Top panel represents the $q^2$ variation of branching ratios of  $\Lambda_b \to p \tau^-  \bar \nu_\tau$ (left panel) and $\Lambda_b \to \Lambda_c^+ \tau^-  \bar \nu_\tau$ (right panel) decay modes  in the presence of only $S_L$ new coefficient. The corresponding plots for forward-backward asymmetries  are shown in the bottom panel. Here red bands are due to  the additional new physics contribution coming from only $S_L$ coefficient. } \label{Br-SL}
\end{figure}

\begin{figure}[htb]
\centering
\includegraphics[scale=0.55]{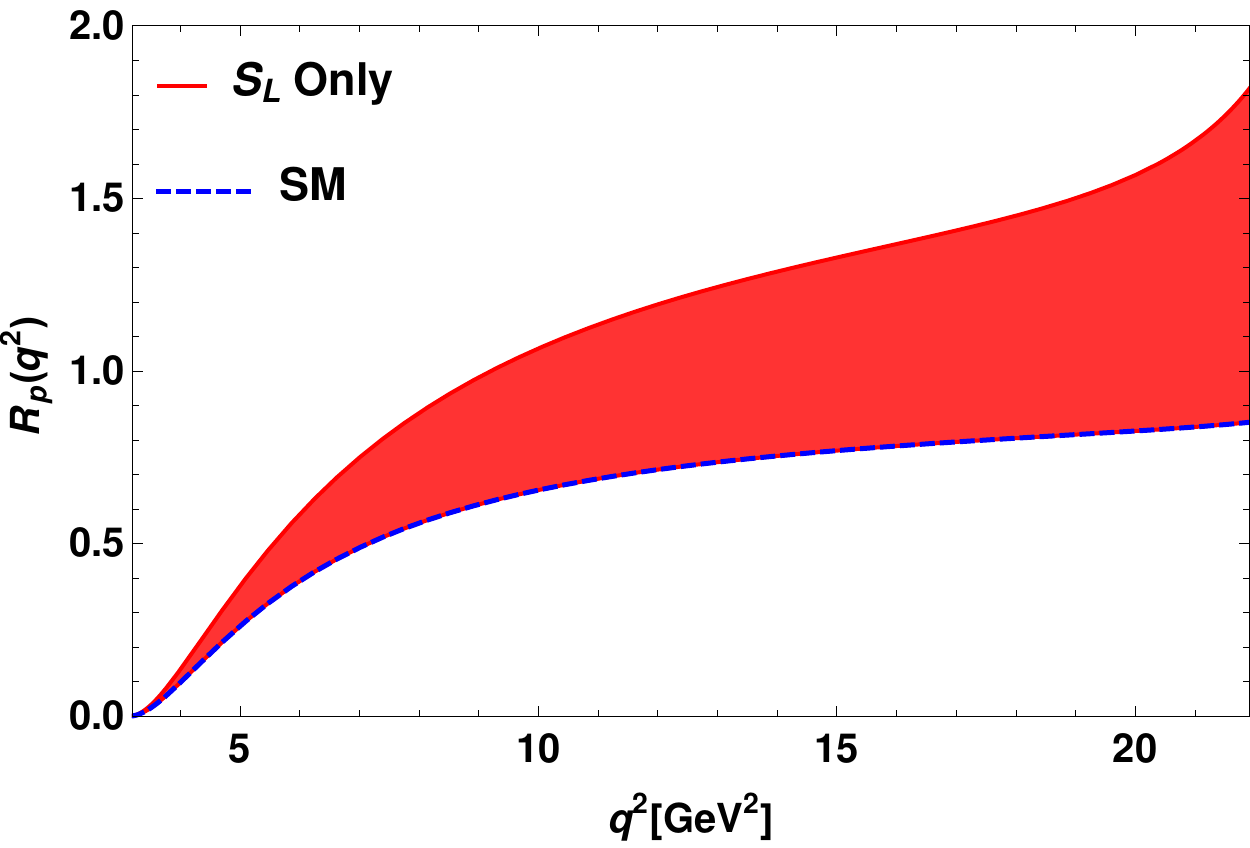}
\quad
\includegraphics[scale=0.55]{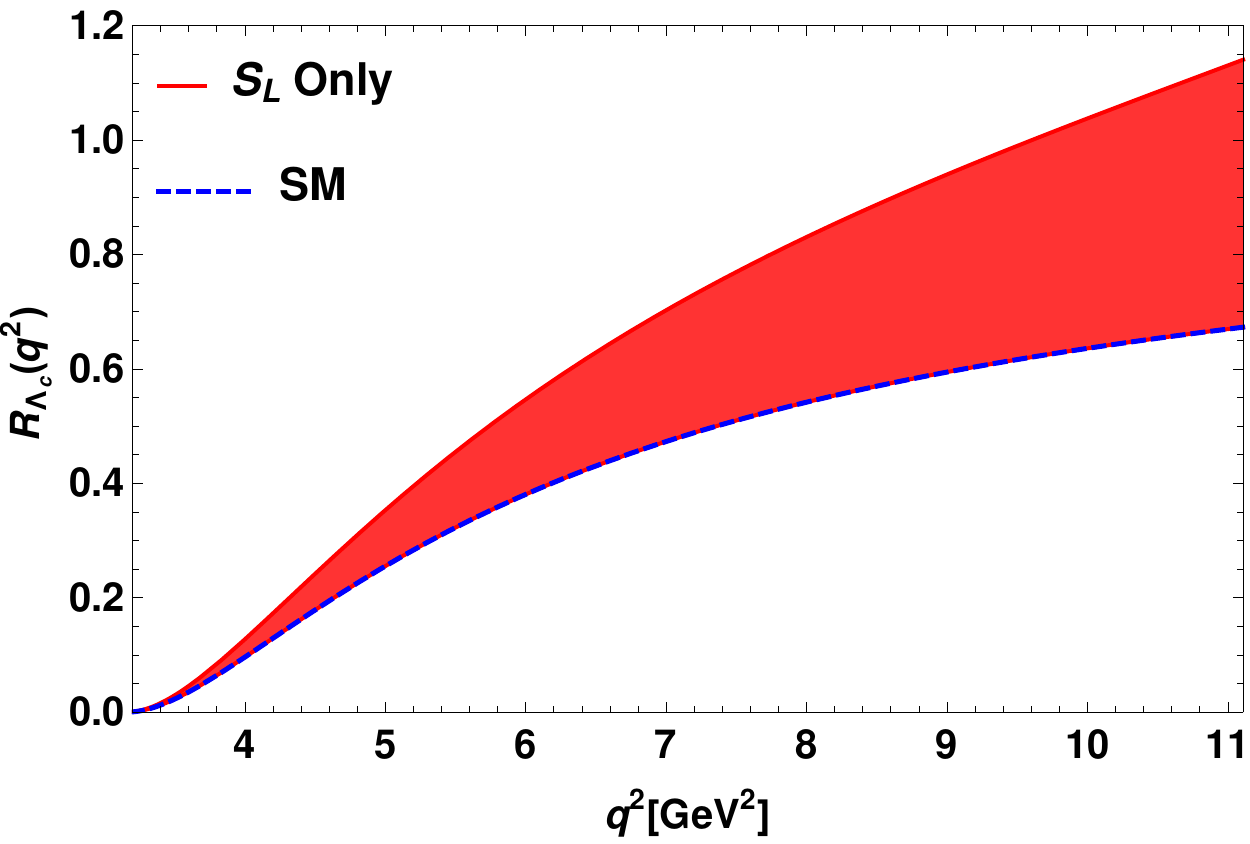}
\caption{The  variation of $R_p$ (left panel) and  $R_{\Lambda_c}$ (right panel) with respect to $q^2$  in the presence of only $S_L$  coefficient.}\label{LNU-SL}
\end{figure}

\begin{figure}[h]
\centering
\includegraphics[scale=0.55]{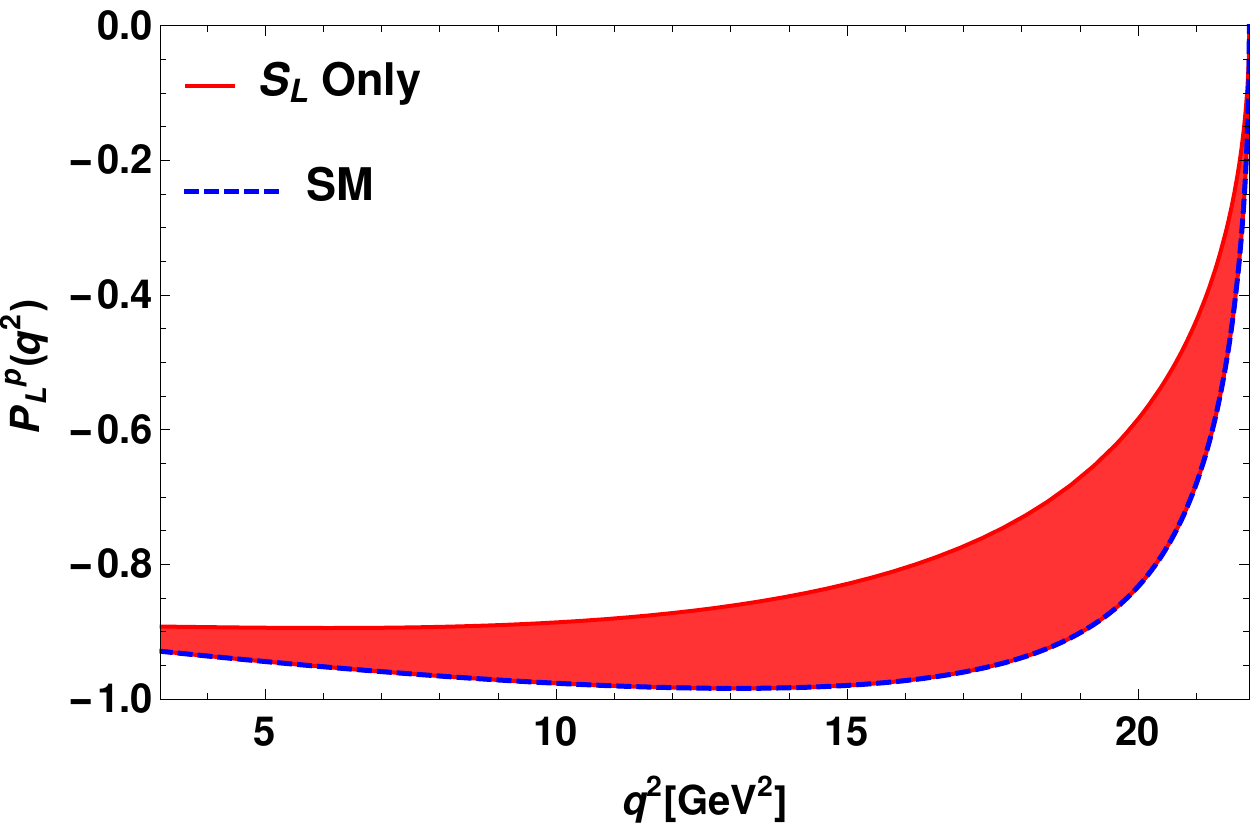}
\quad
\includegraphics[scale=0.55]{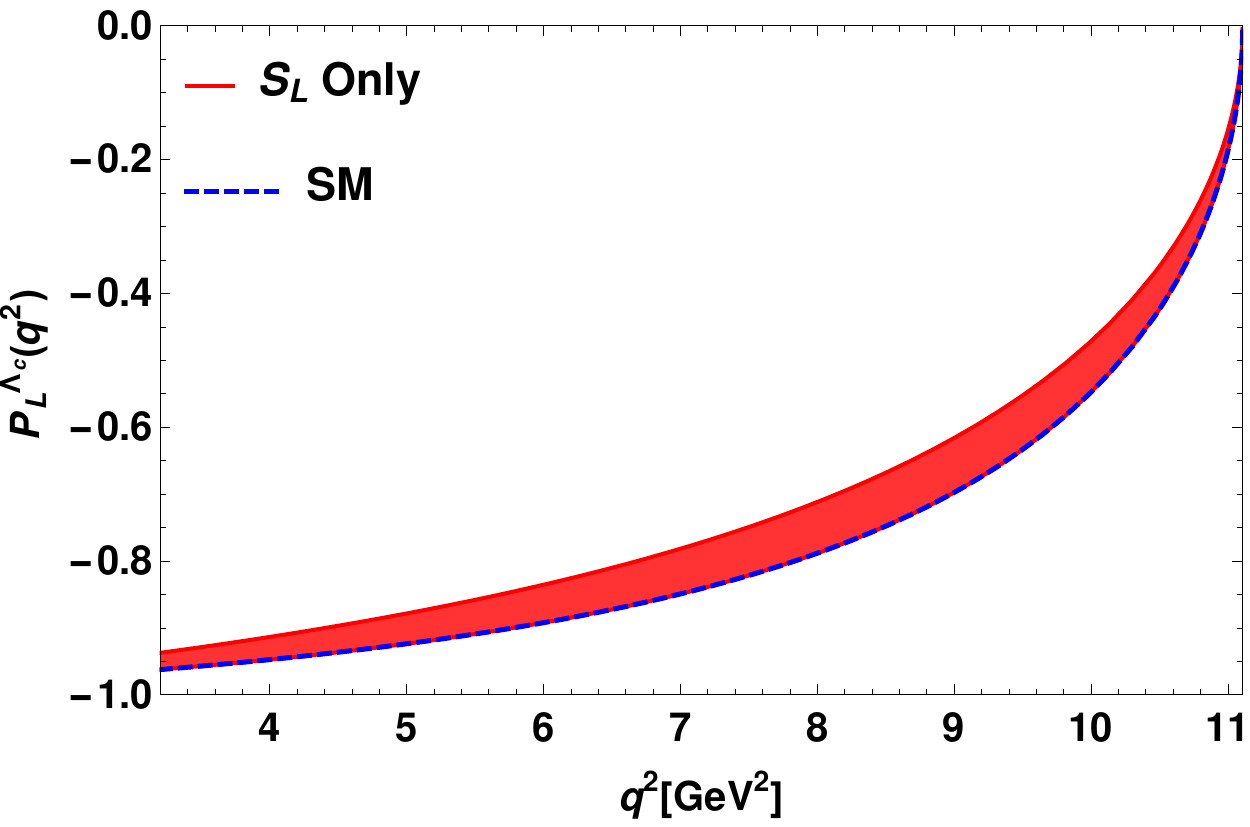}\\
\includegraphics[scale=0.55]{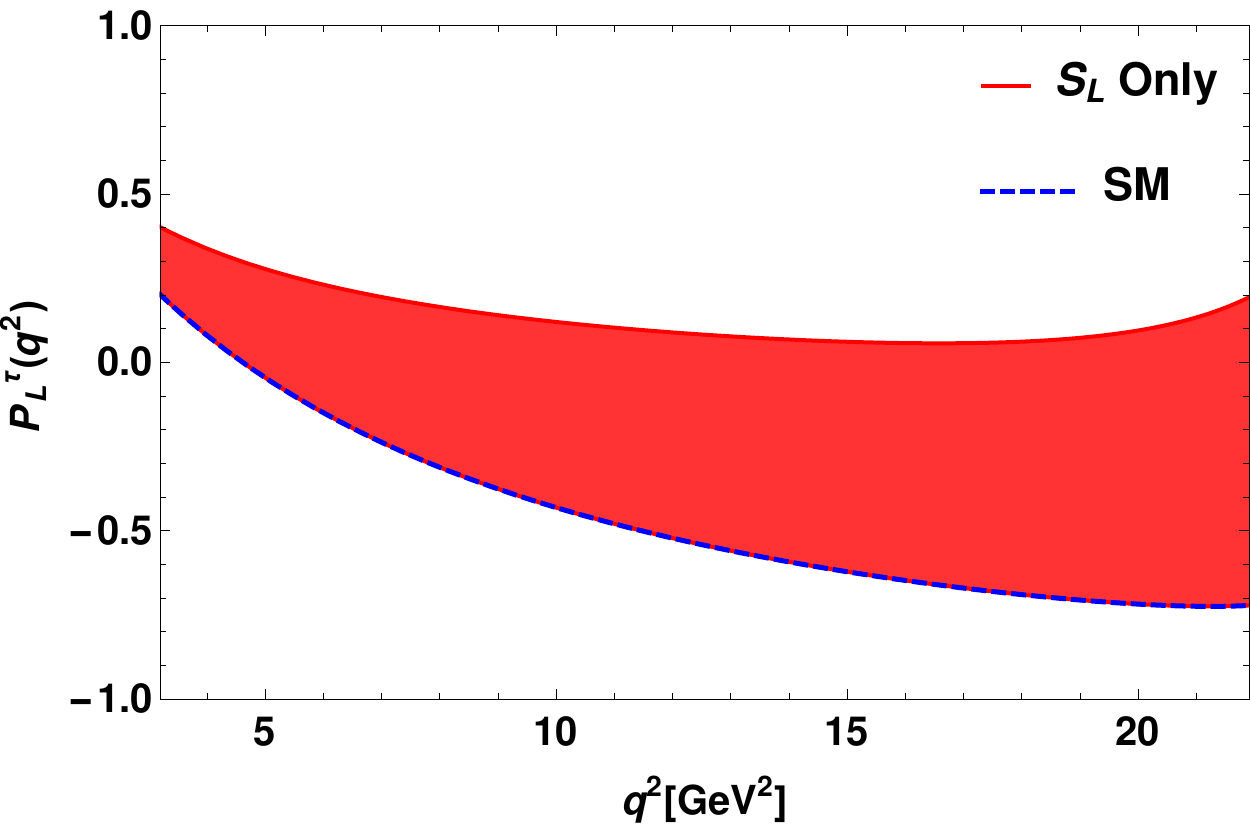}
\quad
\includegraphics[scale=0.55]{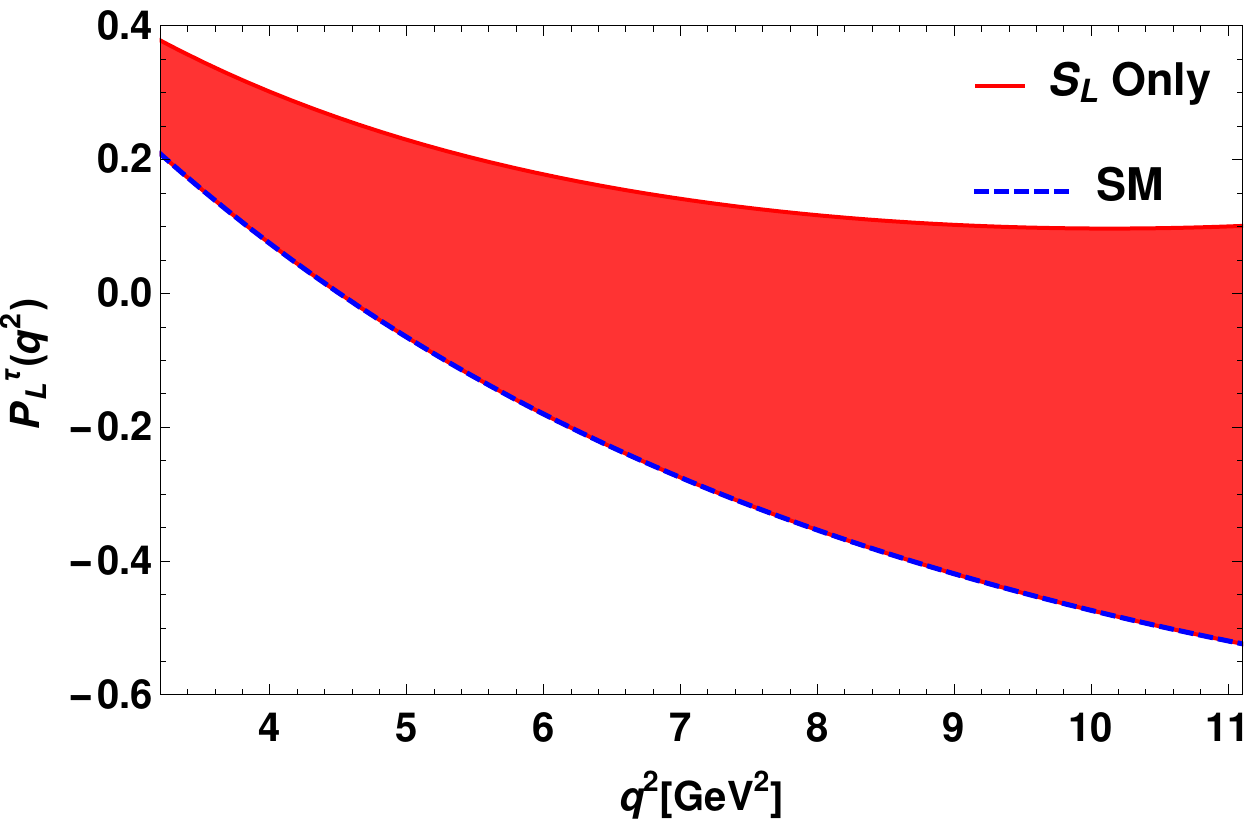}\\
\caption{The plots in the left panel represent the longitudinal  polarizations of daughter light baryon $p$ (left-top panel) and the charged $\tau$ lepton (left-bottom)     with respect to $q^2$ for only $S_L$ coefficient. The corresponding   plots for $\Lambda_b \rightarrow \Lambda_c \tau^- \bar \nu_\tau$ mode are shown in the right panel.} \label{hp-SL}
\end{figure}

Here, we explore the impact of only $S_L$ coefficient on the angular observables of heavy-heavy and heavy-light semileptonic decays of $\Lambda_b$ baryon.   In section III, we discussed  the constraints on the $S_L$ coupling. In the top panel Fig. \ref{Br-SL}\,, we present the plots for the differential branching ratios of $\Lambda_b \to p \tau \bar \nu_\tau$ (left) and $\Lambda_b \to \Lambda_c \tau \bar \nu_\tau$ (right) decay processes with respect to $q^2$ in the presence of  $S_L$ coefficient.  The corresponding plots for the forward-backward asymmetry are shown in the bottom panel.  In these figures, the red bands stand for the NP contribution from $S_L$ coefficient.  The additional  contributions  provide   deviation in the branching ratios and forward-backward asymmetries from their SM values. The $q^2$ variation of the $R_p$ (left panel) and $R_{\Lambda_c}$ (right panel) LNU parameters in the presence of $S_L$ coupling are given in Fig. \ref{LNU-SL}\,. In the presence of only $S_L$ coupling, the longitudinal polarization components of the $p$ (top-left panel) and $\Lambda_c$ (top-right panel) daughter baryons with respect to $q^2$ are presented in the top panel of Fig. \ref{hp-SL}\, and the bottom panel depicts the longitudinal lepton polarization asymmetry parameters for $\Lambda_b \to p (\Lambda_c) \tau \bar \nu_\tau$ processes. The lepton polarization asymmetry parameters provide profound deviation from the SM in comparison to their longitudinal hadron  polarization parameters. The top-left panel of Fig. \ref{LNU-SLSR}\, shows the variation of   $R_{\Lambda_c p}^\tau$ parameter with $q^2$. In Table \ref{SLSR:Tab}\,, we report the numerical values of all these parameters. 

\subsection{Scenario D: Only $S_R$ coefficient}
\begin{figure}[htb]
\centering
\includegraphics[scale=0.55]{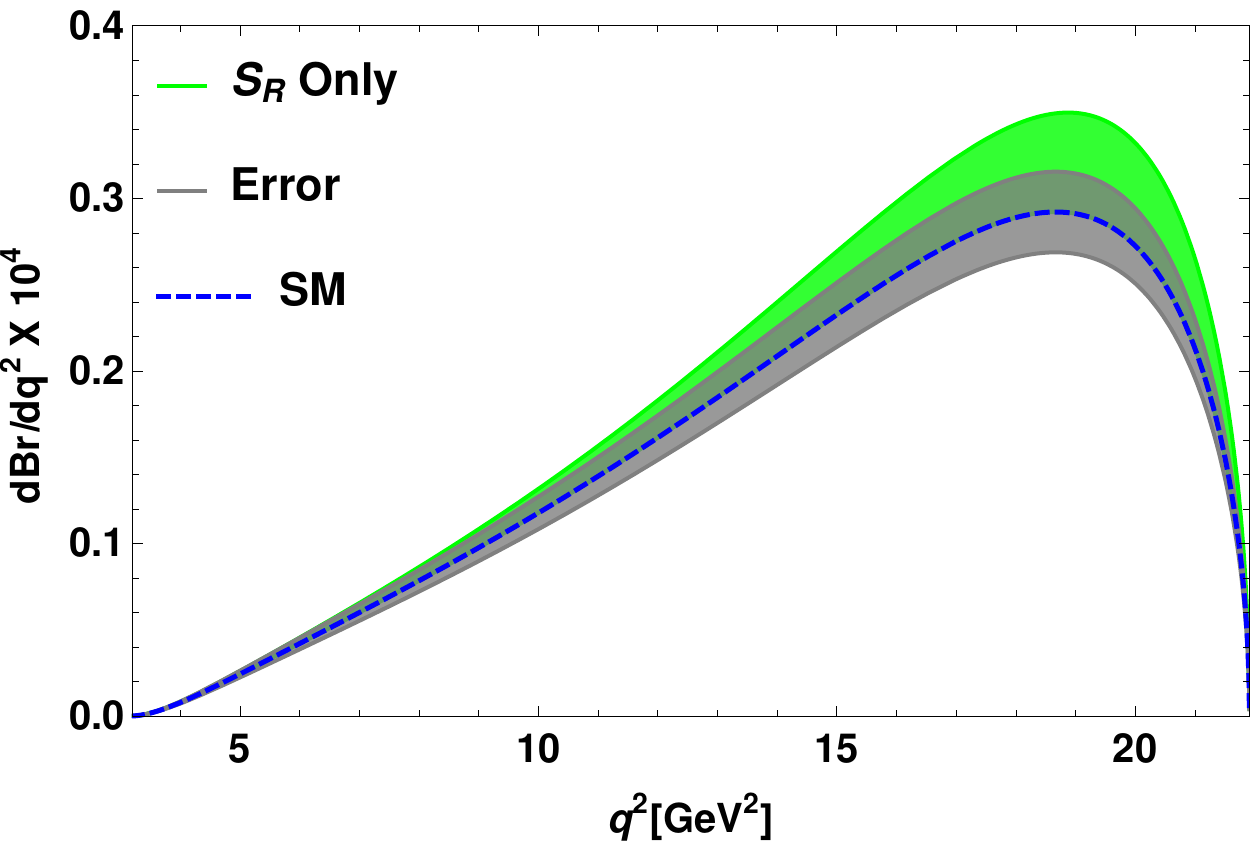}
\quad
\includegraphics[scale=0.55]{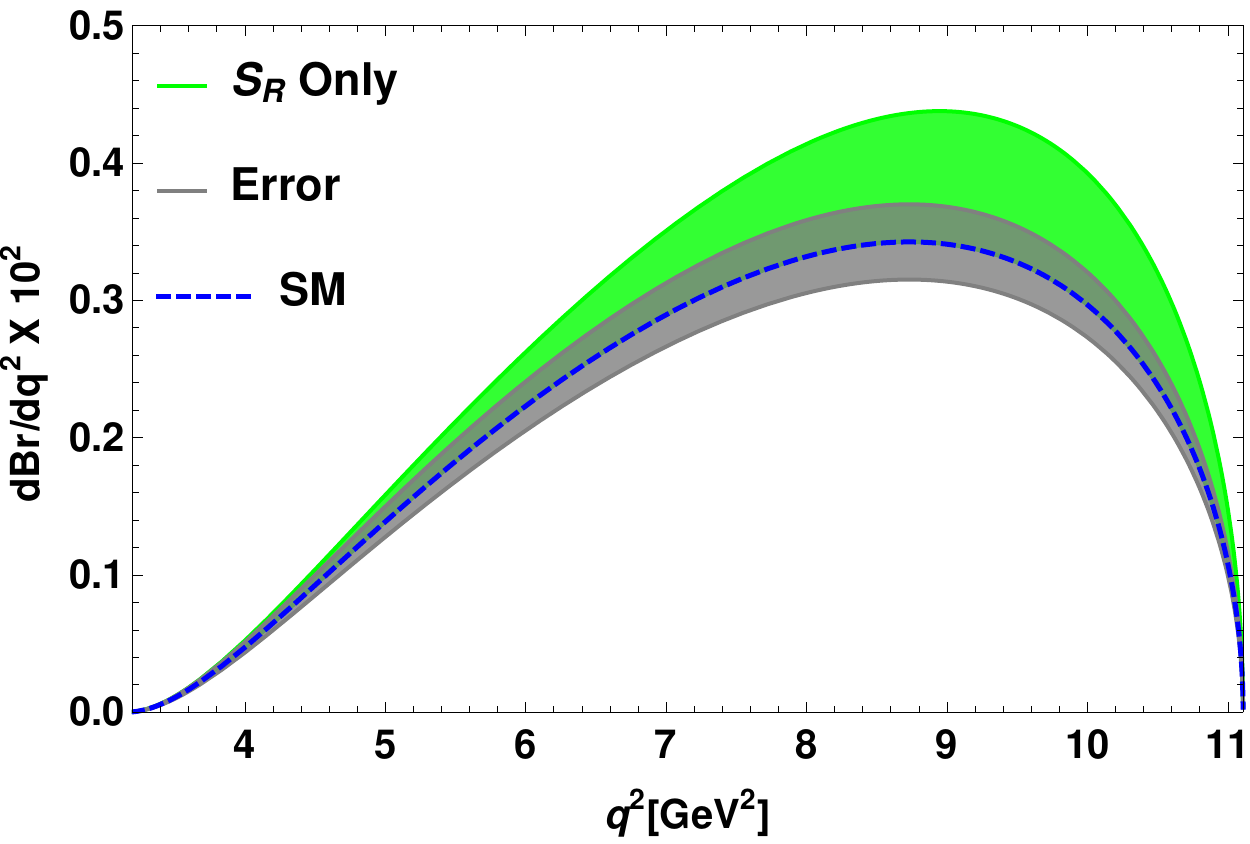}
\quad
\includegraphics[scale=0.55]{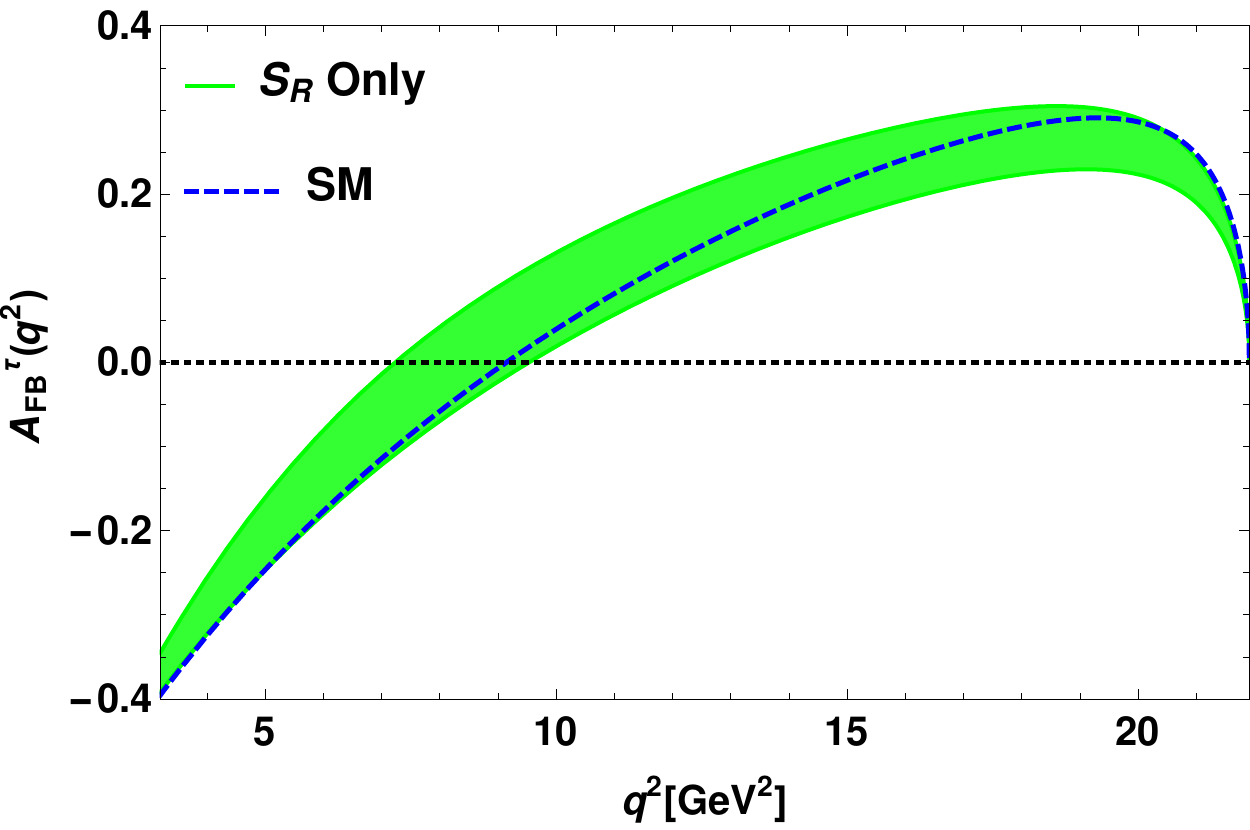}
\quad
\includegraphics[scale=0.55]{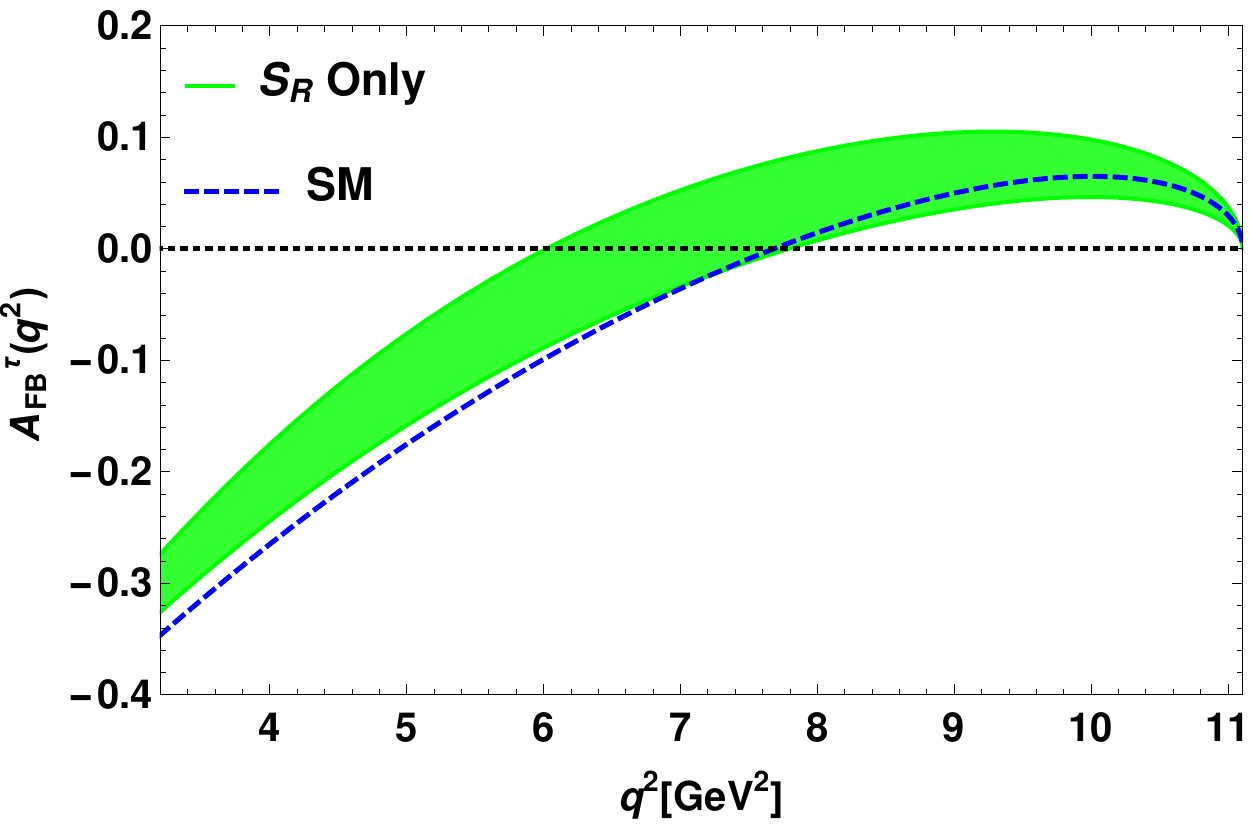}
\caption{Top panel represents the $q^2$ variation of branching ratios of  $\Lambda_b \to p \tau^-  \bar \nu_\tau$ (left panel) and $\Lambda_b \to \Lambda_c^+ \tau^-  \bar \nu_\tau$ (right panel) decay processes in the presence of only $S_R$ coefficient. The corresponding plots for  the forward-backward asymmetries are shown in the bottom panel. Here green bands stand for the additional new physics contribution coming from only $S_R$ coefficient. } \label{Br-SR}
\end{figure}

\begin{figure}[htb]
\centering
\includegraphics[scale=0.55]{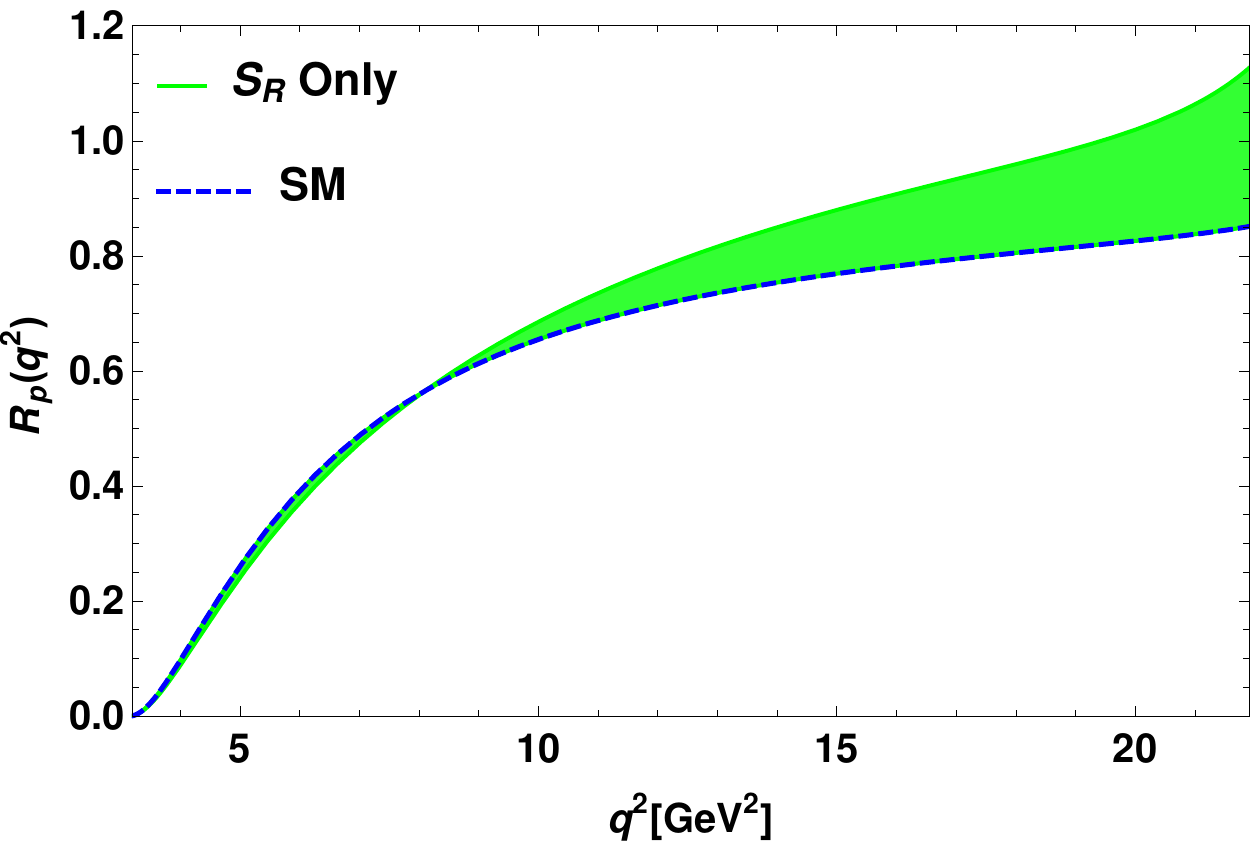}
\quad
\includegraphics[scale=0.55]{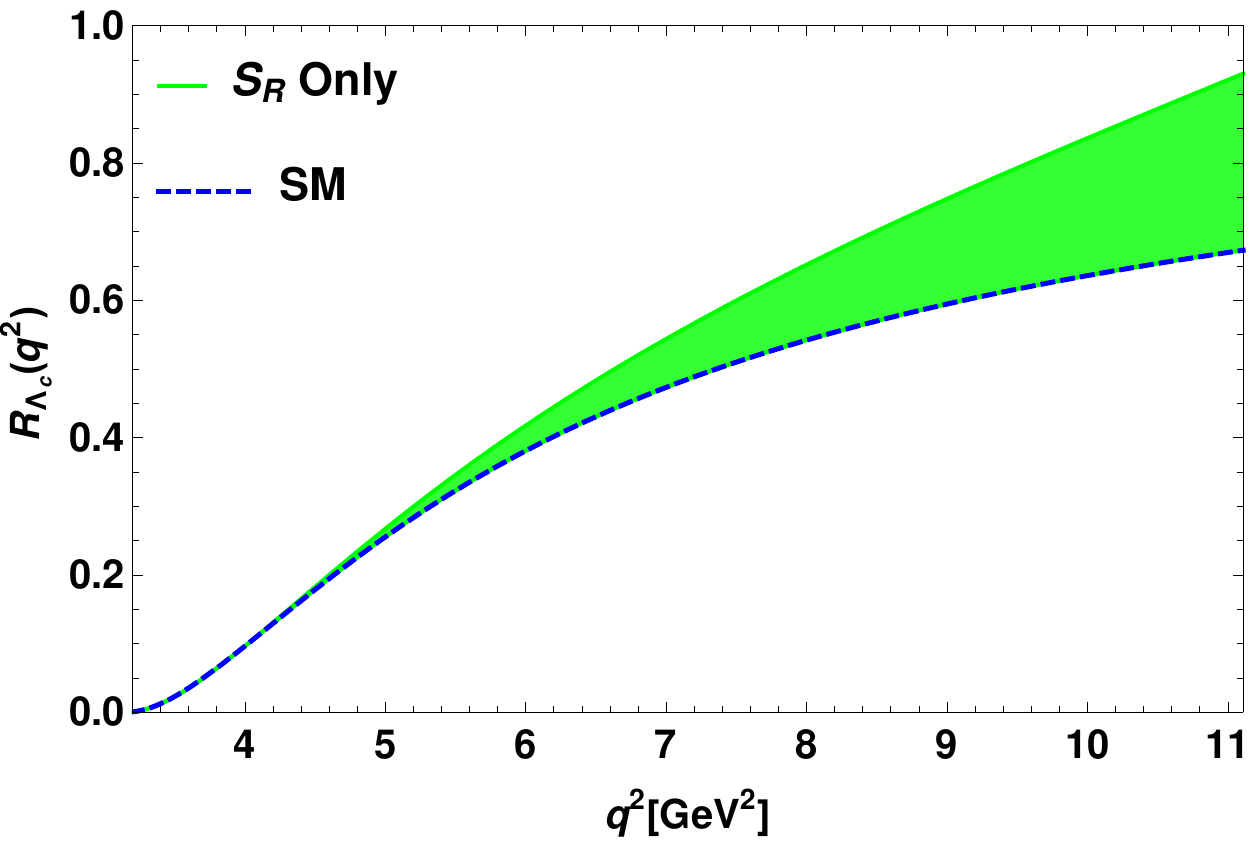}
\caption{The  variation of $R_p$ (left panel) and  $R_{\Lambda_c}$ (right panel) with respect to $q^2$  in the presence of only $S_R$ coefficient.} \label{LNU-SR}
\end{figure}

\begin{figure}[h]
\centering
\includegraphics[scale=0.55]{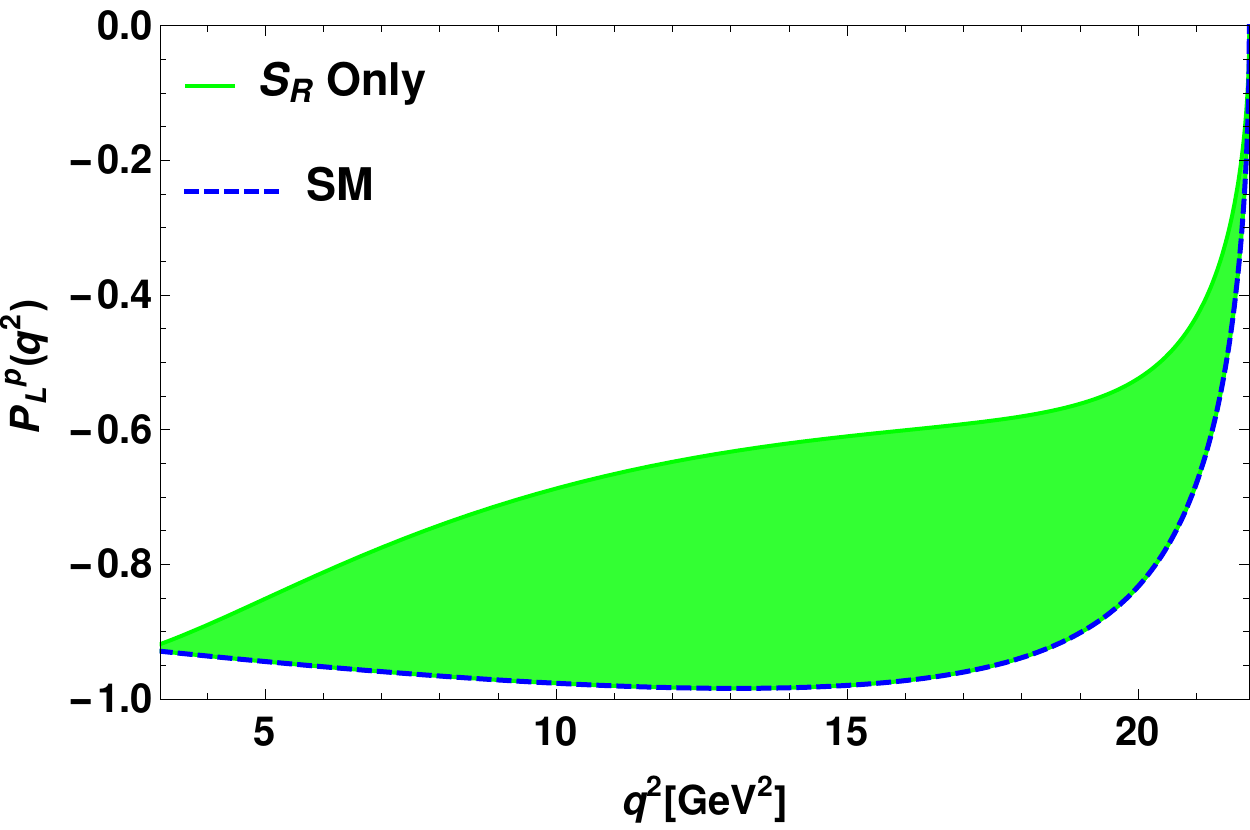}
\quad
\includegraphics[scale=0.55]{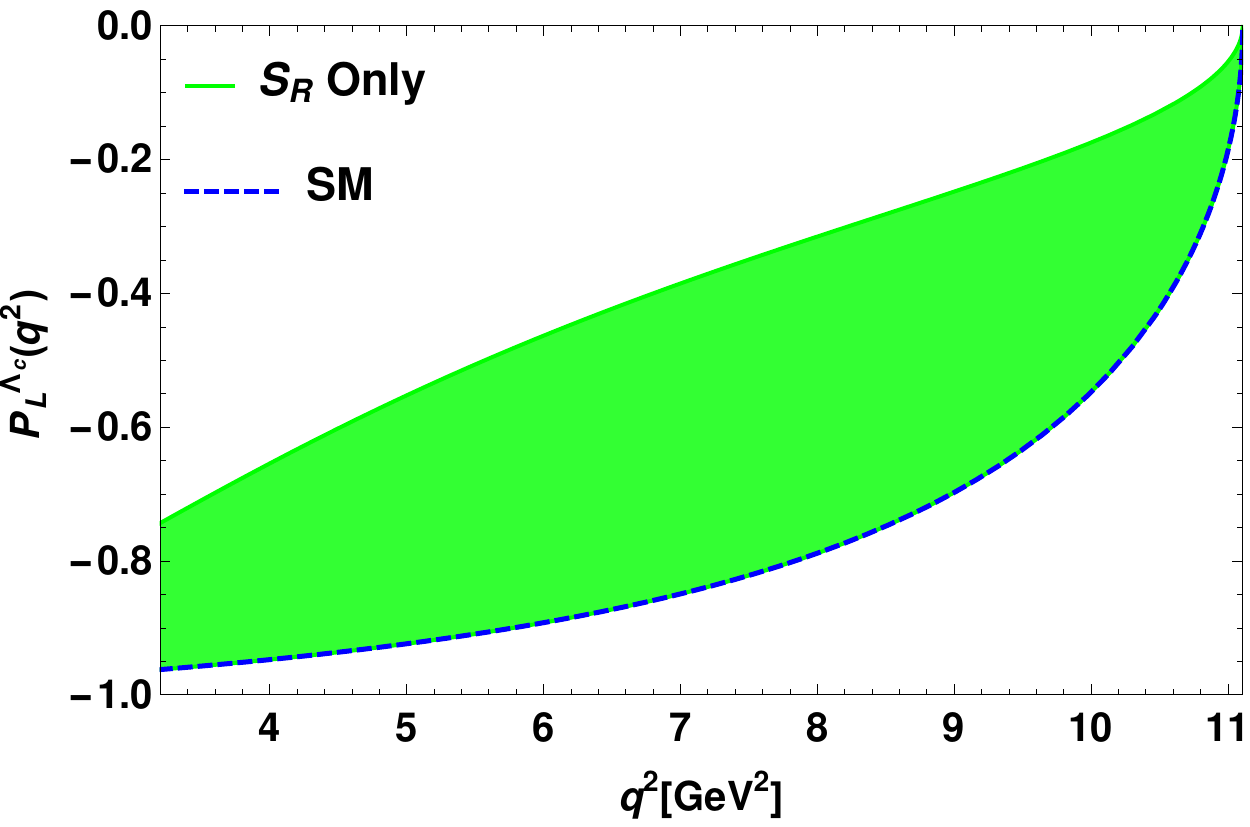}\\
\includegraphics[scale=0.55]{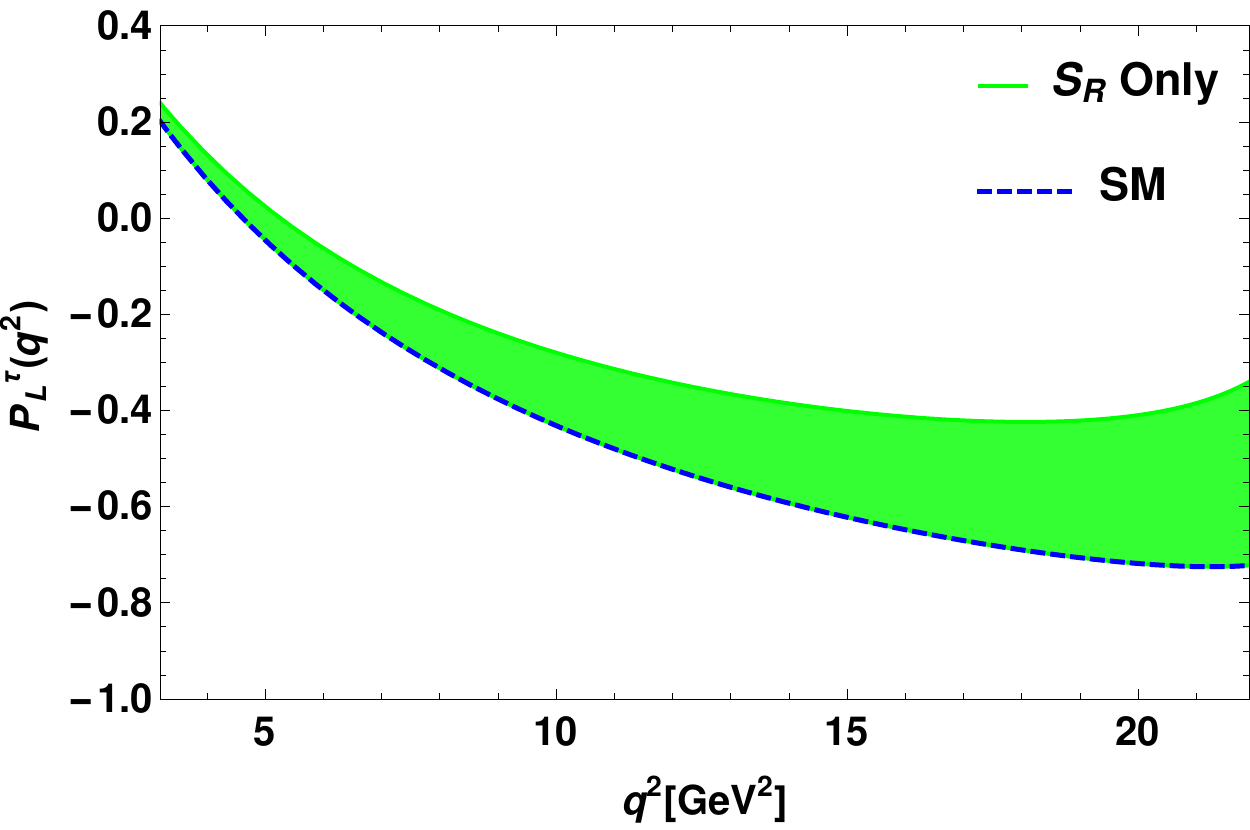}
\quad
\includegraphics[scale=0.55]{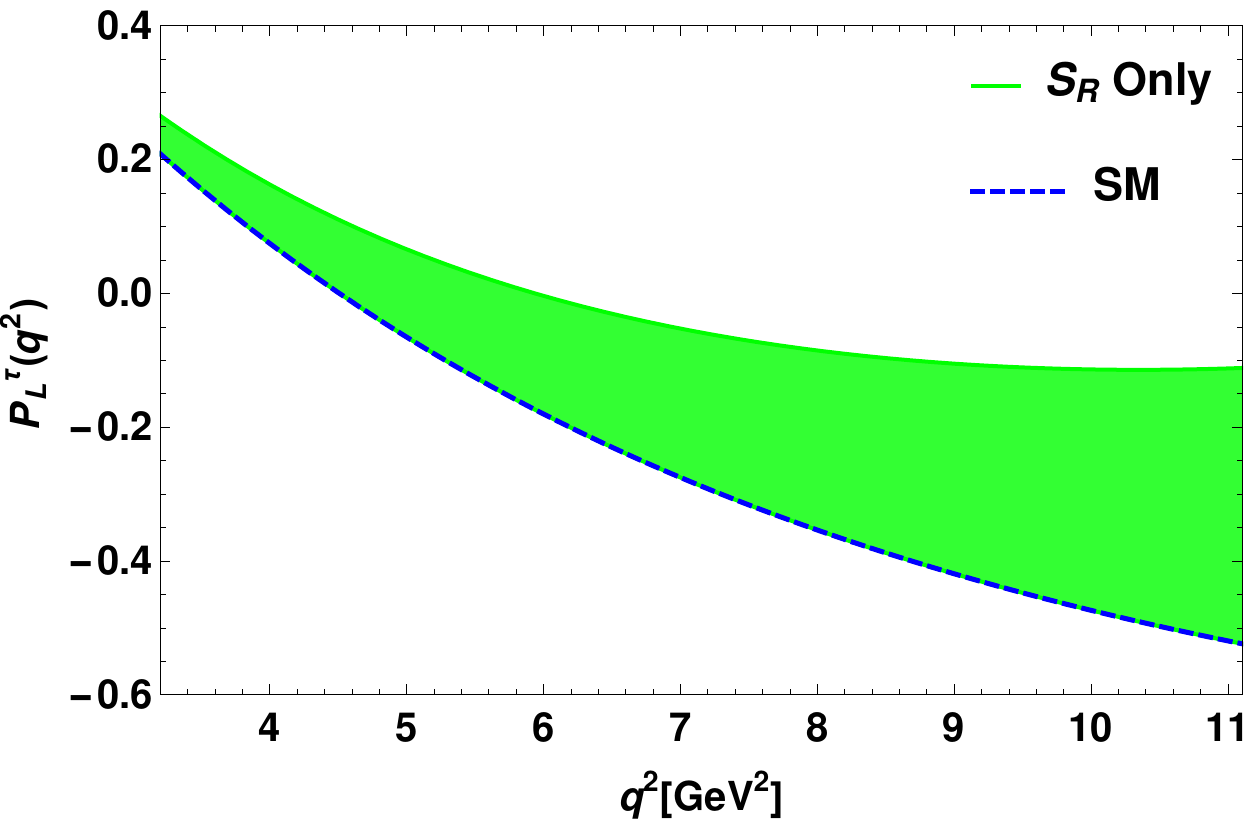}\\
\caption{The plots in the left panel represent the longitudinal  polarizations of daughter light baryon $p$ (left-top panel) and the charged $\tau$ lepton (left-bottom)     with respect to $q^2$ for only $S_R$ coefficient. The corresponding   plots for $\Lambda_b \rightarrow \Lambda_c \tau^- \bar \nu_\tau$ mode are shown in the right panel.} \label{hp-SR}
\end{figure}

In this subsection, we perform an analysis for  semileptonic decay modes of $\Lambda_b$ baryon  with the additional $S_R$ coupling. Using the  allowed ranges of the real and imaginary part of $S_R$ coupling from Table \ref{Tab:con}\,, the branching ratios of $\Lambda_b \to p \tau \bar \nu_\tau$ (left) and $\Lambda_b \to \Lambda_c \tau \bar \nu_\tau$ (right) decay processes with respect to $q^2$  are presented in Fig. \ref{Br-SR}\,.  The bottom panel of this figure represents the  $q^2$ variation of the  forward-backward asymmetry for $\Lambda_b \to p \tau \bar \nu_\tau$ (left) and $\Lambda_b \to \Lambda_c \tau \bar \nu_\tau$ (right).  In these figures, the green bands are due to  the additional new contribution of $S_R$ coefficient to the SM.  We observe profound deviation in the  branching ratios and forward-backward asymmetries of these decay modes from their SM values. Left (right) panel of Fig. \ref{LNU-SR}, show the effect of $S_R$ coupling on the  $q^2$ variation of $R_p$ ($R_{\Lambda_c}$)  parameter.  The longitudinal polarization components of the $p$ (top-left panel) and $\Lambda_c$ (top-right panel) daughter baryons with respect to $q^2$ in the presence of  contribution from only $S_R$ coefficient,  are presented in the top panel of Fig. \ref{hp-SR}\, and the bottom panel depict the longitudinal lepton polarization asymmetry parameters for $\Lambda_b \to p (\Lambda_c) \tau \bar \nu_\tau$ processes. We notice significant deviation of hadron and lepton polarization asymmetries from their corresponding SM values due to additional contribution from $S_R$  coupling.  The plot for the  $R_{\Lambda_c p}^\tau$ parameter with $q^2$ in  the presence of only $S_R$ coefficient is presented in the right panel of Fig. \ref{LNU-SLSR}\,. The numerical values of all these parameters are presented in Table \ref{SLSR:Tab}\,. Since the  convexity parameters are independent of scalar type couplings, the $S_{L,R}$ coefficients play no role for this parameter. 

\subsection{Scenario E: Only $T_L$ coefficient}
\begin{figure}[htb] 
\centering
\includegraphics[scale=0.55]{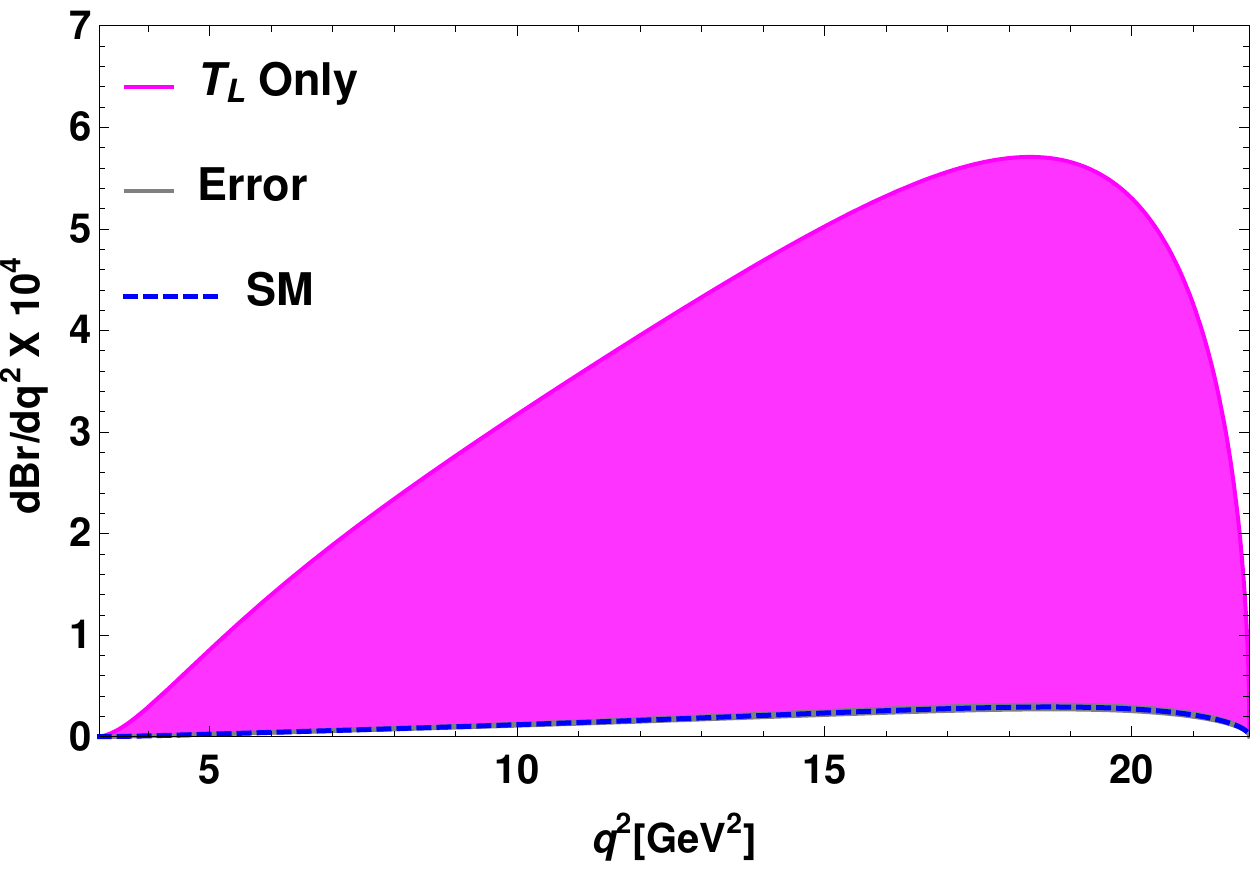}
\quad
\includegraphics[scale=0.55]{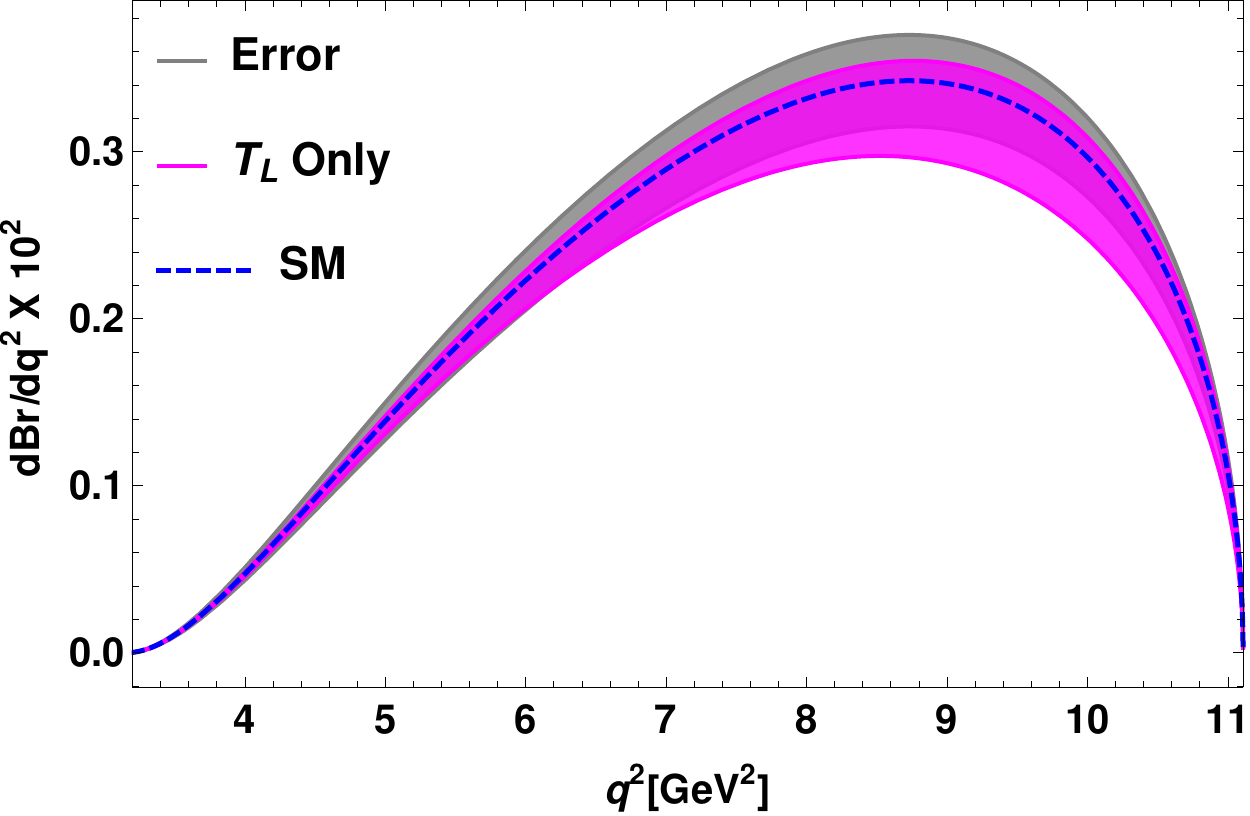}
\quad
\includegraphics[scale=0.55]{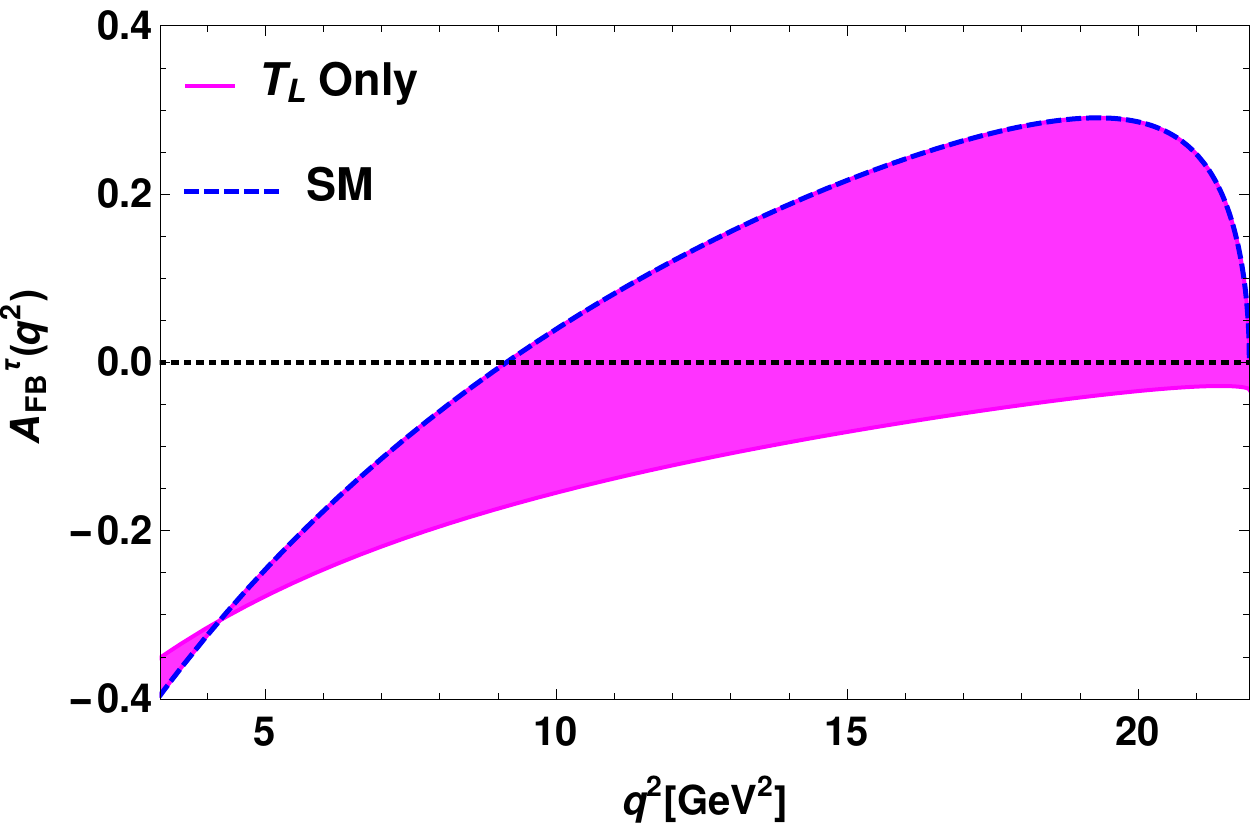}
\quad
\includegraphics[scale=0.55]{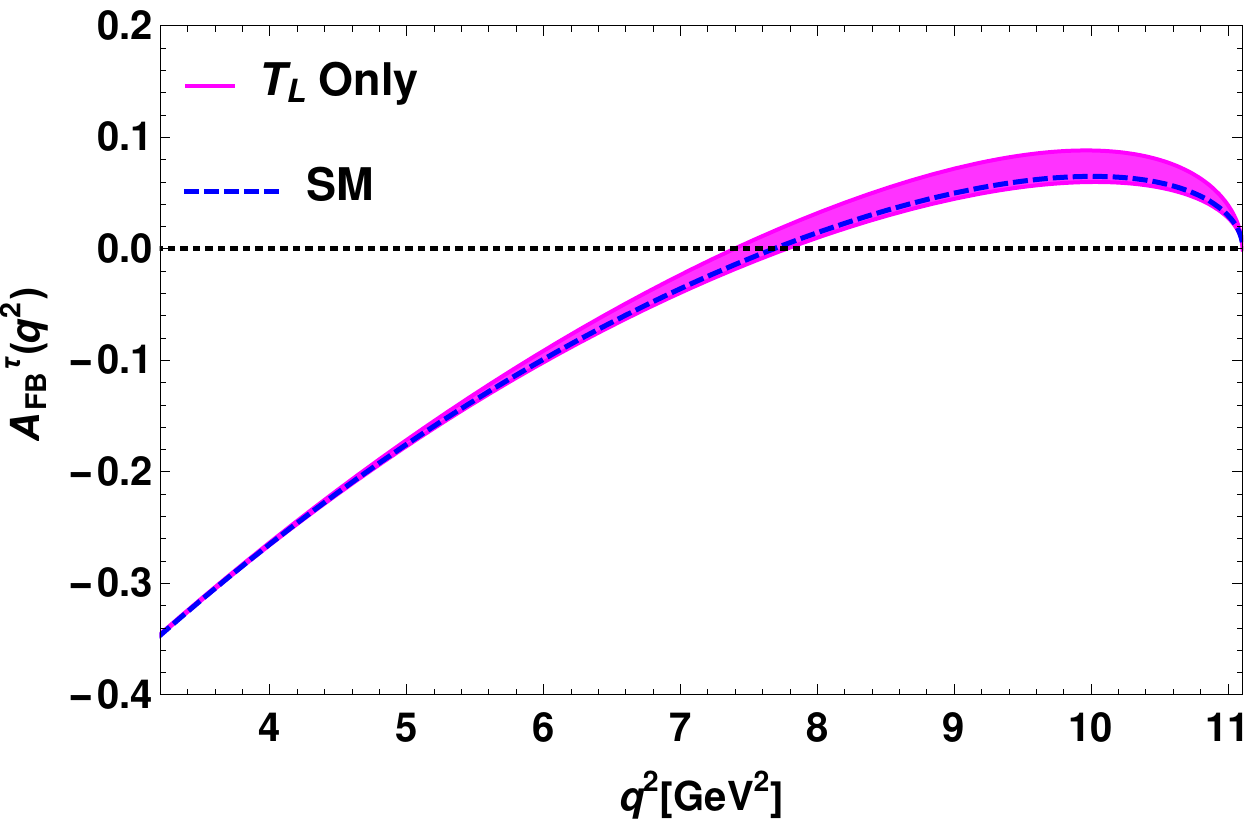}
\quad
\includegraphics[scale=0.55]{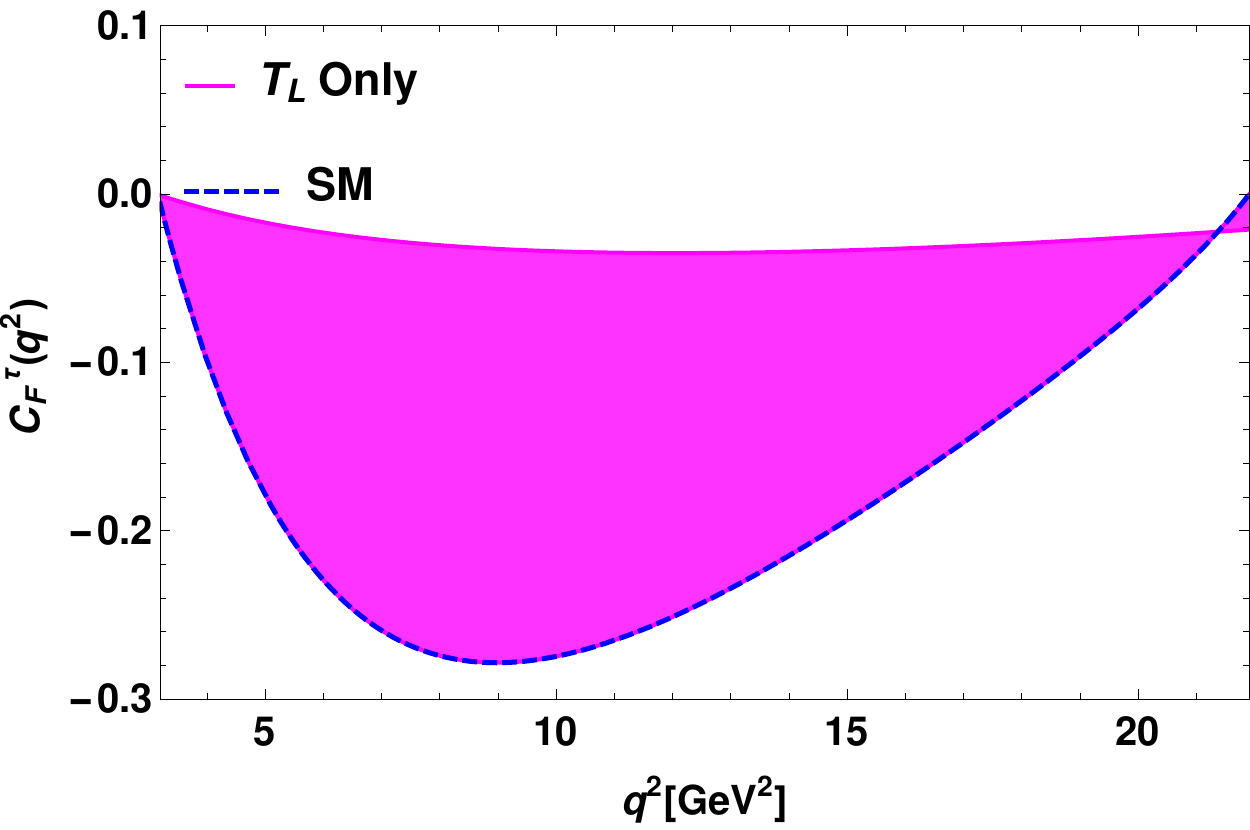}
\quad
\includegraphics[scale=0.55]{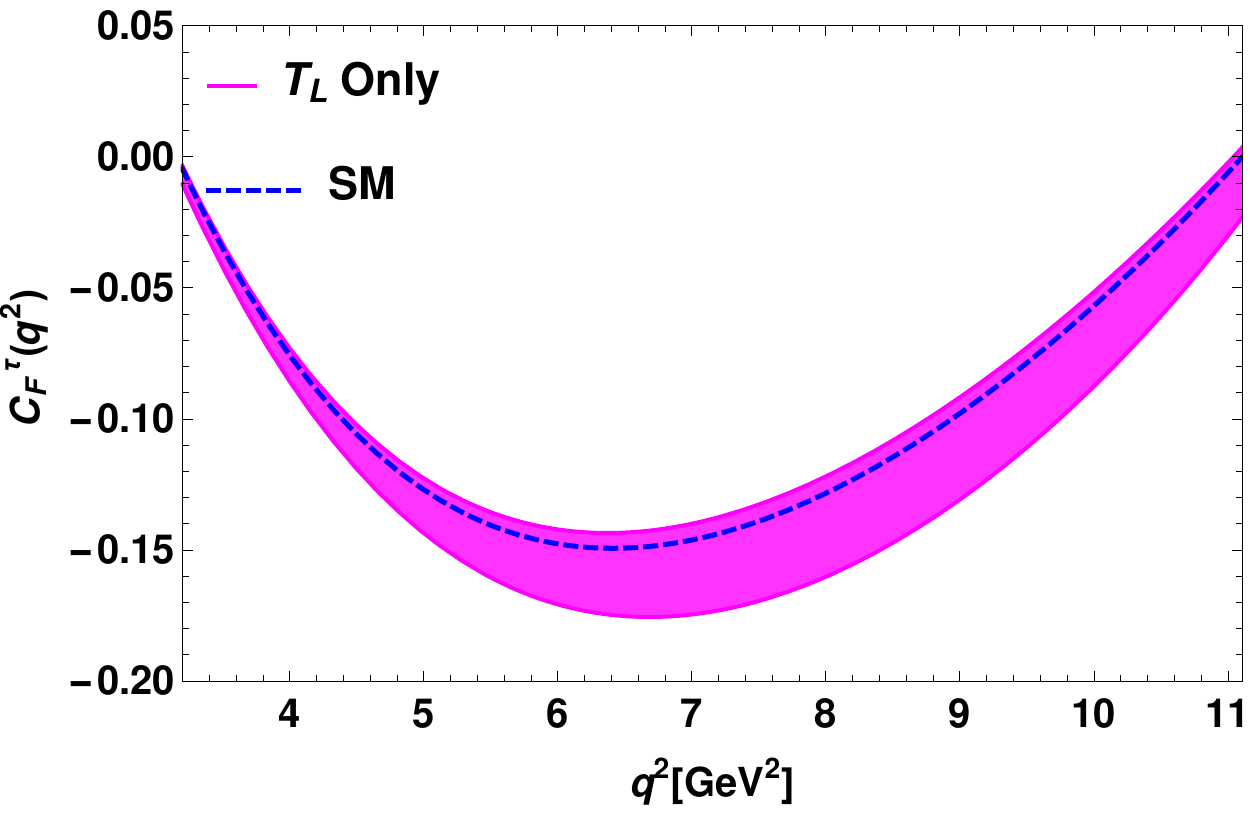}
\caption{Top panel represents the $q^2$ variation of branching ratio of $\Lambda_b \to p \tau^-  \bar \nu_\tau$ (left panel) and $\Lambda_b \to \Lambda_c^+ \tau^-  \bar \nu_\tau$ (right panel) for only $T_L$ new coefficient. The corresponding plots of forward backward asymmetry and the convexity parameters are shown in the middle and bottom panels respectively. Here magenta bands are due to the additional new physics contribution coming from only $T_L$ coefficient. }\label{Br-TL}
\end{figure}

\begin{figure}[htb]
\centering
\includegraphics[scale=0.55]{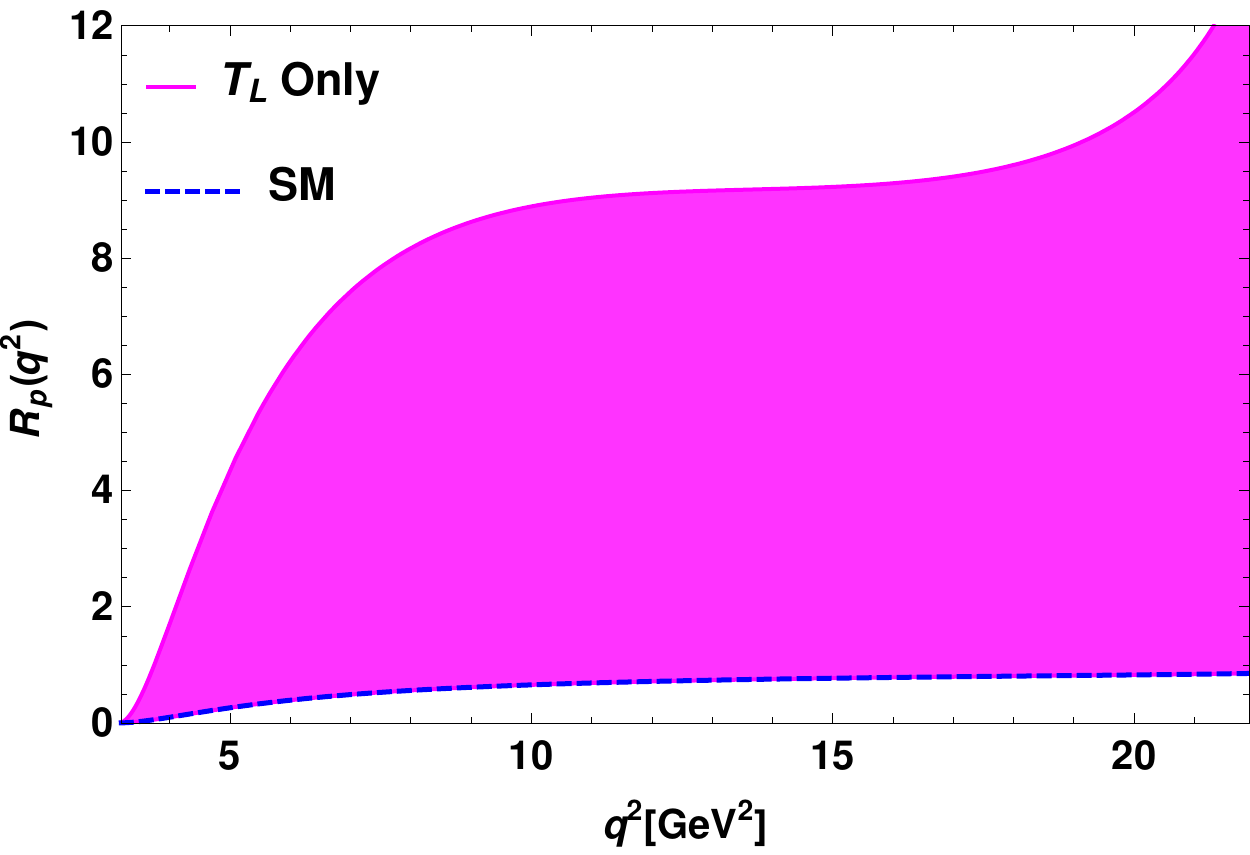}
\quad
\includegraphics[scale=0.55]{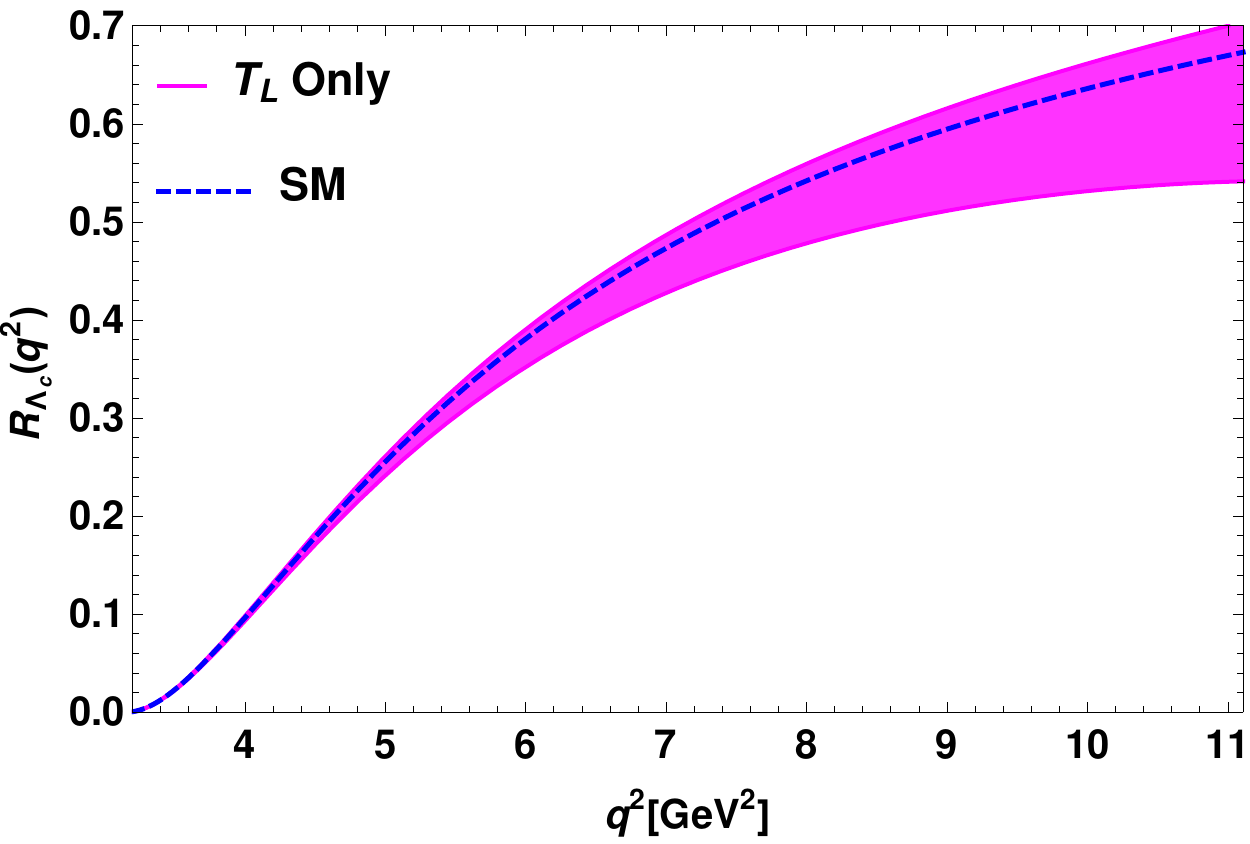}
\caption{The  variation of $R_p$ (left panel) and  $R_{\Lambda_c}$ (right panel) with respect to $q^2$  in the presence of only $T_L$ coefficient.}\label{LNU-TL} 
\end{figure}

\begin{figure}[h]
\centering
\includegraphics[scale=0.55]{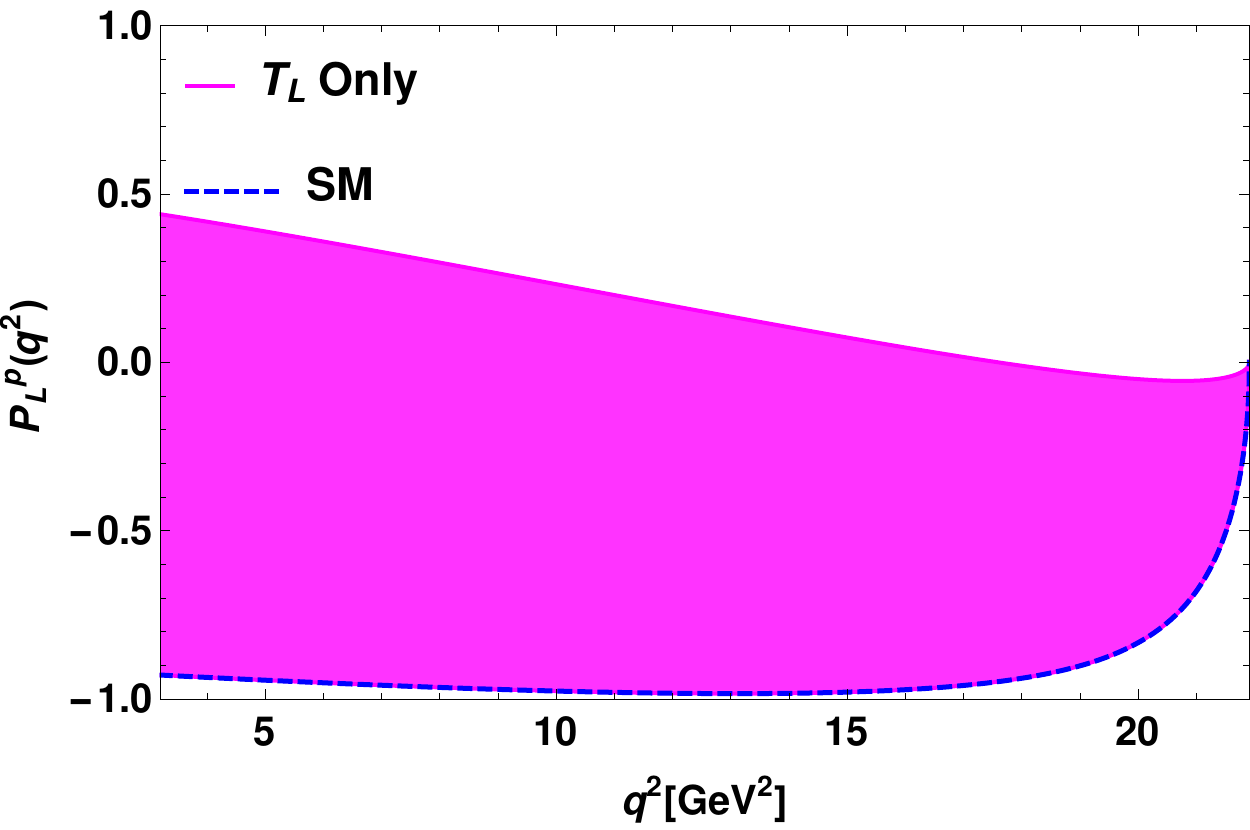}
\quad
\includegraphics[scale=0.55]{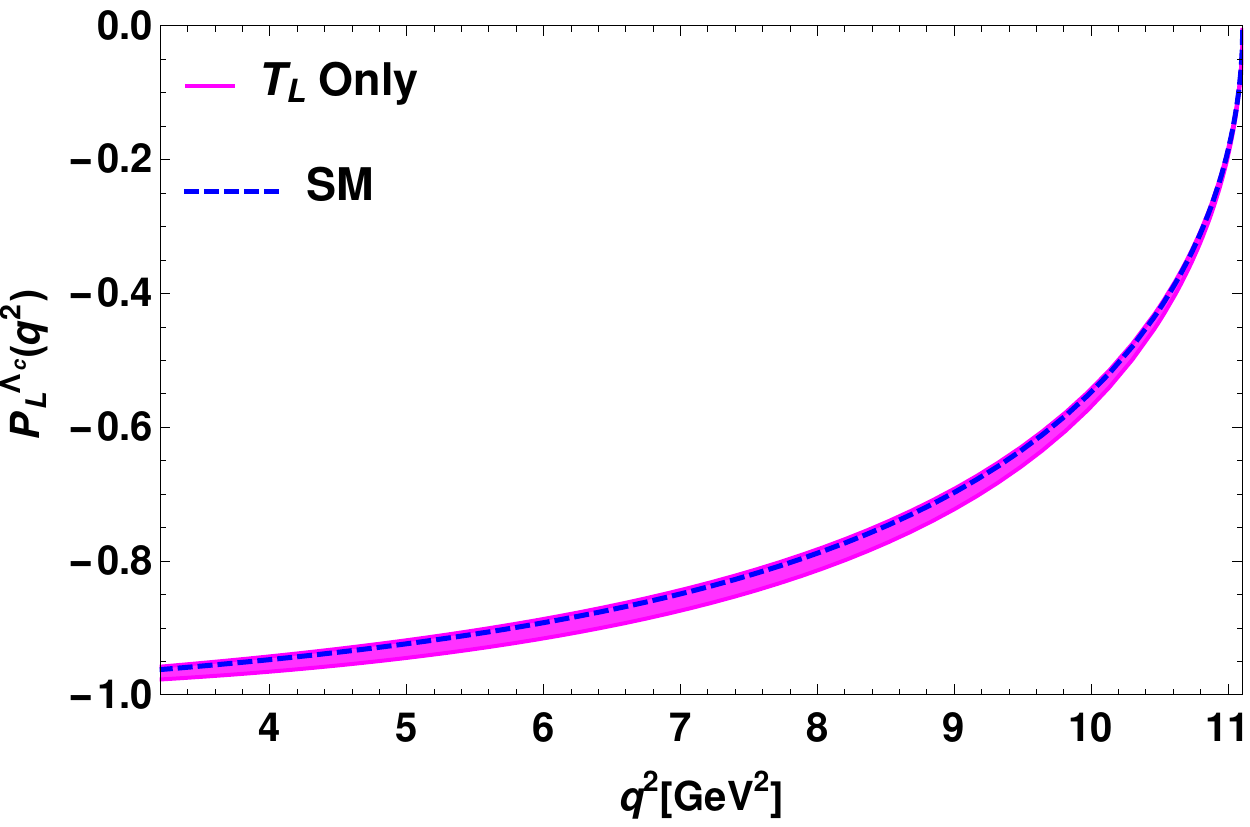}\\
\includegraphics[scale=0.55]{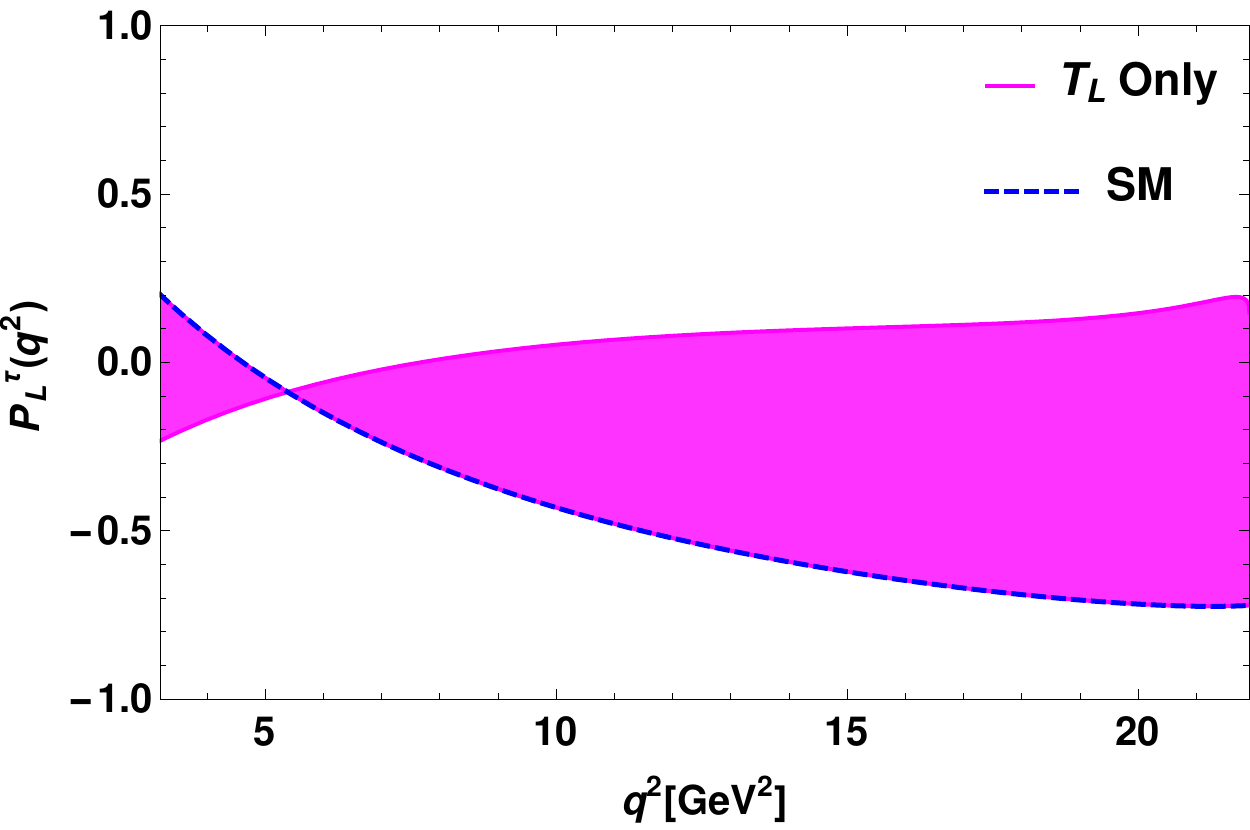}
\quad
\includegraphics[scale=0.55]{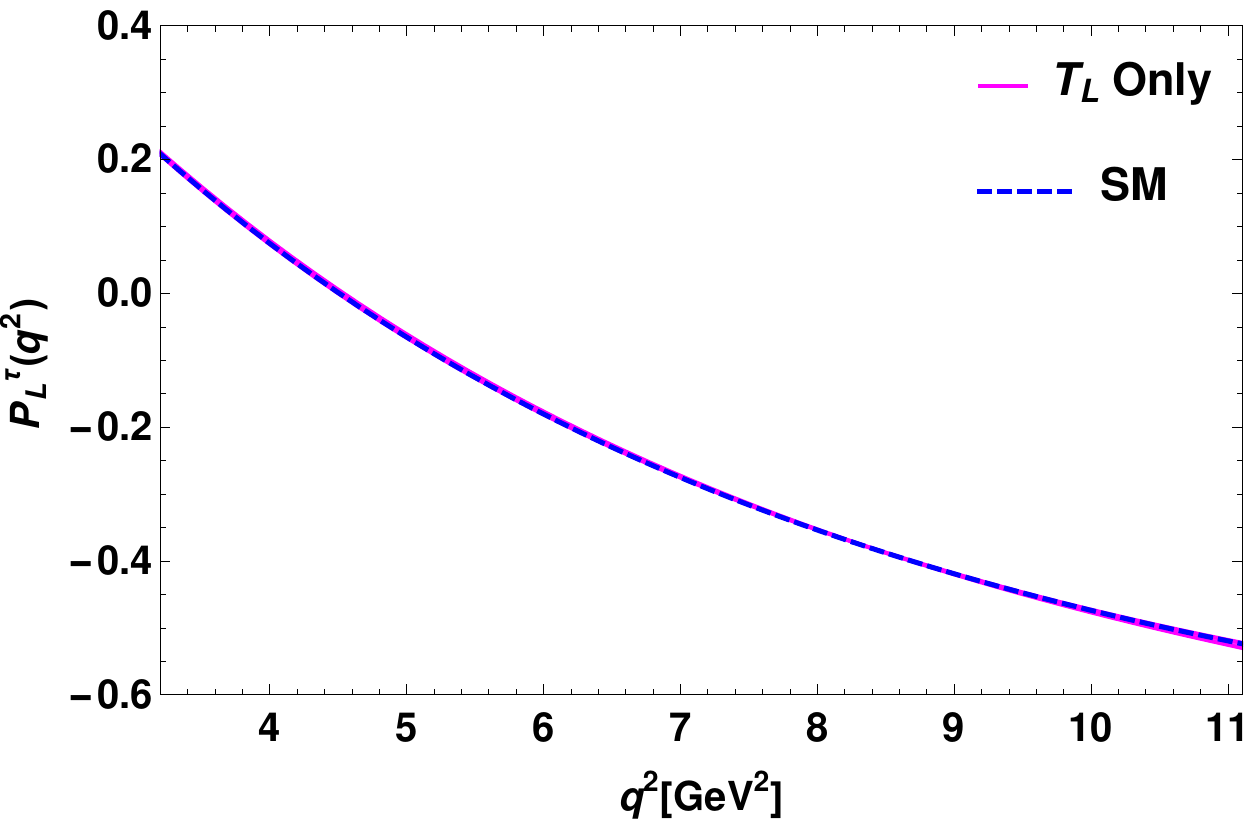}\\
\caption{The plots in the left panel represent the longitudinal  polarizations of daughter light baryon $p$ (left-top panel) and the charged $\tau$ lepton (left-bottom)     with respect to $q^2$ for only $T_L$ coefficient. The corresponding   plots for $\Lambda_b \rightarrow \Lambda_c \tau^- \bar \nu_\tau$ mode are shown in the right panel.} \label{hp-TL}
\end{figure}

\begin{figure}[htb]
\centering
\includegraphics[scale=0.43]{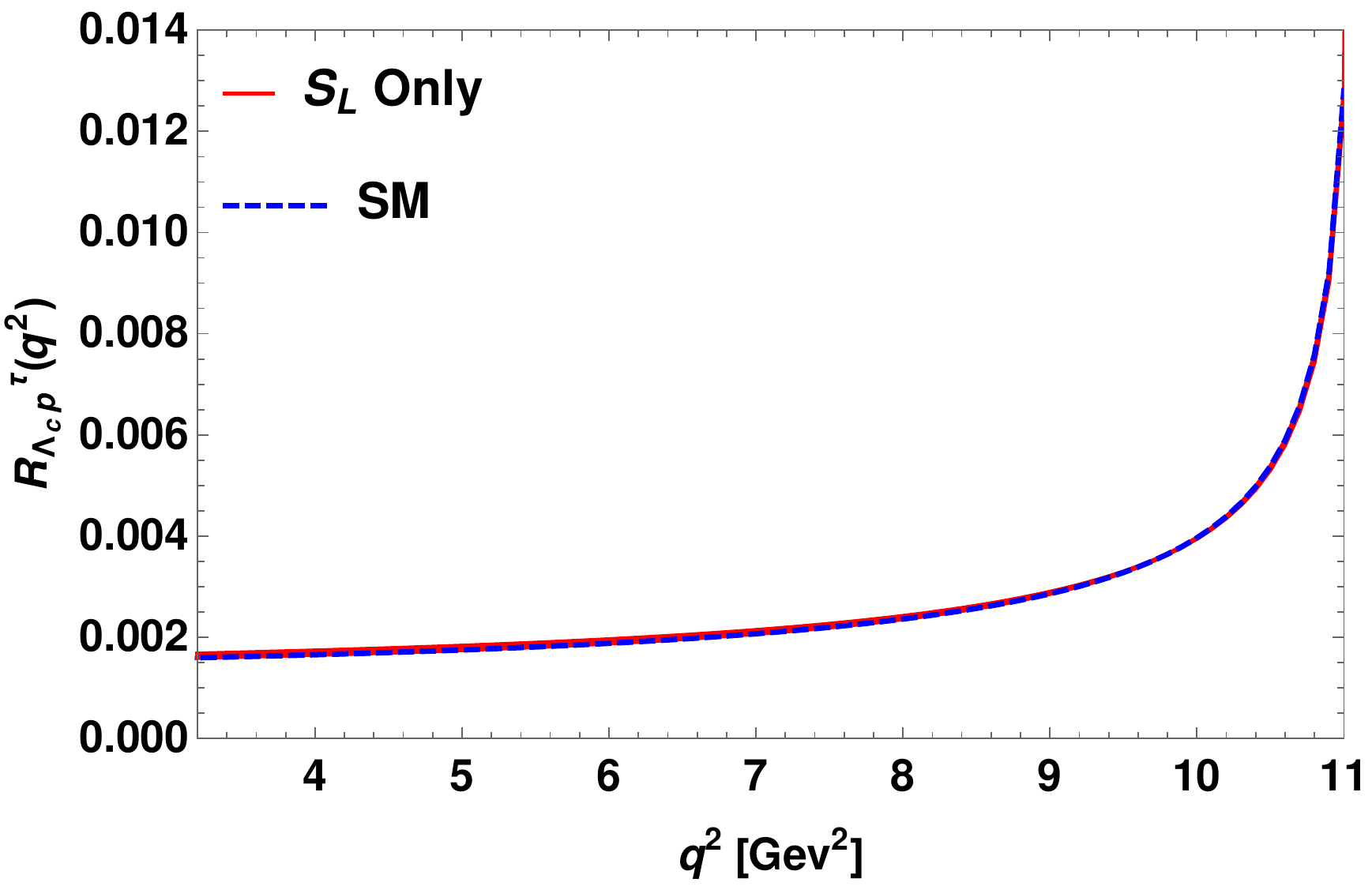}
\quad
\includegraphics[scale=0.43]{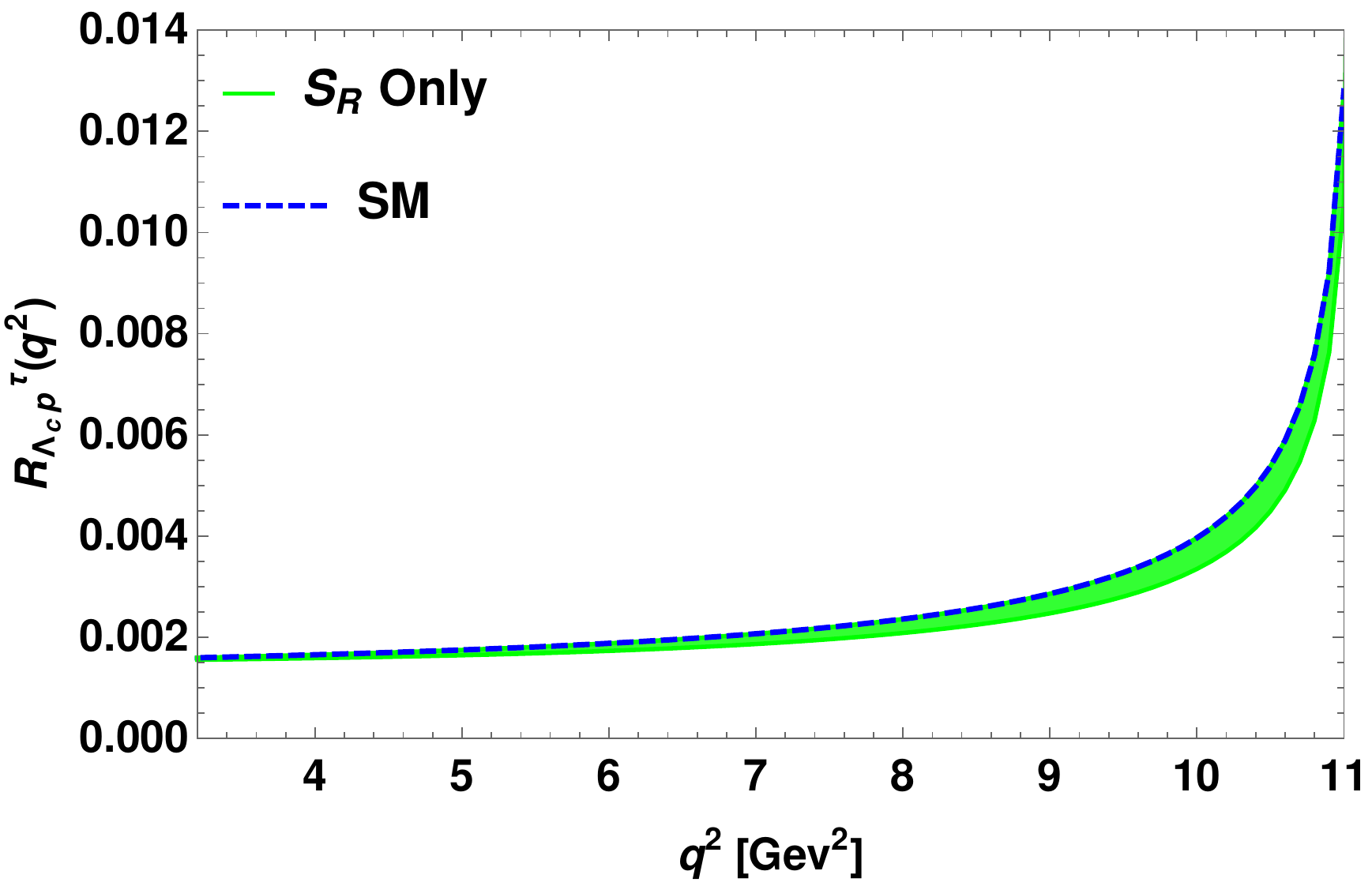}
\quad
\includegraphics[scale=0.43]{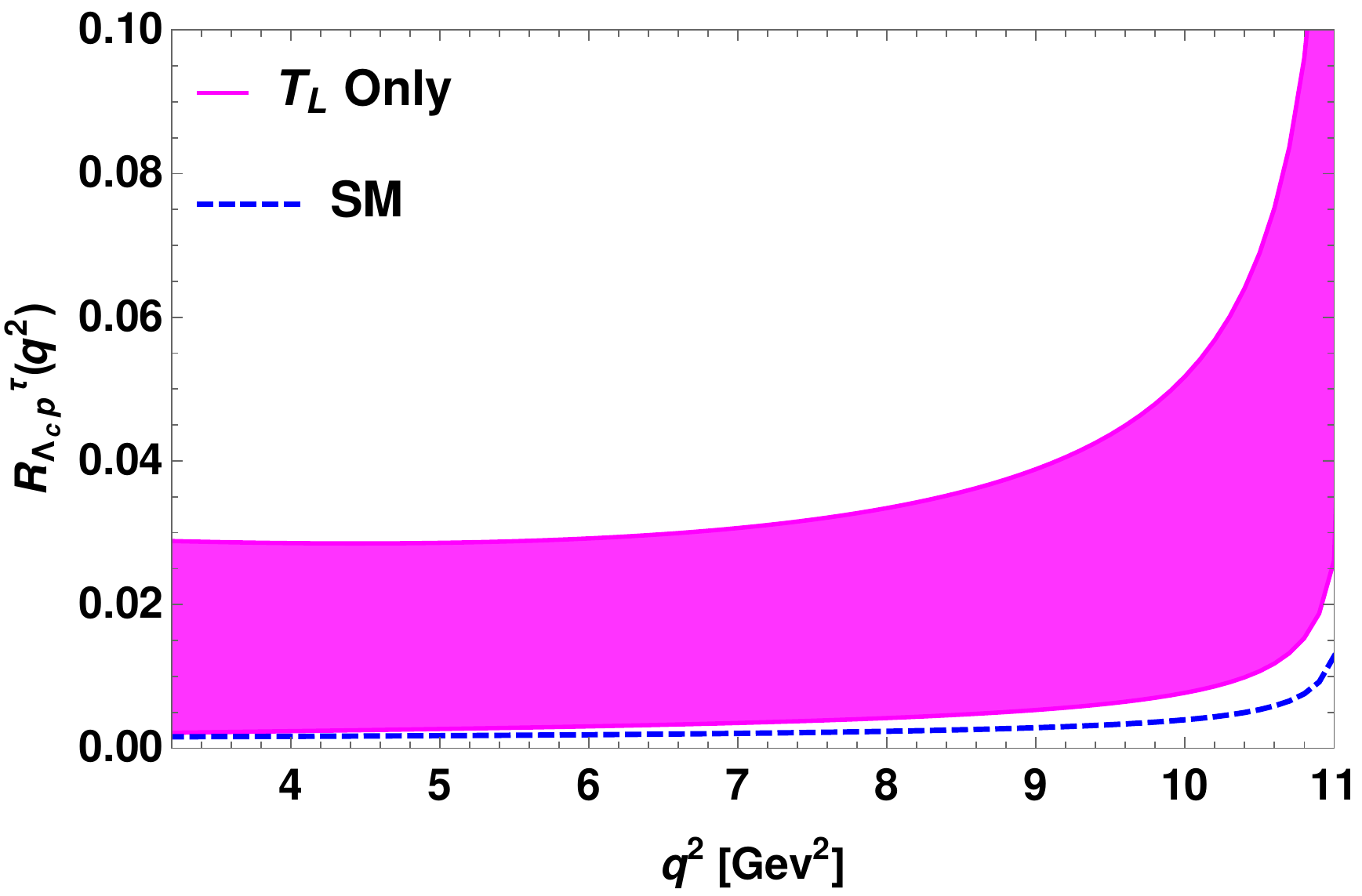}
\caption{The  variation of  $R_{\Lambda_c p}^\tau$ parameter with respect to $q^2$ in the presence of only $S_L$ (top-left panel), $S_R$ (top-right panel) and $T_L$ (bottom panel)  coefficients.} \label{LNU-SLSR}
\end{figure}

 The sensitivity of tensor coupling on various physical observables associated with semileptonic baryonic $b \to (c,u)\tau \bar \nu_\tau$  decay processes will be investigated in this subsection. The allowed region of real and imaginary parts of the tensor coupling are presented in section III.  Using all the input parameters and the constrained new tensor coefficient, we show the $q^2$ variation of branching ratio (left-top panel), forward-backward asymmetry (left-middle panel) and convexity parameter (left-bottom panel) of  $\Lambda_b \to p \tau \bar \nu_\tau$ decay mode in the left panel of Fig. \ref{Br-TL}\,. The right panel of this figure represents the corresponding  plots for  $\Lambda_b \to \Lambda_c \tau \bar \nu_\tau$ process. Here the magenta bands represent the additional contribution coming  from the new $T_L$ coefficient. For $\Lambda_b \to p \tau \bar \nu_\tau$ process,  as the bound on $T_L$ is weak, the branching ratio, forward-backward asymmetry and the convexity parameter deviate significantly from their SM predications compared to the observables for $\Lambda_b \to \Lambda_c \tau \bar \nu_\tau$ process. 
 For $\Lambda_b \to \Lambda_c \tau \bar \nu_\tau$ process, the deviations are quite minimal as the coefficient $T_L$ is severely constrained.  In the presence of $T_L$ coefficient, the numerical values of the convexity parameters are 
\bea
\langle C_{F}^\tau \rangle |_{\Lambda_b \to p}^{T_L} =-0.017\to -0.027\;,  ~~~~
\langle C_{F}^\tau \rangle |_{\Lambda_b \to \Lambda_c}^{T_L} =-0.121\to -0.098\;.
\eea 
The  plots for the lepton nonuniversality parameter $R_p$ (left panel) and $R_{\Lambda_c}$ (right panel) are shown in Fig. \ref{LNU-TL}\,. The top panel of Fig. \ref{hp-TL} represents the hadron polarization asymmetry parameters of $\Lambda_b \to p \tau \bar \nu_\tau$ (left panel) and $\Lambda_b \to \Lambda_c \tau \bar \nu_\tau$ (right panel) process and the corresponding plots for lepton polarization asymmetries are given in the bottom panel of this figure.  We observe that, the LNU parameter, longitudinal hadron and lepton polarization asymmetries of $\Lambda_b \to p \tau \bar \nu_\tau$ process have large deviation from their SM values due to the presence of tensor coupling, whereas  negligible deviations ($R_{\Lambda_c}$ has some deviation from its SM result) are noticed for the observables of  $\Lambda_b \to \Lambda_c \tau \bar \nu_\tau$ decay mode.  The $q^2$ variation of $R_{\Lambda_c p}^\tau$ parameter is depicted in the bottom panel of Fig. \ref{LNU-SLSR}\,. Table \ref{SLSR:Tab} shows the integrated values of all these angular observables. 

\begin{table}[htb]
\centering
\caption{The predicted values of branching ratios,  forward-backward asymmetries, longitudinal hadron ad lepton polarization asymmetries and lepton non-universality parameters of $\Lambda_b \to (\Lambda_c, p) \tau \bar \nu_\tau$  processes in the SM and in the presence of only $S_{L,R}$ and $T_L$ new coefficients.} \label{SLSR:Tab}
\begin{tabular}{|c|c|c|c|}
\hline
Observables~&~Values for $S_L$ coupling~&~Values for $S_R$ coupling~&~Values for $T_L$ coupling~\\
\hline
\hline
${\rm Br}(\Lambda_b \to p \tau^- \bar \nu_\tau)$ ~&~$(2.98-5.25) \times 10^{-4}$ ~&~$(2.98-3.48)\times 10^{-4}$~&~$(0.298-6.68)\times 10^{-3}$~\\
$A_{FB}^\tau$~&~$-0.019\to 0.139$ ~&~$0.086 \to 0.177$~&~$-0.172 \to-0.125$\\
$P_L^p$~&~$-0.896 \to -0.73 $ ~&~$-0.896 \to -0.6$~&~$-0.896 \to 0.337$~\\
$P_L^\tau$~&~$-0.515 \to 0.123$ ~&~$-0.515\to -0.31$~&~$-0.515 \to 0.037$\\
$R_p$~&~$0.692-1.266$ ~&~$0.692-0.81$~&~$0.692-8.8$\\
\hline
${\rm Br}(\Lambda_b \to \Lambda_c^+ \tau^- \bar \nu_\tau)$&$(1.76-2.7)\times 10^{-2}$ &$(1.76-2.2)\times 10^{-2}$ ~&~$(1.553-1.82)\times 10^{-2}$\\
$A_{FB}^\tau$&$-0.121 \to-0.06$&$-0.786\to -0.005$~&~$-0.034\to -0.09$~\\
$P_L^{\Lambda_c}$~&~$-0.796 \to -0.725$ ~&~$-0.796 \to -0.4$~&~$-0.79 \to -0.812$\\
$P_L^\tau$~&~$-0.207 \to 0.178$ ~&~$-0.207 \to -0.0021$~&~$-0.207$\\
$R_{\Lambda_c}$~&~ $0.353-0.539$~&~$0.353-0.44$ ~&~$0.31 \to 0.364$\\
\hline
$R_{\Lambda_c p}$&$(1.693-1.95)\times 10^{-2}$&$(1.582-1.693)\times 10^{-2}$~&~$0.0192-0.367$ \\

\hline
\end{tabular}
\end{table}


%
 \section{Conclusion}
 
In this work, we have  performed a model independent analysis of   baryonic $\Lambda_b \to (\Lambda_c, p) l\bar{\nu_{l}}$ decay processes by considering the generalized effective Lagrangian in the presence of new physics. We considered the new couplings to be complex  in our analysis. In order to constrain  the new couplings, we have assumed that  only one coefficient to be  present  at a time  and constrained the new coefficients by comparing the theoretical predictions of ${\rm Br}(B_{u,c}^+ \to \tau^+ \nu_\tau)$, ${\rm Br}(B \to \pi \tau \bar \nu_\tau)$, $R_\pi^l$, $R_{D^{(*)}}$ and $R_{J/\psi}$ observables with their  measured experimental data. Using the allowed parameter space, we estimated the branching ratios, forward-backward asymmetries, convexity parameters of $\Lambda_b \to (\Lambda_c, p) l \bar \nu_\tau$ decay processes. We also investigated the longitudinal polarization components of the daughter baryon  $(p, \Lambda_c)$ and the final state charged  lepton, $\tau$. The convexity parameter  only depend on the (axial)vector and tensor type couplings and are independent of the  $S_{L, R}, T_L$ coefficients.  Inspired by the observation of lepton non-universality parameters in various $B$ meson decays, we have also scrutinized the lepton universality violating parameters $(R_p, R_{\Lambda_c}, R_{\Lambda_c p}^\tau)$ in the baryonic decay modes. 
 We  found significant deviation in the branching ratios and the $R_p,~R_{\Lambda_c}, ~R_{\Lambda_c p}$ parameters from their corresponding standard model values, in the presence of additional new vector like  coupling ($V_L$ coefficient). However, such coupling does not affect  the  convexity parameter, forward-backward asymmetries, lepton and hadron polarization asymmetries. We further, noticed profound deviation in the branching ratios and  all other angular observables of semileptonic  baryonic $b \to (u,c) \tau \bar \nu_l$ decay processes due to the additional contribution of $V_R$ coupling to the SM. 
The branching ratios,  forward-backward asymmetries, longitudinal hadron and lepton polarization asymmetry parameter and the LNU observables deviate significantly from their corresponding  standard model results in the presence of $S_{L,R}$ coefficients. These coefficients do not have significant effect on $R_{\Lambda_c p}$ parameter.  We have also computed the  branching ratio, forward-backward asymmetry, convexity parameter, hadron and lepton polarization asymmetries and LNU parameter  of $\Lambda_b \to p(\Lambda_c) \tau \bar \nu_\tau$ decay process by using the additional contribution from new tensor $(T_L)$ coupling. All the angular observables of $\Lambda_b \to p \tau \bar \nu_\tau$ process receive significant deviations from their SM values, compared   to the corresponding parameters of $\Lambda_b \to \Lambda_c \tau \bar \nu_\tau$ decay mode.  To conclude, we have explored the effect of individual complex $V_{L,R}$, $S_{L, R}$ and $T_L$ couplings on the angular observables of baryonic decays of $\Lambda_b$ baryon. We found profound deviation from the standard model results due to the presence of these new couplings.  We noticed that the  $V_R$ and $S_L$ couplings significantly  affect  all the  observables and the tensor coupling plays a vital role in the case of $\Lambda_b \to p \tau \bar \nu_\tau$ decay mode. Though there is no experimental measurement on these baryonic $b \to (u,c) \tau \bar \nu_\tau$  decay processes, the study of these modes  are found to be very crucial  in order to shed light on the nature of new physics.
\acknowledgements

RM  would like to thank Science and Engineering Research Board (SERB), Government of India for financial support through grant No. SB/S2/HEP-017/2013. AR   acknowledges  University Grants Commission for financial support.

\appendix
\section{Helicity-dependent differential decay rates}
The expressions for the helicity-dependent differential decay rates required to analyze the longitudinal hadron and lepton  polarization asymmetries are given by \cite{Li:2016pdv} 
\bea
\frac{{\rm d}\Gamma^{\lambda_2=1/2}}{{\rm d}q^2}&=&\frac{m_l^2}{q^2}\Big[\frac{4}{3}\big(H_{\frac{1}{2},+}^2+H_{\frac{1}{2},0}^2+
3H_{\frac{1}{2},t}^2\big)+\frac{2}{3}\big(H^{T^2}_{\frac{1}{2},+,-}+H^{T^2}_{\frac{1}{2},0,t}
+H^{T^2}_{\frac{1}{2},+,0}+H^{T^2}_{\frac{1}{2},+,t}\nn\\&&+2H^T_{\frac{1}{2},+,-}H^T_{\frac{1}{2},0,t}+2H^T_{\frac{1}{2},+,0}H^T_{\frac{1}{2},+,t}\big)\Big]
+\frac{8}{3}\big(H_{\frac{1}{2},0}^2+H_{\frac{1}{2},+}^2\big)+4 H^{SP^2}_{\frac{1}{2},0}\nn\\&&+\frac{1}{3}\big(H^{T^2}_{\frac{1}{2},+,-}+H^{T^2}_{\frac{1}{2},0,t}+H^{T^2}_{\frac{1}{2},+,0}
+H^{T^2}_{\frac{1}{2},+,t}+2H^T_{\frac{1}{2},+,-}H^T_{\frac{1}{2},0,t}+2H^T_{\frac{1}{2},+,0}H^T_{\frac{1}{2},+,t}
\big)\nn\\&&
+\frac{4m_l}{\sqrt{q^2}}\Big[\big(H_{\frac{1}{2},0}H^T_{\frac{1}{2},+,-}
+H_{\frac{1}{2},0}H^T_{\frac{1}{2},0,t}+H_{\frac{1}{2},+}H^T_{\frac{1}{2},+,0}
+H_{\frac{1}{2},+}H^T_{\frac{1}{2},+,t}\big)\nn\\&&+2\big(H_{\frac{1}{2},t}H^{SP}_{\frac{1}{2},0}\big)\Big]\,,\nn\\
\frac{{\rm d}\Gamma^{\lambda_2=-1/2}}{{\rm d}q^2}&=&\frac{m_l^2}{q^2}\Big[\frac{4}{3}\big(H_{-\frac{1}{2},-}^2\!+\!H_{-\frac{1}{2},0}^2+
3H_{-\frac{1}{2},t}^2\big)\!+\!\frac{2}{3}\big(H^{T^2}_{-\frac{1}{2},+,-}\!+\!H^{T^2}_{-\frac{1}{2},0,t}\!
+\!H^{T^2}_{-\frac{1}{2},0,-}\!+\!H^{T^2}_{-\frac{1}{2},-,t}\nn\\&&+2H^T_{-\frac{1}{2},+,-}H^T_{-\frac{1}{2},0,t}+2H^T_{-\frac{1}{2},0,-}H^T_{-\frac{1}{2},-,t}\big)\Big]
+\frac{8}{3}\big(H_{-\frac{1}{2},-}^2+H_{-\frac{1}{2},0}^2\big)+4H^{SP^2}_{-\frac{1}{2},0}\nn\\
&&+\frac{1}{3}\big(H^{T^2}_{-\frac{1}{2},+,-}\!+\!H^{T^2}_{-\frac{1}{2},0,t}\!+\!
H^{T^2}_{-\frac{1}{2},0,-}\!+\!H^{T^2}_{-\frac{1}{2},-,t}\!
+\!2H^T_{-\frac{1}{2},+,-}H^T_{-\frac{1}{2},0,t}\!+\!2H^T_{-\frac{1}{2},0,-}H^T_{-\frac{1}{2},-,t}\big)\nn\\&&
+\frac{4m_l}{\sqrt{q^2}}\Big[\big(H_{-\frac{1}{2},0}H^T_{-\frac{1}{2},+,-}
+H_{-\frac{1}{2},0}H^T_{-\frac{1}{2},0,t}+H_{-\frac{1}{2},-}H^T_{-\frac{1}{2},0,-}
+H_{-\frac{1}{2},-}H^T_{-\frac{1}{2},-,t}\big)\nn\\&&+2\big(H_{-\frac{1}{2},t}H^{SP}_{-\frac{1}{2},0}\big)\Big]\,,\nn \\
\frac{{\rm d}\Gamma^{\lambda_\tau=1/2}}{{\rm d}q^2}&=&\frac{m_l^2}{q^2}\Big[\frac{4}{3}\big(H_{\frac{1}{2},+}^2\!+\!H_{\frac{1}{2},0}^2\!+\!
H_{-\frac{1}{2},-}^2\!+\!H_{-\frac{1}{2},0}^2\big)\!+\!4\big(H_{\frac{1}{2},t}^2\!+\!H_{-\frac{1}{2},t}^2\big)\Big]
+4\big(H^{SP^2}_{\frac{1}{2},0}+H^{SP^2}_{-\frac{1}{2},0}\big)\nn\\&&+\frac{1}{3}\Big[H^{T^2}_{\frac{1}{2},+,-}\!+\!H^{T^2}_{\frac{1}{2},0,t}+H^{T^2}_{\frac{1}{2},+,0}+
H^{T^2}_{\frac{1}{2},+,t}+H^{T^2}_{-\frac{1}{2},+,-}+H^{T^2}_{-\frac{1}{2},0,t}+H^{T^2}_{-\frac{1}{2},0,-}\!+
H^{T^2}_{-\frac{1}{2},-,t}\nn\\
&&+2\big(H^T_{\frac{1}{2},+,-}H^T_{\frac{1}{2},0,t}+H^T_{\frac{1}{2},+,0}H^T_{\frac{1}{2},+,t}+
H^T_{-\frac{1}{2},+,-}H^T_{-\frac{1}{2},0,t}+H^T_{-\frac{1}{2},0,-}H^T_{-\frac{1}{2},-,t}\big)\Big]\nn\\&&+\frac{4m_l}{3\sqrt{q^2}}\Big[6\big(H_{\frac{1}{2},t}H^{SP}_{\frac{1}{2},0}+H_{-\frac{1}{2},t}
H^{SP}_{-\frac{1}{2},0}\big)+\big(H_{\frac{1}{2},0}H^T_{\frac{1}{2},+,-}+
H_{\frac{1}{2},0}H^T_{\frac{1}{2},0,t}\nn\\
&&+H_{\frac{1}{2},+}H^T_{\frac{1}{2},+,0}+H_{\frac{1}{2},+}H^T_{\frac{1}{2},+,t}+
H_{-\frac{1}{2},0}H^T_{-\frac{1}{2},+,-}+H_{-\frac{1}{2},0}H^T_{-\frac{1}{2},0,t}+
H_{-\frac{1}{2},-}H^T_{-\frac{1}{2},0,-}\nn\\
&&+H_{-\frac{1}{2},-}H^T_{-\frac{1}{2},-,t}\big)\Big]\,,\nn\\
\frac{{\rm d}\Gamma^{\lambda_\tau=-1/2}}{{\rm d}q^2}&=&\frac{8}{3}\big(H_{\frac{1}{2},+}^2\!+\!H_{\frac{1}{2},0}^2\!+\!H_{-\frac{1}{2},-}^2\!+
\!H_{-\frac{1}{2},0}^2\big)+\!\frac{2m_l^2}{3q^2}\Big[H^{T^2}_{\frac{1}{2},+,-}\!+\!H^{T^2}_{\frac{1}{2},0,t}\!+
\!H^{T^2}_{\frac{1}{2},+,0}\!+\!H^{T^2}_{\frac{1}{2},+,t}\!\nn\\
&&+\!H^{T^2}_{-\frac{1}{2},+,-}\!+\!H^{T^2}_{-\frac{1}{2},0,t}
+H^{T^2}_{-\frac{1}{2},0,-}\!+\!H^{T^2}_{-\frac{1}{2},-,t}+2\big(H^T_{\frac{1}{2},+,-}H^T_{\frac{1}{2},0,t}
+H^T_{\frac{1}{2},+,0}H^T_{\frac{1}{2},+,t}\nn\\
&&+H^T_{-\frac{1}{2},+,-}H^T_{-\frac{1}{2},0,t}+H^T_{-\frac{1}{2},0,-}
H^T_{-\frac{1}{2},-,t}\big)\Big]\nn\\
&&+\frac{8m_l}{3\sqrt{q^2}}\big(H_{\frac{1}{2},0}H^T_{\frac{1}{2},+,-}+
H_{\frac{1}{2},0}H^T_{\frac{1}{2},0,t}+H_{\frac{1}{2},+}H^T_{\frac{1}{2},+,0}+
H_{\frac{1}{2},+}H^T_{\frac{1}{2},+,t}\nn\\&&+H_{-\frac{1}{2},0}H^T_{-\frac{1}{2},+,-}+H_{-\frac{1}{2},0}H^T_{-\frac{1}{2},0,t}+
H_{-\frac{1}{2},-}H^T_{-\frac{1}{2},0,-}+H_{-\frac{1}{2},-}H^T_{-\frac{1}{2},-,t}\big)\,.
\eea 

\section{Form factors relations}
The relation betwen various form factors are given as \cite{Feldmann:2011xf, Chen:2001zc}
\bea
 f_0 &=& f_1 + \frac{q^2}{M_{\Lambda_b}-m_\Lambda} \, f_3 \,,\qquad  f_+ = f_1 - \frac{q^2}{M_{\Lambda_b}+m_\Lambda} \, f_2  \,,
\qquad  f_\perp = f_1 - (M_{\Lambda_b}+m_\Lambda) \, f_2 \,,\nn \\
g_0 &=& g_1 - \frac{q^2}{M_{\Lambda_b}+m_\Lambda} \, g_3 \,, \qquad  g_+ = g_1 + \frac{q^2}{M_{\Lambda_b}-m_\Lambda} \, g_2  \,,\qquad
 g_\perp = g_1 + (M_{\Lambda_b}-m_\Lambda) \, g_2
\,,\nn \\
 h_+ &=& f_2^T - \frac{M_{\Lambda_b}+m_\Lambda}{q^2} \, f_1^T  \,,
\qquad
 h_\perp = f_2^T - \frac{1}{M_{\Lambda_b}+m_\Lambda} \, f_1^T
\,,\nn \\
 \tilde h_+ &=& g_2^T + \frac{M_{\Lambda_b}-m_\Lambda}{q^2} \, g_1^T  \,,
\qquad
 \tilde h_\perp = g_2^T + \frac{1}{M_{\Lambda_b}-m_\Lambda} \, g_1^T
\,,
\eea
with 
\bea
f_2^T&=&f_T-f_T^S q^2,\qquad f_1^T=\left(f_T^V+f_T^S(M_{B_1}-M_{B_2}) \right)q^2, \qquad f_1^T=-\frac{q^2}{M_{B_1}-M_{B_2}} f_3^T, \nn \\
g_2^T&=&g_T-g_T^S q^2,\qquad g_1^T=\left(g_T^V+g_T^S(M_{B_1}+M_{B_2}) \right)q^2, \qquad g_1^T=\frac{q^2}{M_{B_1}+M_{B_2}} g_3^T\,.
\eea 
\bibliography{BL}

\end{document}